\documentclass[letterpaper]{usbthesis}

\usepackage{amsmath}
\usepackage{amssymb}
\usepackage{color}
\usepackage{mathrsfs}
\usepackage{graphics}
\usepackage{graphicx}
\usepackage{bm}

\allowdisplaybreaks

\author{Kristian Rabenstein}

\title{Decoherence and Quantum Measurement of
Josephson-Junction Qubits }

\month{December} \year{2005}

\director{Dmitri Averin}{SBU}

\chairman{James Lukens}{SBU }

\fstmember{Vladimir Korepin}{Professor, Department of Physics, SBU
}

\outmember{Serge Luryi}{Distinguished Professor, Department of
Electrical and Computer Engineering, SBU}

\begin{document}

\pagenumbering{roman} \maketitle \makeapproval

\begin{abstract}
\singlespacing
This work deals with two pressing issues in the design and
operation of Josephson qubits -- loss of coherence and
measurement.

The first part is devoted to the problem of decoherence in single
and double qubit systems. We present a quantitative description of
quantum coherent oscillations in the system of two coupled qubits
in the presence of weak noise that in general can be correlated
between the two qubits. It is shown that in the experimentally
realized scheme the waveform of the excited oscillations is not
very sensitive to the magnitude of decoherence correlations.
Modification of this scheme into a potentially useful probe of the
degree of decoherence correlations at the two qubits is suggested.
Furthermore, an explicit non-perturbative expression for
decoherence of quantum oscillations in a single and double qubit
systems by low-frequency noise is presented. Decoherence strength
is controlled by the noise spectral density at zero frequency
while the noise correlation time $\tau$ determines the time $t$ of
crossover from the $1/\sqrt{t}$ to the exponential suppression of
coherence. The first part is concluded by introduction of a useful
Monte-Carlo-based numerical technique for  simulations of
realistic qubit dynamics subject to general noise, time dependent
bias and tunneling, and general environment temperature.

The second part addresses the  problem of quantum measurement of
qubit through discussion of a versatile new type of the magnetic
flux detector which can be optimized to perform several
measurement schemes in quantum-limited regime. The detector is
based on manipulation of ballistic motion of individual fluxons in
a Josephson transmission line (JTL), with the output information
contained in either probabilities of fluxon
transmission/reflection, or time delay associated with the fluxon
propagation along the JTL. The equations for conditional evolution
of the measured system both in the transmission/reflection and
time-delay regimes are derived for this detector. Combination of
the quantum-limited detection with control over individual fluxons
should make the JTL detector suitable for implementation of
non-trivial quantum measurement strategies, including conditional
measurements and feedback control schemes.

\end{abstract}

\begin{acknowledgements}
It is my pleasure to thank all people who provided the foundation
and the guidance in completing this work.

Among them, first and foremost, I am grateful to my adviser
Professor Dmitri Averin without whose patience and expertise this
work would not be possible. I would also like to thank  Professor
Vasili Semenov and Viktor Sverdlov who co-authored with me on
several publications, as well to Professor James Lukens for many
useful discussions.

In the end I thank my Parents and my uncle Bruno for the
never-ending support.

\end{acknowledgements}
\begin{dedication}
\textsl{To my parents and grandparents.}
\end{dedication}
\tableofcontents 
\newpage
\listoffigures
\listoftables
 \pagenumbering{arabic} \

\chapter{Introduction}
\emph{Any algorithmic process can be simulated efficiently} (i.e.
in polynomial time) \emph{using a Turing Machine} is the
celebrated Church-Turing  thesis that for a long time instilled
the belief that the efficiency of computation is independent of
its underlying physical process \cite{Tur36,Chu36}. For this
reason, the efficiency of computer architecture built on
irreversible binary language was not questioned at first. Rather,
one of the important topics in the earlier research of
computational devices was the energy dissipation. One such study
\cite{Lan61} showed that the erasure of information is the only
source of energy dissipation. Later, it was suggested \cite{Ben73}
that the fundamental limit of energy dissipation, $k_BTln2$ per
one erased bit of information, can be avoided by employing the
computation that is both physically and logically reversible since
reversible system can always be "rolled back" to the zero state
along constant energy path.

A particular way to employ reversible computation is by using
quantum mechanics as the underlying physical process of
computation since all closed quantum systems are reversible. In
addition to the vanishing energy dissipation, a quantum system
provides a possibility of preparing a superposition of its
different states. This speeds up the information processing since
it allows operating on many states of the computational basis
simultaneously.

The advantage of this so-called quantum parallelism and the
reversibility of quantum systems was for the first time fully
realized by David Deutsch. He formulated quantum computation as a
new field of computation based on quantum rather than classical
dynamics \cite{Deu85}.  The same work featured the first quantum
algorithm that was in certain cases more efficient than any Turing
machine algorithm for the same problem.  In 1992 together with
Jozsa, Deutsch improved on this algorithm so that it was always
more efficient than any classical one \cite{Deu92}. Further
discoveries of quantum algorithms that are more efficient than the
respective classical algorithms by Simon \cite{Sim94}, Shor
\cite{Sho94}, and Grover \cite{Gro97} propelled quantum
computation into the forefront of scientific research.

The fundamental element of quantum computation is the quantum bit
or \emph{qubit} for short.  It is a two-level unitary system where
the unitarity is manifested through the presence of  the quantum
coherent oscillations. The oscillations represent periodically
evolving probability of a qubit occupying some mixture of its two
states. For this reason reconstructing previous states of the
qubit is always possible. Decoherence, i.e. the loss of coherence,
suppresses the quantum coherent oscillations and with it the
reversibility of the qubit and in the process degrades its ability
to preform the tasks needed in quantum computation \cite{Div98}.

An isolated quantum system does not exhibit decoherence.  For it
to happen, the quantum system needs to interact with other systems
in a such way that its dynamics becomes dependent on the
surrounding systems. Classical systems in such circumstances
undergo energy relaxation as described by the
fluctuation-dissipation theorem \cite{Cal51}. Quantum systems
exhibit energy relaxation and decoherence. While quantum
relaxation has its counterpart in classical dynamics, decoherence
is a purely quantum phenomenon. It is caused by quantum phase
interference or entanglement of two systems. It is not conditional
on the systems being in contact all the time, rather if entangled
in the past, the two systems generally stay entangled.

For computational purposes the qubit needs to be initialized,
coupled with other qubits, and in the end measured.  This requires
entanglement of the qubit with the other qubits and the
surroundings.  Thus to have a qubit completely isolated from the
environment is impossible. Rather, it is necessary to find ways to
design physical qubits so that they exhibit prolonged coherent
oscillations while integrated in larger structures.

The leading candidate for achieving such goals are superconducting
Josephson junctions.  They are scalable and integrable solid state
devices where Cooper pairs of electrons move in phase and behave
as a single quantum particle of much larger size than than the
elementary quantum particles. Additionally Josephson junctions are
ideal for nano-fabrication and integration into larger circuits
needed for additional tasks such as measurement.

Measurement  can be seen as an extraction of information from the
computational device and its transcription into a classical
signal.  The rate of the information extraction from the qubit is
at the most as fast as its rate of decoherence due to the
back-action of the measuring device \cite{Ave03}. Furthermore, the
"no-cloning" theorem (eg. see \cite{Nie00}) prevents us from
making multiple copies of the same qubit and repeating the same
measurement to recover any lost information. Therefore, when
devising the measurement of a qubit it is imperative to be as
close as possible to this ideal limit, otherwise the non-ideal
measurement of qubit will result in the loss of information.

We see that decoherence is  present and measurement is needed in
the design of quantum computers. Understanding these processes is
imperative for the further advancement of the field.

\section{Quantum Mechanics of a Two-Level System}

We can think of a qubit as a quantum system where two of its
neighboring levels with similar energies are separated by big
energy gaps from the remaining levels, so that they can be
considered independent. This can be accomplished with a mesoscopic
Josephson junction consisting of a small superconducting island or
a box separated by a thin insulating barrier from a larger
superconducting electrode so that Cooper pairs can tunnel to and
from the box \cite{Ave85,But87}. If the superconducting gap of the
box is much larger than the thermal energy $k_BT$ and the charging
energy of the box $E_c\equiv2e^2/C$, then its dynamical properties
are solely determined by the excess number of Cooper pairs in the
box. The potential difference $V_q$ between the island and the
electrode induces continuous polarization charge $CV_q$ inside the
Cooper pair box of capacitance $C$. Because of the small size of
the box, its capacitance is small, and  the energy spectrum of the
box is dominated by its charging energy. The quantum tunneling of
Cooper pairs between the box and the reservoir is then a
perturbative effect described by Josephson tunneling amplitude
$-E_J/2$.

\begin{figure}
\begin{center}
\includegraphics[scale=0.7]{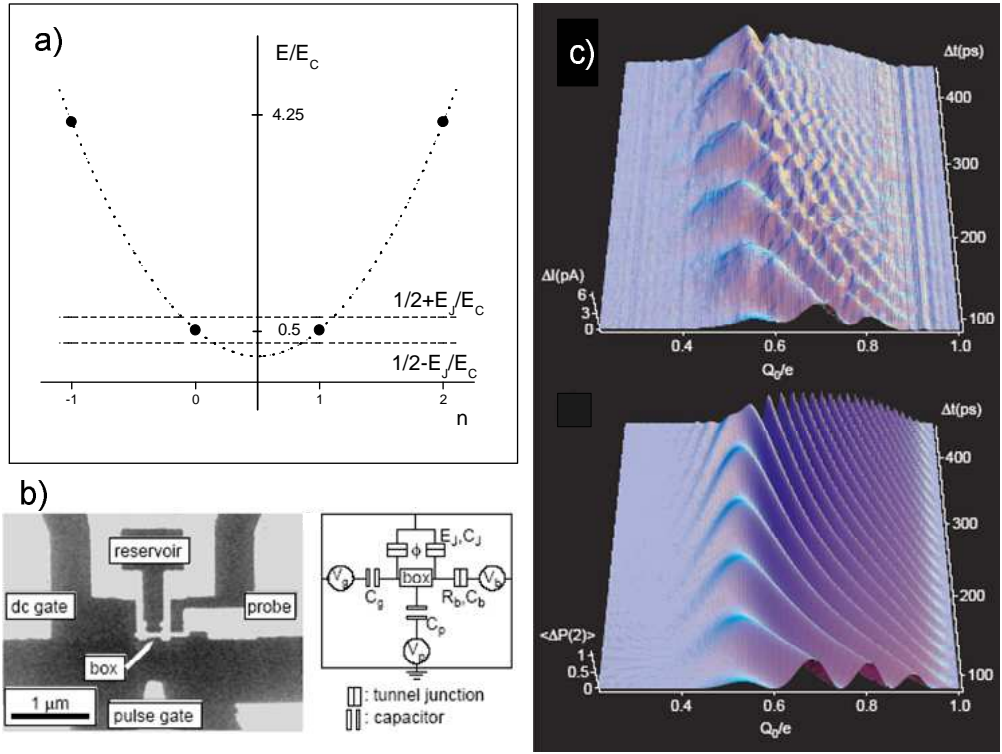}\\
\caption{a) Energy spectrum of Josephson junction operating in
charge regime with dots representing energy values of different
charge states and dashed lines representing two lowest quantum
levels with charging energy degeneracy removed by Josephson
tunneling ($E_C/E_J=8, q=1/2$). b). Picture and scheme of the
first operational charge qubit. c) The quantum coherent
oscillations observed by Nakamura et. al. with the device pictured
above \cite{Nak99}}\label{fig1.5}.
\end{center}
\end{figure}

In the basis spanned by the number of Cooper pairs in the box
$|n\rangle$, the Hamiltonian of the system consists of the
charging and tunneling parts which in that order yield the
Hamiltonian as:
\begin{equation}\label{i0}
    H=E_c(n-n_q)^2-\frac{E_J}{2}\left(|n\rangle\langle n+1|
    +|n+1\rangle\langle n|\right),
\end{equation}
where $n$ is the excess number of the Cooper pairs in the box, and
$n_q\equiv CV_q/2e$ is the induced static charge expressed as the
number of Cooper pairs, $n_q \in [0,1)$. If the applied potential
$V_q$ is adjusted  that $n_q\approx1/2$,  the gap between the
system's two lowest charge states $|n=0\rangle\equiv |R\rangle$
and $|n=1\rangle\equiv |L\rangle$ is small (i.e. $E_{10}=
E_C(1-2n_q)\approx0$), while the gap to the other states is on the
order of $E_C$ and considerably larger. If $k_BT\ll E_C$, the two
lowest levels are separated from the rest of the levels (figure
\ref{fig1.5}a) and as experimentally demonstrated in
\cite{Fle97,Bou99} they can be considered independent.

After truncation of the higher energy levels, the Hamiltonian
(\ref{i0})  can be written in the form of a general qubit
Hamiltonian

\begin{equation}\label{i1}
    H=-\frac{1}{2}(\varepsilon\sigma_z+\Delta\sigma_x),
\end{equation}
where $\sigma_{x,z}$ are Pauli matrices,
$\varepsilon=E_C(1-2n_q)/2$, and $\Delta=|E_J|$ \footnote{In the
case of a two-level system, an individual $\Delta$ can always be
made real by a unitary transformation.}.  The eigenvectors
$|0\rangle$ and $|1\rangle$ of this Hamiltonian define the energy
basis of the qubit with eigenenergies $\pm\Omega/2$. The qubit
Hamiltonian is diagonal in this basis, $H_{diag}=-\Omega/2
\:\sigma_z$, and the vectors spanning the two bases are related as
\begin{equation}\label{i3}
    |L\rangle=x_-|0\rangle +x_+|1\rangle, \quad
    |R\rangle=x_+|0\rangle -x_-|1\rangle,
\end{equation}
where $x_{\pm}=\sqrt{(\Omega\pm\varepsilon)/2\Omega}$ and
$\Omega=\sqrt{\varepsilon^2+\Delta^2}$.

The state of the qubit is fully specified by its wave function
$|\Psi(t)\rangle$, while its time evolution is determined by
unitary transformation on $|\Psi\rangle$,
\begin{equation}\label{i4}
    |\Psi(t)\rangle=\exp\left[-\frac{i}{\hbar}\int_{t_0}^tH(t')dt'\right]|\Psi(t_0)\rangle.
\end{equation}
Given the normalized  wave function of the qubit as
$|\Psi(0)\rangle=a|0\rangle+b|1\rangle$, in time $t$ it evolves to
$|\Psi(t)\rangle=ae^{i\Omega t/2}|0\rangle+be^{-i\Omega
t/2}|1\rangle$ expressed in the energy basis, or equivalently to
\begin{equation}\label{i6}
  |\Psi(t)\rangle = \left(ax_+e^{i\Omega t/2}-bx_-e^{-i\Omega t/2}\right)|L\rangle
  +\left(ax_-e^{i\Omega t/2}+ bx_+e^{-i\Omega
  t/2}\right)|R\rangle,
\end{equation}
if expressed in the charge basis. As it can be seen above, the
probability of finding the qubit in one of its eigenstates is
stationary, while the probability of finding the qubit in one of
its charge states (\ref{i6}) exhibits coherent oscillations with
period $2\pi/\Omega$.

Quantum coherent oscillations are signatures of both the
parallelism and reversibility of quantum systems. Thus, the
observation of coherent quantum oscillations in any qubit design
is imperative. The time of coherence expressed as the multiple of
time interval that takes to perform a single operation on a qubit
define the quality factor of a qubit.

The first successful solid state qubit design  \cite{Nak99} was a
Cooper pair box qubit (figure \ref{fig1.5}b).  It has demonstrated
both the energy band structure of Cooper pair box and the
existence of the quantum coherent oscillations of Cooper pairs
(figure \ref{fig1.5}c).  The quality factor of the qubit was low
due to the decoherence induced by the proximity of the measurement
electrode. The subsequent experiments \cite{Vio02,Pas03,Wal04}
have improved on the shortcomings of the early experiments, but
the same experiments have labelled the suppression of decoherence
as the most pressing problem in building a qubit suitable for
realistic quantum computation.

\section{Josephson Junctions}

The Cooper pair box introduced above is an example of a Josephson
Junction that in general refers to any two superconducting
electrodes separated by a weak link and characterized by the
presence of a zero-voltage supercurrent as postulated by Brian
David Josephson in 1962 \cite{Jos62,Jos64}. The supercurrent $I_s$
across the junction is related to the difference $\Delta\phi$
between the wave function phases of Cooper pair condensates of the
two electrodes:
\begin{equation*}
    I_s=I_c\sin(\Delta\phi),
\end{equation*}
where $I_c$ is the critical current. Zero voltage persists across
the junction as long as the total current across it is smaller
than the critical current.  In this case, the phase difference
remains stationary and the junction is in S-state. Increasing the
current above the critical current switches junction into a
running or R-state, finite voltage difference $V$ across the
junction appears, and the phase difference starts to evolve in
time as
\begin{equation*}
    \frac{d(\Delta\phi)}{dt}=\frac{2e}{\hbar}V.
\end{equation*}

\subsection{Small Junction}

For  a junction small enough to assume that the phase along it is
constant, and in the limit of low temperatures $(T\rightarrow
0K)$, the two equations above yield quantum mechanical description
of small Josephson Junctions (e.g. see \cite{Lik84}). In the
quantum description, the phase difference $\Delta\phi$ is replaced
by a more general gauge invariant phase $\phi$  conjugate  with
the number of Cooper pairs $n$ transmitted across the junction,
$\left[\phi,n\right]=i$. The junction Hamiltonian is determined by
junctions' charging energy $E_C$, and Josephson energy $E_J$:
\begin{eqnarray}\label{i8}
    H_{JJ}=E_C(n-n_q)^2-E_J(1-\cos(\phi)), \quad \\
    E_C=\frac{2e^2}{C},\quad E_J=\frac{\Phi_0 I_c}{2\pi},\quad \Phi_0=\frac{h}{2e}, \nonumber
\end{eqnarray}
where $C$ is the junction capacitance, $n_q$ is the externally
induced continuous polarization charge, and $\Phi_0$ is the
magnetic flux quantum.

\begin{figure}
  \includegraphics[scale=0.5]{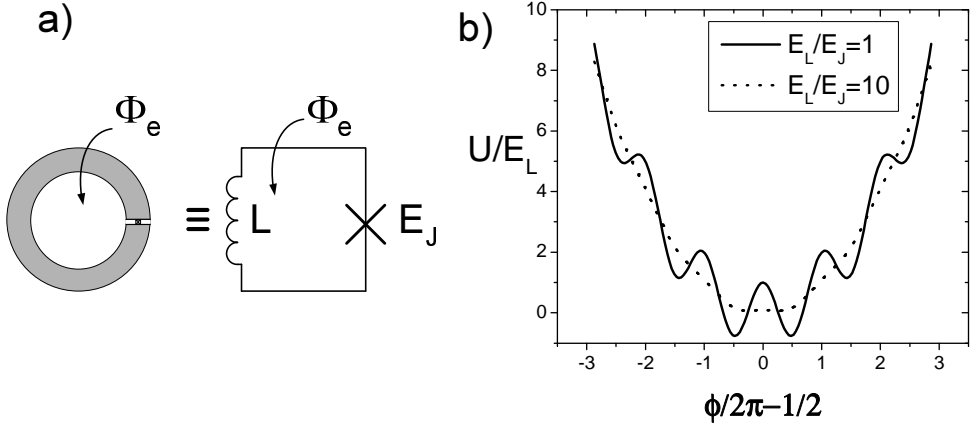}\\
  \caption{a) rf-SQUID and its equivalent circuit. b) Its potential energy as
  a function of phase for two different ratios of inductive $E_L$ and Josephson $E_J$ energies
  ($E_L=\Phi_0^2/2L$,  $\Phi_e/\Phi_0=0.5$).}\label{fig1.3}
\end{figure}

Depending on the ratio between charging and Josephson energies the
dynamics of Josephson junctions can easily be analyzed in two
respectfully extreme regimes. The limit $E_C\gg E_J$ is the one of
the Cooper pair box and it was discussed earlier. In the opposite
limit, $E_J\gg E_c$, the phase difference across the junction
becomes a good quantum number that describes the quantized
magnetic flux that is passing between the two electrodes of the
junction.  After the dropping of the constant $E_J$ term from
(\ref{i8}), the Hamiltonian can be expressed in in the
"coordinate" representation $(n\rightarrow -i\partial_{\phi})$ as
\begin{equation}\label{i9}
    H=E_c(i\partial_{\phi}+n_q)^2-E_J\cos(\phi).
\end{equation}
The periodic potential enables expressing the solutions of this
Hamiltonian in the form of Bloch functions
\begin{equation*}
    |\Psi\rangle=u(\phi)e^{-in_q\phi},
\end{equation*}
where the continuous nature of $n_q$ gives rise to the existence
of energy bands, while $u(\phi)$ has period of $2\pi$ and can be
expressed in the terms of Mathew Functions \cite{Abr70}.

The use of the Cooper pair dynamics for design of charge qubits
was obvious after one junction electrode was made into a
superconducting dot that trapped Cooper pairs.  The Josephson flux
dynamics can also be used for design of qubits by trapping the
flux that passes through the junction on one of its sides  as
proposed in \cite{Leg80}.  This requires inserting the junction
into a superconducting ring as shown in figure \ref{fig1.3}a.  The
flux that passes through the junction from the right gets trapped
inside the superconducting ring and in this process it induces a
persistent current in the ring.  This way shorted junction is
called \emph{rf-SQUID}\footnote{SQUID is short for Superconducting
QUantum Interference Device. The device is a good flux to voltage
transducer and it is often used for precise flux measurements.
These measurements require monitoring the device with tank circuit
operating at radio frequencies, thus the prefix rf.}.

\begin{figure}
\begin{center}
\includegraphics[scale=0.7]{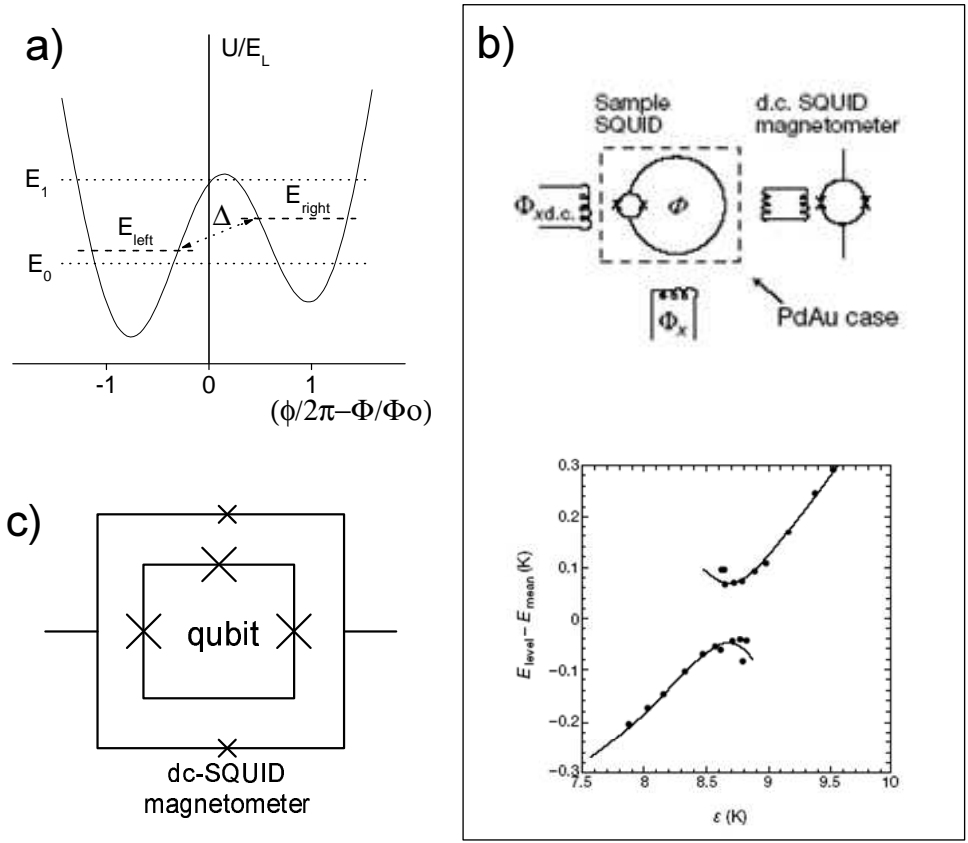}\\
\caption{a) Energy levels for rf-SQUID for ${E_L/E_C=3}$,
${\Phi/\Phi_o=11/8\pi}$, $ \Delta/E_L=0.06$.  b) Experimental
setup used in observing first superposition of quantum flux in a
qubit and the plot of measured energy relative to the mean level
energy  as a function of bias $\varepsilon$.  The plot clearly
shows energy splitting caused by the existence of tunneling. For
more see \cite{Fri00} c) Three-junction flux qubit and imbedded
dc-SQUID magnetometer used for the first observation of flux qubit
oscillations.}\label{fig1.6}
\end{center}
\end{figure}

Putting a junction into a superconducting loop of inductance $L$
gives rise to the inductive part in the Hamiltonian (\ref{i9}).
The loop shorts the junction and the induced charge $n_q$ becomes
irrelevant since it can always be screened out. Generally, some
external flux $\Phi_{ext}$ can be applied to the loop  and thus
inductively  coupled with the Josephson phase. Then the  rf-SQUID
Hamiltonian is
\begin{equation*}
H=-E_c\partial_{\phi}^2-E_J\cos(\phi)+\frac{\Phi_0^2}{2L}(\phi/2\pi-\Phi_{ext}/\Phi_0)^2.
\end{equation*}
The periodicity of (\ref{i9}) is clearly broken.  Depending on the
ratio of inductive energy $E_L=\Phi_0^2/2L$ and the Josephson
tunneling $E_J$, the shape of the potential varies from one with a
single minimum to one with  many local potential minima (figure
\ref{fig1.3}b). If $E_L/E_J\approx 3$ and
$\Phi_{ext}/\Phi_0\approx 1/2$ the potential is characterized by
two local minima (figure \ref{fig1.6}a) each corresponding to a
different flux state that can mutually tunnel through the barrier,
and whose relative energy spacing can be controlled by outside
applied flux. This way optimized rf-SQUID can be used as qubit.
The experiments\footnote{These experiments used rf-SQUID with the
single junction replaced by dc-SQUID in order to control the
tunneling in addition to the energy splitting of the rf-SQUID. For
more on dc-SQUID see \cite{Lik84,Rug90,Thi96}.} \cite{Rou95,Fri00}
have observed macroscopic quantum tunneling and the superposition
of macroscopic persistent current states (figure \ref{fig1.6}b),
but the need for the large inductance $L$ has hampered the
observation of coherent oscillations since large inductance makes
the qubit very sensitive to the fluctuations in the environment.
As argued in \cite{Moo99}, the need for the large inductance loop
can be overcome by building a three-junction SQUID (figure
\ref{fig1.6}c) characterized by smaller external inductance and
thus less sensitive to the outside fluctuations. The  ability of
the three-junction SQUID to preform as a flux qubit was
demonstrated in \cite{Chi03}.

\subsection{Long Junctions}

If the junction interface is long, the passing of a fluxon between
the electrodes cannot be considered instantaneous but rather as a
real time event.  In the certain limit of the junction parameters,
the long junction supports the propagation of fluxons along the
junction where the dynamics of the fluxons is affected by the
injected or outside induced junction current. This way optimized
long junction is often referred to as Josephson transmission line
(JTL). At low temperatures these fluxons exhibits quantum dynamics
\cite{Wal04}, and they can be considered as individual quantum
particles propagating in one or two-dimensional space, whose
metric and the potential are fully determined by the junction
parameters and the inserted current density respectively.

\begin{figure}
  \begin{center}
  \includegraphics[scale=0.6]{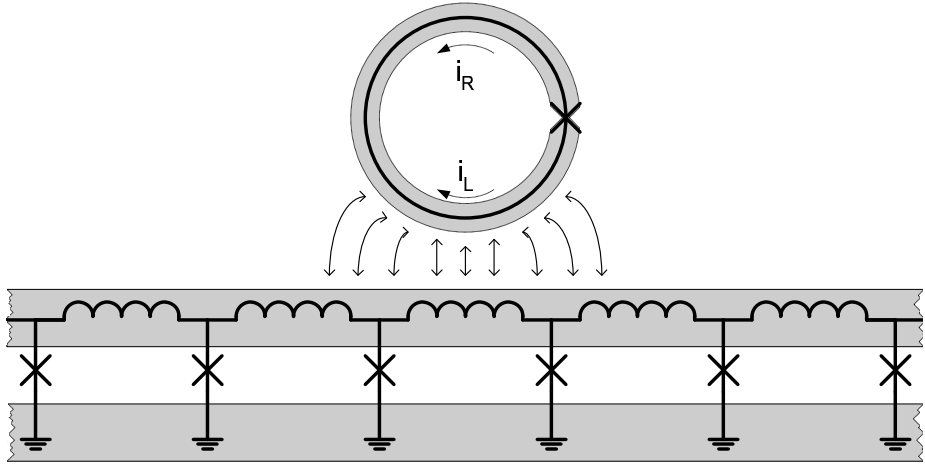}\\
\caption{Long Josephson junction, depicted in the bottom of the
drawing,  can be viewed as distributed structure consisting of
small Josephson junctions connected in parallel by inductors.   In
the certain limit of the junction parameters, the long junction
supports propagation of fluxons in between its interfaces. It can
perform as a detector, since the near-by flux qubit couples to it
inductively and modulates the scattering matrix of the propagating
fluxons.}\label{fig1.8}
  \end{center}
\end{figure}

The varying phase of a long junction can be analyzed by treating
the long junction as a distributed structure. In the case of a
long junction with a varying superconducting phase along only one
if its interfaces, the junction can be subdivided into strips with
the strips treated as  small junctions connected in parallel by
inductors as shown in the lower part of figure \ref{fig1.8}. The
Hamiltonian of this one-dimensional structure is given by
\begin{equation}\label{i10}
    H=\sum_{n=1}^N\left\{\frac{q^2_n}{2C_n}-E_{Jn}\cos(\phi_n)\right\}
    +\sum_{n=1}^{N-1}E_{Ln}\left(\phi_n-\phi_{n+1}\right)^2,
\end{equation}
where $q_n$ and $\phi_n$ are conjugate phase and charge variables
of strip $n$ that satisfy $[\phi_n,q_m]=2ei\delta_{n,m}$, while
$C_n$, $E_{Jn}$ and $E_{Ln}$ are respectively the capacitance,
Josephson tunneling, and inductive energy of the strips.

The equation (\ref{i10}) can be expressed in the continuous
approximation if the limit where the strip size is much smaller
than the length of the junction but larger than the
superconducting penetration depth and the spacing between the
electrodes.  The continuous Lagrangian density of the uniform
junction with length expressed in the units of Josephson
penetration depth $\lambda_J$ and time in the units of plasma
frequency $\omega_p$, reduces to the sine-Gordon Lagrangian

\begin{eqnarray}\label{i11}
    {\cal{L}}_{SG}&=&\sqrt{\epsilon_J \epsilon_L}\left\{
    \frac{1}{2}(\partial_{\tau}\phi)^2
    -\frac{1}{2}(\partial_{x}\phi)^2+\cos(\phi)\right\}\\
    \nonumber \epsilon_J=\Phi_0J_c/2\pi,& &
    \epsilon_L = (\Phi_0/2\pi)^2/l_0, \qquad
    \omega_p^2=\epsilon_j/c_0,\qquad
    \lambda_J^2=\epsilon_L/\epsilon_J,
\end{eqnarray}
where $J_c$, $c_0$ and $l_0$ are  Josephson current density,
junction capacitance density and inductance density respectively.

In the framework of semi-classical quantization of sine-Gordon
Lagrangian \cite{Raj82}, the propagating fluxon is represented as
a relativistic soliton of size $\pi\lambda_J$. The average
position and velocity of the soliton are specified by its two
collective coordinates, while the other coordinates correspond to
the internal quantum excitations of the soliton.  At low
temperatures ($k_BT\ll\sqrt{\epsilon_J\epsilon_L}$), and as long
as the velocity of fluxon is not on the order of
$\lambda_J\omega_p$, the fluxon cannot excite these internal
degrees of freedom.  It behaves as a relativistic quantum particle
of mass $M=\sqrt{2\epsilon_J\epsilon_L}/\lambda_J^2\omega_p^2$
propagating free in one-dimensional relativistic space with "speed
of light" given as $c=\lambda_J\omega_p$.

Coupling a flux qubit with the JTL (figure \ref{fig1.8}) creates
potential in the JTL, and the Hamiltonian (\ref{i10}) and
Lagrangian (\ref{i11}) get a source term whose explicit form
depends on the type and strength of coupling.  In any case, the
source term is localized as long as the qubit size is much smaller
than the JTL, and the freely propagating fluxons will now scatter
off the potential caused by the localized source term  and the
transmission properties of the 1-D channel formed by the optimized
JTL will depend on the qubit state. This setup can be used for
determining the state of the qubit, but if not designed properly
the propagating fluxons along the JTL will interact with the qubit
and cause decoherence.

\section{Modelling Decoherence}

The qubits introduced above are affected by the changes in the
environment. This is not just caused by the measurement of the
qubit, rather it can be contributed to many other factors. For
example, 1/f charge fluctuations induce variations of the static
Coopers in the pair box, while the inductive loop of the rf-SQUID
easily couples the flux qubit with the outside electromagnetic
fluctuations that vary the bias of the qubit. The coupling of the
qubit with the outside noise sources adds additional degrees of
freedom to the system whose random nature smear the coherent phase
of the condensates and decohere the qubit. In many cases the
microscopic origin of the decoherence is unknown.  Therefore, to
study the decoherence it is necessary to devise a model that
treats the qubit as a quantum system in contact with some general
environment which in the appropriate limit reduces to a stochastic
process.

System-plus-bath or system-plus-reservoir is the most natural way
to  approach this problem.  It considerers the weak entanglement
between the system of interest $S$, and the much larger
surroundings that act as a heat bath or a reservoir
\cite{Nak58,Sen60,Pri61}. Reducing the general equations of motion
by tracing-out the environmental degrees of freedom leaves the
equations dependent explicitly only on $S$. In this process the
environment is assumed classical, static, and unaffected by $S$.
Thus any energy that moves from $S$ is not stored, but lost in the
environment. This results in slow, irreversible transfer of the
energy from $S$ to the environment that is reflected in the
reduced equations of motion. In the further works
\cite{Fey63,Zwa73,Cal81}, the general reservoir was replaced by
infinite number of harmonic oscillators. This took into an account
that the environment by itself is a quantum system and relaxed the
earlier condition of strictly classical environment.

\emph{Spin-Boson}\footnote{The name originates from the analogy
between two-level system and spin coupled to many oscillators that
act as bosons.} is the common name for the model that precipitates
after the system-plus-reservoir model is applied to a quantum
two-level system \cite{Wei99}.  Explicitly, it consists of the
qubit Hamiltonian (\ref{i1}) linearly coupled to a bath consisted
of a large number of harmonic oscillators
\begin{equation}
    H_{SB}=H+\frac{\hbar}{2}\sigma_z\sum_{i=1}^N\lambda_i(c_i+c^{\dag}_i)+
    \sum_{i=1}^N\hbar\omega_ic^{\dag}_ic_i,
\end{equation}
where $c_i$ and $c^{\dag}_i$ are creation and annihilation
operators, and $\lambda_i$ is the coupling between the $i$-th
oscillator and the qubit.  The environmental effects are
phenomenologically represented by the noise spectral density
$S(\omega)$ of the coupling
\begin{equation}
    S(\omega)=\sum_{i=1}^N\lambda_i^2\delta(\omega-\omega_i).
\end{equation}

For weak coupling, the qubit will slowly couple to the oscillators
in the bath, and in the large oscillator bath limit
$(N\rightarrow\infty)$, the energy dissipated by the qubit will
never come back.  As a consequence, the qubit will relax to some
energy equilibrium state determined by the qubit-bath coupling.
This equilibrium state will effectively become the ground state of
the qubit and for this reason the qubit will become static and
stop exhibiting coherent oscillations.

\section{Quantum Measurement}

In order to measure the qubit it is necessary to entangle it with
a detector whose classical output signal is correlated with the
state of the qubit.  Because of the entanglement, a definite state
of the output signal localizes the qubit in one of the states
determined by the detector-qubit coupling.  This is the same
process as the reservoir induced decoherence discussed earlier.
The major difference is that the extracted information from the
qubit is not all lost, but that a part or preferably all of it is
reflected in the detector output.

Single-shot measurement is characterized by perfect correlation
between the qubit and the detector output.  This requires strong
coupling between the detector and the qubit.  Thus measuring the
qubit extracts all the information and instantaneously collapses
the qubit into the state correlated with the observed output.

As an example we can consider the setup as shown in figure
\ref{fig1.8}, where the JTL is situated close to the flux qubit
consisting of appropriately optimized rf-SQUID.  In the case of
negligible  decoherence and strong inductive coupling between the
qubit and JTL, a fluxon propagating from the left along the
transmission line is totally reflected if the qubit is in
$|R\rangle$ state, and totally transmitted if the qubit is in the
$|L\rangle$ state. So after the scattering, the transmitted fluxon
is entangled with the $|L\rangle$ and vice-versa. Detecting
weather the fluxon was reflected of transmitted  will collapse the
qubit into the state entangled with the scattering outcome.

In order to recover the statistical properties of the
superposition state of the qubit, it is necessary to repeat the
single-shot measurement on a certain number of the identically
prepared and time evolved qubits \cite{Woo81}.  The observation of
the quantum coherent oscillations of the qubit requires further
measurement of the superpositions at different times.

Coupling the qubit weakly with the detector does not result in
instantaneous wave function collapse, and for this reason the
measurement provides only limited information about the qubit.
This is reflected in the detector output which is not perfectly
correlated with the qubit, rather it is characterized by existence
of the overlap between the two signals that represent the
corresponding qubit states.

If the qubit and JTL in the earlier example are weakly coupled,
the fluxon scattering matrix is only affected partially by the
state of the qubit. The transmitted or reflected fluxon is
entangled with both states of the qubit.  Detecting the scattering
outcome in this case does not collapse the qubit, but provides
information in which state the qubit is more likely to be.

The advantage of the weakly coupled detector is its ability to
monitor qubit continuously for some time. During the measurement
process, the qubit will slowly localize under the influence of the
detector back-action which does not necessarily keep the qubit
coherent, rather the off-diagonal density matrix element in the
flux basis is limited from above as
$|\sigma_{LR}(t)|\le\sqrt{\sigma_{LL}(t)\sigma_{RR}(t)}$.

The back-action dephasing and the continuous extraction of
information are two  parallel dynamical processes where the rate
of back-action induced dephasing is never smaller that the rate of
information extraction. In the ideal case of the two rates being
equal, the detector back-action  keeps the qubit coherent, and the
detector will extract information form the qubit as fast as it is
localizing it. Any deviation from the ideal measurement will
result in loss of information since the decoherence rate will be
larger than the information extraction rate. This is a general
result that can be obtained from linear response theory
\cite{Ave03a} applied to a Hamiltonian $H=H_q+H_d+xf$, where $H_q$
and $H_d$ are general quantum system and detector respectively
that are linearly connected through a general quantum system
coordinate $x$ and some detector "force" $f$. The detector
provides output signal $o(t)$ that consists of the detector nose
$q(t)$ and response $\lambda x(t)$.

Assuming that the detector is static, the  dynamics of the
measurement is determined by three correlators  of detector
parameters that describe detector back-action $\langle
f(0)f(t)\rangle$, detector response $\langle f(0)q(t)\rangle$, and
the output nose $\langle q(0)q(t)\rangle$.  In this framework, the
condition of the ideal measurement is generalized  of having
detector satisfy $Re(\langle q(0)f(t)\rangle)=0$.  Furthermore, it
can be shown that in the case of the ideal measurement of coherent
oscillations the detector signal-to-noise ratio is limited to four
from above \cite{Kor01}.

This limitation is a direct consequence of the detector
back-action and it can be traced back to the nature of the
detector-qubit coupling.  If $x$ in the coupling is to be made a
constant of motion of the measured system, then the detector would
preform quantum non-demolition (QND) measurement that has no upper
limit on the signal to noise ratio \cite{Ave02}. To illustrate QND
measurement consider fluxons propagating along the JTL detector
weakly coupled to the flux qubit.  If the fluxon insertion
frequency equals to the qubit frequency, the passing fluxons will
"kick" the qubit in exactly the same way and in this process exert
no back-action on the qubit.  This detector will thus be able to
determine the flux state of the qubit, but it will also make the
qubit static in the basis of detector coupling \cite{Jor05}.

\part{Quantum Decoherence}

\chapter{Weakly Coupled High-Frequency Noise}

The purpose of this chapter is to develop theoretical description
of decoherence in the case of two coupled qubits subject to high
frequency (\emph{high-f}) noise. It follows a standard, Markovian
approximation approach for description of weak decoherence based
on the perturbation expansion of the density matrix evolution
equation. The result obtained within this simple scheme is useful
as a starting point for noise treatment of coupled qubit systems
and as a benchmark for more elaborate models.

\section{The Model}

The model is a system-plus-bath model where an arbitrary number of
quantum operators $Q_i$ of the quantum system $H_0$ are each
weakly connected with a different bath $H_{B}^{(i)}$ though a
generalized bath force $f_i$.  The baths can be correlated.
Furthermore, they are assumed to be much larger that the quantum
system and initially in an equilibrium. The large size and the
weak coupling makes the baths unaffected by the qubit. They remain
in equilibrium and can be separated out from the total density
matrix of the system. Explicitly, $\rho_{tot}(t)=\rho(t)\rho_B$,
where the two density matrices on the right respectively describe
the tensor product of quantum system and the total
"bath-described" environment.

If the system is initialized at $t=0$, its Hamiltonian
\begin{equation}\label{hf1}
    H=H_0+\sum_i H_{B}^{(i)}+\sum_{i}f_{i}Q_i,
\end{equation}
expanded to the second order  in the coupling yields differential
equation that describes the evolution of the qubit's density
matrix in the interaction picture as:

\begin{eqnarray}\label{hf2}
    \dot{\rho}(t)&=&-\frac{i}{\hbar}\sum_{i}tr_B\{\left[f_{i}(t)Q_i(t),\rho(0)\rho_B\right]\}
    \\ \nonumber
    &-&\frac{1}{\hbar^2}\sum_{ij}\int_{0}^{t}dt'tr_B\{\left[f_{i}(t)Q_i(t),\left[f_{j}(t')Q_j(t'),
    \rho(t')\rho_B\right]\right]\}.
\end{eqnarray}

In the  equation (\ref{hf2}), the force operators can be factored
out and grouped into their expectation values $\langle
f_i(t)\rangle$ and correlations $\langle f_i(t)f_j(t')\rangle$,
where $\langle\cdots\rangle\equiv tr_B\{\rho_B\cdots\}$. Since
$H_0$ can always be renormalized to absorb the non-diagonal force
operators ($\langle f_i(t)\rangle\ne0$), it can be assumed that
$\langle f_i(t)\rangle\equiv0$ and that the environment is
represented only by the force correlators. According to the
central limit theorem (eg. see \cite{Roe01}) and the the Wick
theorem \cite{Neg88,Wei99} using the second order correlators to
describe the noise created by a large system is sufficient. The
central limit theorem implies that the macroscopic size of the
environment makes the noise generated by it Gaussian, while the
higher order correlators of Gaussian noise are always expressible
by Wick's theorem in the terms of second order correlators.
Therefore, the correlators $\langle f_i(t)f_j(t')\rangle$ or
equivalently their spectral densities
\begin{equation}\label{hf3}
    S_{ij}(\omega)=\int_{-\infty}^{\infty}\langle f_i(0)f_j(t)\rangle
    e^{i\omega t}dt,
\end{equation}
fully describe the environment in our case.

The further assumption of Markovian \footnote{Evolution of a
Markovian system does not depend on its history.}  nature of the
system, implies that $\dot{\rho}(t)$ depends only on $\rho(t)$ and
justifies replacing $\rho(t')\rightarrow \rho(t)$ in the integral
part of the equation (\ref{hf2}). The high frequency nature of the
noise makes the bath response time instantaneous and the
integration limit in (\ref{hf2}) can be extended to infinity. The
differential equation (\ref{hf2}) simplifies to
\begin{eqnarray}\label{hf4}
    \dot{\rho}(t)=-\frac{1}{\hbar^2}\sum_{ij}\int_{0}^{\infty}dt'
    \left\{\left[Q_i(t),Q_j(t-t')\rho(t)\right]\langle
    f_i(t')f_j(0)\rangle\right. \nonumber \\
    -\left.\left[Q_i(t),\rho(t)Q_j(t-t')\right]
    \langle    f_j(0)f_i(t')\rangle\right\}.
\end{eqnarray}

Expressing the operators $Q_i(t)$ in the interaction
representation, (i.e.
$Q_i(t)=\exp(it\hat{H}_0/\hbar)Q_i(0)\exp(-it\hat{H}_0/\hbar)$)
and then recognizing that all but  non-oscillatory terms of the
equation (\ref{hf4}) average out, yields differential matrix
equation that describes the relaxation of the quantum system. The
equation is in general solvable but progressively more complicated
with the increasing energy spectrum of $H_0$.

\section{Single Qubit}

Long and extensive history of studying two-level  quantum systems
in optics and NMR calls for the discussion about the decoherence
in a single qubit to be very short. The more detailed derivation
of the results can be found in standard literature, i.e.
\cite{Blu81}.

We start by specifying the two-level system coupled to a bath
through its vertical polarization
\begin{equation*}
    H=-\frac{\hbar}{2}\left[\varepsilon\sigma_z+\Delta\sigma_x\right]
    -\frac{\hbar f}{2}\sigma_z+H_B,
\end{equation*}
with the bias $\varepsilon$, tunnelling $\Delta$ and noise $f$ all
having units of frequency.  The choice of the coupling corresponds
to the electrostatic interaction through finite coupling
capacitance for charge qubits, or magnetic interaction for flux
qubits. The interaction part of the Hamiltonian in the interaction
basis is:
\begin{equation*}
    H_{int}^{diag}=
    -\frac{\hbar f}{2\sqrt{\varepsilon^2+\Delta^2}}
    \left[\varepsilon\sigma_z-\Delta\sigma_x\right].
\end{equation*}
After following all the steps outlined above, it is easy to arrive
to standard two-level relaxation result that implies longest
coherence times and shortest relaxation times for qubits operating
at zero-bias, ($\epsilon=0$), point. The relaxation rate at this
optimal point is $\Gamma\equiv 1/T_1=S(\Delta)/2$ while the
decoherence rate is $\gamma\equiv1/T_2=S(\Delta)/4$. Consistent to
Fermi's golden rule they are different by factor-of-2.

\section{Double Qubit}

Motivation for studying decoherence in coupled qubits is provided
by the first successful double charge qubit experiment
\cite{Pas03}, where it was found that the decoherence rate for
quantum coherent oscillations in two qubits at the optimal bias
point is with good accuracy factor-of-four larger than the
decoherence rate in effectively decoupled qubits. An interesting
question for theory is whether this factor-of-four increase of the
decoherence rate is a numerical coincidence, or it reflects some
basic property of the decoherence mechanisms in charge qubits. As
will become clear from the discussion below, the theory developed
in this section favors ``numerical coincidence'' point of view.
Other aspects of decoherence in coupled qubits have been studied
before numerically in \cite{Gov01,Tho02,Sto03}.

In general, it is well understood that decoherence rates of
different states of two coupled qubits can be quite different if
the random forces created by the qubit environments responsible
for decoherence are completely or partially correlated. Most
importantly, in the case of complete correlation, the qubit system
should have a ``decoherence-free subspace'' (DFS) spanned by the
states $|01\rangle$, $|10\rangle$ \cite{Blu81,Ave03,Gov01}, since
completely correlated external environments can not distinguish
these states. In contrast, the subspace spanned by $|00\rangle$
and $|11\rangle$ experiences decoherence that is made stronger by
the correlations between environmental forces acting on the two
qubits. So the role of the quantitative theory in description of
decoherence in the dynamics of coupled qubits is to see to what
extent subspaces with different decoherence rates participate in
the qubit oscillations for different methods of their excitation.

\subsection{The model and environmental correlations}

\begin{figure}[h]
\begin{center}
\includegraphics[scale=0.4]{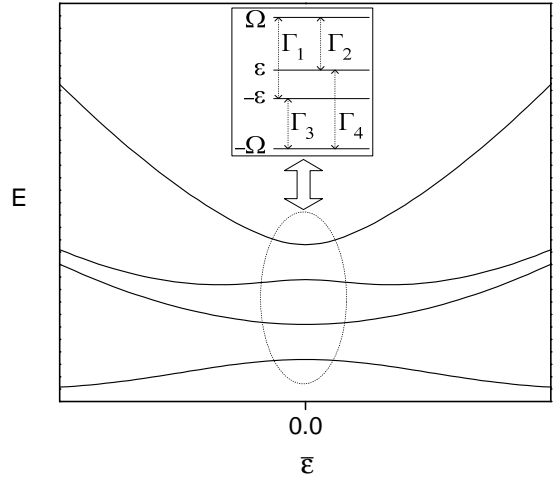}\\
\caption{ Schematic structure of the energy levels of the two
coupled qubits as functions of the common bias of the qubits. The
inset shows the diagram of the decoherence-induced transitions
between the levels at ``co-resonance'' point where the bias
vanishes.}\label{fig2.1} \end{center}
\end{figure}

The Hamiltonian of the system of two qubits coupled directly by
interaction between the basis-forming degrees of freedom is:
\begin{equation}
H_0 = \hbar\sum_{j=1,2} (\varepsilon_j \sigma_{z}^{(j)} + \Delta_j
\sigma_{+}^{(j)}+\Delta_j^* \sigma_{-}^{(j)}) +\hbar\nu
\sigma_{z}^{(1)} \sigma_{z}^{(2)} \, , \label{hf5} \end{equation}
where $\sigma$'s denote the Pauli matrices, $\nu$ is the qubit
interaction energy, $\Delta_j$ is the tunnel amplitude and
$\varepsilon_j$ is the bias of the $j$-th qubit. Four energy
levels of the Hamiltonian (\ref{hf5}) are shown schematically in
figure \ref{fig2.1} as functions of the common bias
$\bar{\varepsilon} \equiv \varepsilon_1=\varepsilon_2$ of the two
qubits. This work considers quantum coherent oscillations in the
qubits biased at the ``co-resonance'' point \cite{Pas03}, where
$\varepsilon_1= \varepsilon_2 =0$. Such bias conditions are
optimal for the oscillations.

It can be shown explicitly that the occupation probabilities of
the qubit basis states (that are of interest for us) are
insensitive to the phases of the qubit tunnel amplitudes
$\Delta_j$, so without the loss of generality we will assume that
$\Delta_j$'s are real. The Hamiltonian (\ref{hf5}) at the
co-resonance reduces then to
\begin{equation}
H_0 = \hbar\sum_{j=1,2} \Delta_j \sigma_{x}^{(j)} +\hbar\nu
\sigma_{z}^{(1)} \sigma_{z}^{(2)} \, . \label{hf6} \end{equation}
In the basis composed of eigenstates of the $\sigma_{x}^{(j)}$
operators, the Hamiltonian (\ref{hf6}) can be diagonalized easily.
Eigenenergies and eigenstates are:
\begin{eqnarray}
E_1=\hbar\Omega, & |\psi_1\rangle =\frac{1}{2} [(\gamma+\eta)
(|00\rangle +|11\rangle) +(\eta-\gamma) (|01\rangle +|10\rangle)
],\nonumber \\
E_2=-\hbar\Omega, & |\psi_2\rangle =\frac{1}{2} [(\gamma-\eta)
(|00\rangle +|11\rangle) +(\gamma+\eta) (|01\rangle +|10\rangle)
] , \label{hf7}  \\
E_3=\hbar\epsilon, & |\psi_3\rangle =\frac{1}{2} [(\alpha+\beta)
(|00\rangle -|11\rangle) +(\alpha-\beta) (|10\rangle -|01\rangle)
] , \nonumber \\
E_4=-\hbar\epsilon, & |\psi_4\rangle =\frac{1}{2} [(\beta-\alpha)
(|00\rangle -|11\rangle) +(\alpha+\beta) (|10\rangle -|01\rangle)
] , \nonumber
\end{eqnarray}
where
\begin{eqnarray*}
\Omega=(\Delta^2+\nu^2)^{1/2},\qquad  \epsilon=(\delta^2+
\nu^2)^{1/2},\qquad \\ \alpha, \beta =
\frac{1}{\sqrt{2}}\left(1\pm \frac{\delta} {\epsilon}\right)^{1/2}
\, , \;\;\; \eta,\gamma= \frac{1}{\sqrt{2}} \left(1\pm
\frac{\Delta}{\Omega}\right)^{1/2}
\end{eqnarray*}
and $\Delta\equiv \Delta_1+\Delta_2$, $\delta \equiv \Delta_1-
\Delta_2$. The states $|kl\rangle$ with $\{ k,l\} =\{ 0,1\}$ in
equations~(\ref{hf7}) are the eigenstates of the operators
$\sigma_{z }^{(1,2)}$ in the natural notations: $\sigma_{z}^{(1)}
|kl\rangle = (-1)^k|kl\rangle$ and $\sigma_{z}^{(2)}|kl\rangle =
(-1)^l|kl\rangle$.

We assume that external environments responsible for the
decoherence couple to the basis-forming degrees of freedom of the
qubits. The interaction Hamiltonian is then:
\begin{equation}\label{hf8}
H_i= \hbar\sum_{j=1,2} f_j(t) \sigma_{z}^{(j)}  \, .
\end{equation} The random forces $f_{1,2}(t)$ acting on the
qubits are in general correlated. To describe the weakly
dissipative dynamics of the system in the basis of states
(\ref{hf7}) induced by the interaction (\ref{hf8}) with the
reservoirs, we use the standard equation (\ref{hf4}). Proceeding
in the usual way, we keep in equation (\ref{hf4}) only the terms
that do not oscillate, and therefore lead to changes in $\rho$
that accumulate over time. Equations for the matrix elements
$\rho_{nm}$, $n,m=1,...,4$, of $\rho$ in the basis (\ref{hf7}) are
transformed then as follows:
\begin{eqnarray}\label{hf10}
\dot{\rho}_{nm} & = & \sum_{j,j'=1,2}\big[-\rho_{nm}
(\sigma_{mm}^{(j)}-\sigma_{nn}^{(j)})(\tilde{S}_{jj'}^*(0)
\sigma_{mm}^{(j')} -\tilde{S}_{jj'}(0)\sigma_{nn}^{(j')})
\nonumber \\
& & -\rho_{nm}\sum_{k}(\sigma_{nk}^{(j)}\sigma_{kn}^{(j')}
\tilde{S}_{jj'}(\epsilon_{n}-\epsilon_{k})
+\sigma_{mk}^{(j')}\sigma_{km}^{(j)}\tilde{S}_{jj'}^*
(\epsilon_{m} -\epsilon_{k}))
\nonumber \\
& & +\delta_{nm}\sum_{k}\rho_{kk}(\sigma_{nk}^{(j')}
\sigma_{kn}^{(j)}\tilde{S}_{jj'}(\epsilon_{k}-\epsilon_{n})+
\sigma_{nk}^{(j)} \sigma_{kn}^{(j')}\tilde{S}_{jj'}^*
(\epsilon_{k}-\epsilon_{n})
\nonumber \\
& & + \sum_{(k,l)}\rho_{kl}\sigma_{nk}^{(j')} \sigma_{lm}^{(j)}
(\tilde{S}_{jj'}(\epsilon_{l}-\epsilon_{m})+
\tilde{S}_{j'j}^*(\epsilon_{l}-\epsilon_{m})) \big].
\end{eqnarray} Here $\sigma_{nm}^{(j)}$ denote the matrix elements
$\langle n|\sigma_z^{(j)}|m\rangle$, the last sum is taken over
the pairs $(k,l)$ of states that satisfy the ``resonance''
condition:
\[ \epsilon_{k}-\epsilon_{l}= \epsilon_{n}-\epsilon_{m} \, , \;\;\;
(k,l) \neq (n,m) \, , \] and
\[ \tilde{S}_{jj'}(\omega)=\int_{0}^{\infty}dt\langle f_{j}(t)
f_{j'}(0)\rangle e^{i\omega t} \, . \]

The first term in equation (\ref{hf10}) represents ``pure
dephasing'' that exists when the system operators that couple it
to the environment have non-vanishing average values in the
eigenbasis. As one can see explicitly from equations\ (\ref{hf7}),
the average values $\sigma_{nn}^{(j)}$ of $\sigma_z^{(j)}$ are
vanishing in all states, so that there is no pure dephasing term
in the evolution of the density matrix in our case. The fact that
$\sigma_{nn}^{(j)}$ are vanishing can be related to the vanishing
slope of the system energies with respect to variations of the
bias in the vicinity of the co-resonance point -- see figure
\ref{fig2.1}. Since all coefficients in the eigenfunctions
(\ref{hf7}) are real, the matrix elements $\sigma_{nm}^{(j)}$ are
also real. For real $\sigma_{nm}^{(j)}$, imaginary parts of the
noise correlators $\tilde{S}_{j'j}$ in the second term on the
right-hand-side of equation (\ref{hf10}) represent the
decoherence-induced shifts of the systems' energy levels. These
shifts do not affect decoherence and we will neglect them in our
discussion. With these simplifications, equation (\ref{hf10})
takes the form
\begin{eqnarray}\label{hf11}
& &\dot{\rho}_{nm}  =  \\& &\sum_{j,j'=1,2}\big[ -(\rho_{nm} /2)
\sum_{k}(\sigma_{nk}^{(j)}\sigma_{kn}^{(j')} \mbox{Re} S_{jj'}
(\epsilon_{n}-\epsilon_{k})+\sigma_{mk}^{(j')}\sigma_{km}^{(j)}
\mbox{Re} S_{jj'} (\epsilon_{m} -\epsilon_{k})) \nonumber \\
\nonumber & & +\delta_{nm}\sum_{k}\rho_{kk} \sigma_{nk}^{(j')}
\sigma_{kn}^{(j)}\mbox{Re} S_{jj'}(\epsilon_{k}-\epsilon_{n}) +
\sum_{(k,l)}\rho_{kl}\sigma_{nk}^{(j')} \sigma_{lm}^{(j)}
S_{jj'}(\epsilon_{l}-\epsilon_{m}) \big],
\end{eqnarray} where
\begin{equation}\label{hf12}
S_{jj'}(\omega)=\int_{-\infty}^{\infty}dt\langle f_{j}(t)
f_{j'}(0)\rangle e^{i\omega t} \, .  \end{equation}

The function $S_{12}$ characterizes the correlations between the
environmental forces  acting on the two qubits. For instance, if
the two qubits interact with different environments and $v_1$,
$v_2$ are uncorrelated, $S_{12}=0$, whereas
$S_{12}=S_{11}=S_{22}$, if the qubits are acted upon by the force
produced by a single environment coupled equally to the two
qubits. While the correlators $S_{11}$ and $S_{22}$ are strictly
real, $S_{12}$ can be imaginary, and $S_{21}^*=S_{12}$.
Non-vanishing imaginary part of $S_{12}$ corresponds to the
non-vanishing commutator $[f_1, f_2]$ and implies that the two
qubits are coupled to the two non-commuting variables of the same
reservoir. While this is probably not very likely for qubits with
the basis-forming degrees of freedom of the same nature (which in
a typical situation should be coupled to the same set of
environmental degrees of freedom), the non-vanishing
$\mbox{Im}S_{12}$ should be typical if the qubits have different
basis-forming variables,(e.g. see system proposed in
\cite{Fri02}). Using the spectral decomposition of the correlators
$S_{jj'}(\omega)$ and Schwartz inequality, one can prove \footnote
{Similar proof in slightly  different context of linear quantum
measurements can be found in \cite{Ave03}.} that for arbitrary
stationary reservoirs the correlators satisfy the inequality that
imposes the constraint on $S_{12}(\omega)$:
\begin{equation}
S_{11}(\omega)S_{22}(\omega)\geq |S_{12}(\omega)|^2 \, .
\label{hf13} \end{equation} If the reservoirs are in equilibrium
at temperature $T$, the correlators also satisfy the standard
detailed balance relations:
\begin{equation}
S_{jj'}(-\omega) =e^{-\frac{\hbar\omega}{k_BT}} S_{j'j}(\omega) \,
. \label{hf14} \end{equation} Equation (\ref{hf11}) with the noise
correlators (\ref{hf12}) govern weakly dissipative time evolution
of the two coupled qubits in a generic situation. Below we use
them to determine decoherence properties of quantum coherent
oscillations of the qubits.

\subsection{Quantum coherent oscillations in coupled qubits}

One of the most direct ways of excitation of quantum coherent
oscillations  in individual or coupled qubits that will be
discussed in this work is based on the abrupt variation of the
bias conditions \cite{Nak99,Pas03}. If the qubits are initially
localized in one of their basis states, e.g. $|00\rangle$, and
abrupt variation of the bias brings them to the co-resonance
point, the probabilities for the qubits to be in other basis
states start oscillating with time.

In the simplest detection scheme (realized, for instance, in
experiment \cite{Pas03}) the probability for each qubit to be in
the state $|1\rangle$ is measured independently of the state of
the other qubit. Quantitatively, these probabilities $p_j$ are
obtained from the projection operators $P_j$:
\begin{equation}\label{hf14z}
    p_j=\mbox{Tr}\{\rho P_j \}\, ,\;\;\; P_1= \sum_{k=1,2}
|1k\rangle \langle k1| \, , \;\;\; P_2= \sum_{k=1,2} |k1\rangle
\langle 1k|  \, .
\end{equation}

Finding explicitly the matrix elements of $P_j$ from the wave
functions (\ref{hf7}), one gets:
\begin{eqnarray}
p_{1}(t)=\frac{1}{2}& + & (\alpha\eta+\beta\gamma) \mbox{Re}
[e^{-i\omega_- t}(\rho_{42}(t)-\rho_{13}(t))]\nonumber \\
& + & (\alpha \gamma - \beta\eta)\mbox{Re} [e^{-i\omega_+ t}
(\rho_{14}(t)+ \rho_{32}(t))]\label{hf15a} \, , \\ \nonumber
p_{2}(t)=\frac{1}{2} & - & (\alpha\gamma + \beta\eta) \mbox{Re}
[e^{-i\omega_- t} ( \rho_{13}(t)+\rho_{42}(t))]
\\ & + & (\alpha\eta-\beta\gamma) \mbox{Re} [e^{-i\omega_+ t}
(\rho_{14}(t)-\rho_{32}(t))] \, , \label{hf15} \end{eqnarray}
where $\omega_{\pm} \equiv \Omega \pm \epsilon $, and as in the
equation (\ref{hf11}), the matrix elements of the density matrix
are taken in the interaction representation. Equations
(\ref{hf15a}), (\ref{hf15}), and (\ref{hf10}) show that the
waveform of the coherent oscillations in coupled qubits is
determined by the time evolution of the two pairs of the matrix
elements of $\rho$:
\begin{equation}
\dot{\rho}_{13}=-\Gamma_{13}\rho_{13}+u_-\rho_{42} \, , \;\;\;
\dot{\rho}_{42}=-\Gamma_{42}\rho_{42}+u_+\rho_{13} \, ,
\label{hf16} \end{equation}
\begin{equation}
\dot{\rho}_{14}=-\Gamma_{14}\rho_{14}+v_- \rho_{32} \, , \;\;\;
\dot{\rho}_{32}=-\Gamma_{32}\rho_{32}+v_+ \rho_{14} \, .
\label{hf17} \end{equation}

The decoherence rates in these equations are determined by the
rates of transitions between different energy eigenstates:
\begin{eqnarray}
\Gamma_{13}=\frac{1}{2}\big(\Gamma_1^{(+)}+ \Gamma_2^{(-)}+
\Gamma_2^{(+)}+\Gamma_4^{(+)}\big), \nonumber \\ \Gamma_{14}=
\frac{1}{2}\big(\Gamma_1^{(-)}+\Gamma_1^{(+)}+\Gamma_2^{(+)}+
\Gamma_3^{(+)}\big), \nonumber \\
\Gamma_{32}=\frac{1}{2}\big(\Gamma_2^{(-)}+ \Gamma_3^{(-)}+
\Gamma_4^{(-)}+\Gamma_4^{(+)}\big), \nonumber \\ \Gamma_{42}=
\frac{1}{2}\big(\Gamma_1^{(-)}+ \Gamma_3^{(-)}+\Gamma_3^{(+)}+
\Gamma_4^{(-)}\big)\, . \label{hf18} \end{eqnarray} where
labelling of the transitions is indicated in the inset in figure
\ref{fig2.1}. Transition rates are:
\begin{eqnarray}
\Gamma_{1}^{(\pm)}=\mbox{Re} \sum_{j,j'}S_{jj'}(\pm \omega_+)
\sigma_{14}^{(j)}\sigma_{41}^{(j')} \, , \nonumber \\
\Gamma_{2}^{(\pm)}=\mbox{Re} \sum_{j,j'}S_{jj'}(\pm \omega_-)
\sigma_{13}^{(j)}\sigma_{31}^{(j')} \, , \nonumber \\
\Gamma_{3}^{(\pm)}=\mbox{Re} \sum_{j,j'}S_{jj'}(\pm\omega_-)
\sigma_{24}^{(j)}\sigma_{42}^{(j')}\, , \nonumber \\
\Gamma_{4}^{(\pm)}=\mbox{Re}\sum_{j,j'}S_{jj'}(\pm \omega_+)
\sigma_{23}^{(j)}\sigma_{32}^{(j')} \, . \label{hf19}
\end{eqnarray} The superscripts $\pm$ refer here to transitions in
the direction of decreasing (+) or increasing (-) energy. After
finding matrix elements $\sigma_{nm}$ from the wave functions
(\ref{hf7}), we see explicitly that transitions between the states
1 and 2, as well as 3 and 4 are suppressed.  Since the
corresponding matrix elements are zero,  the rates (\ref{hf19})
are:
\begin{eqnarray}
\Gamma_{1}&=& \frac{1}{2} S_{11} \left[1-\frac{\delta\Delta
+\nu^2}{ \epsilon \Omega}\right] +\frac{1}{2} S_{22}
 \left[1+\frac{\delta\Delta -\nu^2} {\epsilon \Omega}\right] -\mbox{Re} S_{12}
 \frac{\nu\omega_-}{\Omega\epsilon}\, , \nonumber \\
\Gamma_{2}&=& \frac{1}{2} S_{11}  \left[1+\frac{\delta\Delta
+\nu^2}{ \epsilon \Omega}\right] +\frac{1}{2} S_{22}
 \left[1-\frac{\delta\Delta -\nu^2} {\epsilon \Omega}\right] +\mbox{Re} S_{12}
\frac{\nu\omega_+}{\Omega\epsilon} \, , \nonumber \\
\Gamma_{3}&=& \frac{1}{2} S_{11}  \left[1+\frac{\delta\Delta
+\nu^2}{ \epsilon \Omega}\right] +\frac{1}{2} S_{22}
 \left[1-\frac{\delta\Delta -\nu^2} {\epsilon \Omega}\right] -\mbox{Re} S_{12}
\frac{\nu\omega_+}{\Omega\epsilon} \, , \label{hf20}  \\
 \Gamma_{4}&=& \frac{1}{2} S_{11}  \left[1-\frac{\delta\Delta +\nu^2}{
\epsilon \Omega}\right] +\frac{1}{2} S_{22}
\left[1+\frac{\delta\Delta -\nu^2} {\epsilon \Omega}\right]
+\mbox{Re} S_{12}
 \frac{\nu\omega_-}{\Omega\epsilon} \, . \nonumber
\end{eqnarray}
The transfer ``rates'' $u$, $v$ in equations~(\ref{hf16}) and
(\ref{hf17}) are:
\begin{equation}
u_{\pm}=\sum_{j,j'=1,2}\sigma_{14}^{(j')}\sigma_{32}^{(j)}
S_{jj'}(\pm\omega_+)\, , \;\;\;
v_{\pm}=\sum_{j,j'=1,2}\sigma_{13}^{(j')}\sigma_{42}^{(j)}
S_{jj'}(\pm\omega_-) \, . \label{hf21} \end{equation} Explicitly:
\begin{eqnarray}
u = \frac{1}{2} S_{11} \left[1-\frac{\delta\Delta +\nu^2}{
\epsilon \Omega}\right] -\frac{1}{2} S_{22}
\left[1+\frac{\delta\Delta -\nu^2} {\epsilon \Omega}\right]
-i\mbox{Im} S_{12} \frac{\nu\omega_-}{\Omega\epsilon} \, ,\nonumber \\
v = -\frac{1}{2} S_{11} \left[1+\frac{\delta\Delta +\nu^2}{
\epsilon \Omega}\right] +\frac{1}{2} S_{22}
\left[1-\frac{\delta\Delta -\nu^2} {\epsilon \Omega}\right]
-i\mbox{Im} S_{12} \frac{\nu\omega_+}{\Omega\epsilon}\, .
\label{hf22}
\end{eqnarray} Equations (\ref{hf20}) and (\ref{hf22}) do not show
the frequency dependence of noise correlators, which is the same,
respectively, as in the equations~(\ref{hf19}) and (\ref{hf21}).

Each pair, (\ref{hf16}) and (\ref{hf17}), of coupled equations can
be solved directly by diagonalization of the matrix of the
evolution coefficients with a non-orthogonal transformation. In
this way we obtain for the pair of equations (\ref{hf16}):
\begin{eqnarray}
\rho_{13}(t) =\frac{\big[ \rho_{13}(0)(u_+u_-e^{-
\gamma_+t}+c^2e^{- \gamma_-t}) +cu_-\rho_{42}(0)(e^{-\gamma_+t}-
e^{-\gamma_-t}) \big]}{u_+u_-+c^2}\, ,\nonumber \\
\rho_{42}(t)= \frac{\big[ \rho_{42}(0)(u_+u_-e^{-
\gamma_-t}+c^2e^{- \gamma_+t}) +cu_+\rho_{13}(0)(e^{-\gamma_+t}-
e^{-\gamma_-t})\big]}{u_+u_-+c^2}\, . \label{hf23} \end{eqnarray}
where
\begin{eqnarray}
\gamma_{\pm} \equiv (\Gamma_{13}+\Gamma_{42})/2 \pm \big[
(\Gamma_{13}-\Gamma_{42})^2/4+u_+u_-\big]^{1/2} \, , \nonumber \\
c\equiv (\Gamma_{13}-\Gamma_{42})/2 - \big[ (\Gamma_{13}-
\Gamma_{42})^2/4+u_+u_-\big]^{1/2} \, , \label{hf24}
\end{eqnarray} and $\rho_{13}(0)$, $\rho_{42}(0)$ are the initial
values of the density matrix elements that depend on preparation
of the initial state. If, as in the experiment \cite{Pas03}, the
qubits are abruptly driven to co-resonance maintaining the state
$|00\rangle$, these initial values are:
\begin{eqnarray}
\rho_{13}(0)&=&\frac{1}{4}(\gamma+\eta)(\alpha+\beta)\, , \;\;\;
\rho_{32}(0)=\frac{1}{4}(\gamma-\eta)(\alpha+\beta)\, , \nonumber
\\ \rho_{14}(0)&=&\frac{1}{4}(\gamma+\eta)(\beta-\alpha)\, , \;\;\;
\rho_{42}(0)=\frac{1}{4}(\gamma-\eta)(\alpha-\beta)\, .
\label{hf25} \end{eqnarray} Another type of initial conditions
that will be discussed in this work is starting the oscillations
from the state $|10\rangle$. In this case:
\begin{eqnarray}
\rho_{13}(0)&=&\frac{1}{4}(\gamma-\eta)(\alpha-\beta)\, , \;\;\;
\rho_{32}(0)=\frac{1}{4}(\gamma+\eta)(\alpha-\beta)\, ,\nonumber
\\ \rho_{14}(0)&=&\frac{1}{4}(\gamma-\eta)(\alpha+\beta)\, , \;\;\;
\rho_{42}(0)=\frac{1}{4}(\gamma+\eta)(\alpha+\beta)\, .
\label{hf26} \end{eqnarray} Equations (\ref{hf21}) and
(\ref{hf22}) follow directly from the wavefunctions (\ref{hf7}):
$\rho_{nm}(0)=\langle n|i\rangle \langle i|m\rangle$, where
$|i\rangle$ is the initial state.

Solution of the other pair (\ref{hf17}) of coupled equations is
given by the same equations~(\ref{hf19}) and (\ref{hf24}) with
obvious substitutions: $u_{\pm} \rightarrow v_{\pm}$, $\Gamma_{13}
\rightarrow \Gamma_{14}$, $\Gamma_{42} \rightarrow \Gamma_{32}$.
In this work, we are mostly interested in the low-temperature
regime $k_BT/\hbar\ll \epsilon, \Omega$, when transitions  up in
energy can be neglected. In this regime, $u_-,v_- \rightarrow 0$,
and equations for the evolution of the density matrix elements are
simplified. For instance, for $u_- \rightarrow 0$, $c\simeq
u_+u_-/(\Gamma_{13}- \Gamma_{42})$, and equations~(\ref{hf23}) are
reduced to:
\begin{eqnarray}
\rho_{13}(t)&=& \rho_{13}(0)e^{-\Gamma_{13}t}\, , \nonumber
\\ \rho_{42}(t)&=&
\rho_{42}(0)e^{-\Gamma_{42}t}+\frac{u_+}{\Gamma_{13}-
\Gamma_{42}}\rho_{13}(0)(e^{-\Gamma_{13}t}-e^{-\Gamma_{42}t}) \, ,
\label{hf27} \end{eqnarray} where now $\Gamma_{13}=(\Gamma_1+
\Gamma_2+ \Gamma_4)/2$ and $\Gamma_{42}= \Gamma_3/2$.

Time evolution (\ref{hf27}) of the density matrix elements
together with the rates (\ref{hf20}) and (\ref{hf22}), and initial
conditions (\ref{hf25}) and (\ref{hf26}) determines the shape of
the coherent oscillations in two coupled qubits. In the next
section, we discuss this shape in several specific situations.

\subsection{Results and Conclusions}

The shape of coherent oscillations determined in the previous
section depends on the large number of parameters: temperature,
the degree of asymmetry of qubit tunnel energies and couplings to
the environments, frequency dependence of the decoherence, and
strength and nature of the correlations between the two
reservoirs. We analyze some of these dependencies below.

\subsubsection{Experimentally-motivated case}

We start by considering the situation that is close to the
experimentally realized case of oscillations in coupled charge
qubits \cite{Pas03}. As argued above, the correlations between
environments in this case should be real: $\mbox{Im} S_{12}=0$.
The oscillations are excited by driving the system to co-resonance
in the initial state $|00\rangle$. Equations of the previous
section give in this case the following expression for the shape
of the oscillations:
\begin{eqnarray}
p_{1}(t)=\frac{1}{2}-\frac{1}{8}\big[A e^{-\Gamma_{42}t} +B
(e^{-\Gamma_{13}t}-\frac{u_+}{\Gamma_{13}- \Gamma_{42}}
(e^{-\Gamma_{13}t}-e^{-\Gamma_{42}t}))\big] \cos \omega_-t
\nonumber \\
-\frac{1}{8}\big[C e^{-\Gamma_{32}t} +D
(e^{-\Gamma_{14}t}+\frac{v_+}{\Gamma_{14}- \Gamma_{32}}
(e^{-\Gamma_{14}t}-e^{-\Gamma_{32}t}))\big] \cos \omega_+t \, ,
\quad \label{hf28} \end{eqnarray} where \begin{eqnarray*}
A=1+\frac{\delta\Delta +\nu^2}{ \epsilon \Omega}-
\frac{\nu}{\Omega}-\frac{\nu}{\epsilon} \, , \;\;\;
B=1+\frac{\delta\Delta +\nu^2}{\epsilon \Omega}
+\frac{\nu}{\Omega}+ \frac{\nu}{\epsilon} \, , \\
C=1-\frac{\delta\Delta +\nu^2}{ \epsilon \Omega}-
\frac{\nu}{\Omega}+ \frac{\nu}{\epsilon} \, , \;\;\;
D=1-\frac{\delta\Delta + \nu^2}{\epsilon \Omega} +
\frac{\nu}{\Omega}- \frac{\nu}{\epsilon} \, . \end{eqnarray*}

Equation for $p_2(t)$ is the same with signs in front of $\delta$,
$u_+$ and $v_+$ reversed. As a simplifying assumption we take
$S_{11}=S_{22}\equiv S$. The decoherence rates in equation
(\ref{hf28}) then are:
\begin{eqnarray}
&&\Gamma_{13} = S(\omega_+) \left[1-\frac{\nu^2}{ \epsilon
\Omega}\right] +\frac{1}{2} S(\omega_-)
\left[1+\frac{\nu^2}{\epsilon \Omega}\right]
+\frac{1}{2}S_c(\omega_-)
\frac{\nu\omega_+}{\Omega\epsilon} \, , \;\;\; \nonumber \\
\nonumber &&\Gamma_{14}= S(\omega_-) \left[1+\frac{\nu^2}{
\epsilon \Omega}\right] +\frac{1}{2} S(\omega_+)
\left[1-\frac{\nu^2}{\epsilon \Omega}\right]
-\frac{1}{2}S_c(\omega_+) \frac{\nu\omega_-}{\Omega\epsilon}
 \, , \;\;\;  \label{hf29} \\
&&\Gamma_{32}= \frac{1}{2} S(\omega_+)
\left[1-\frac{\nu^2}{\epsilon \Omega}\right] +\frac{1}{2} S_c
(\omega_+)
\frac{\nu\omega_-}{\Omega\epsilon} \, , \;\;\; \\
\nonumber &&\Gamma_{42}= \frac{1}{2} S(\omega_-)
\left[1+\frac{\nu^2}{\epsilon \Omega}\right] -\frac{1}{2} S_c
(\omega_-)
\frac{\nu\omega_+}{\Omega\epsilon} \, , \nonumber \\
\nonumber &&u_+= - S(\omega_+) \frac{\delta \Delta}{\epsilon
\Omega} \, , v_+= - S(\omega_-) \frac{\delta \Delta}{\epsilon
\Omega}\, ,
\end{eqnarray}
where $S_c(\omega)\equiv \mbox{Re} S_{12}(\omega)$. To enable
comparison of these rates to those of individual qubits, we note
that the rate of decoherence of oscillations in a single qubit
with vanishing bias is equal to $S(\Delta_j)/2$ for the $j$th
qubit.

\begin{figure}
\begin{center}
  \includegraphics[scale=0.5]{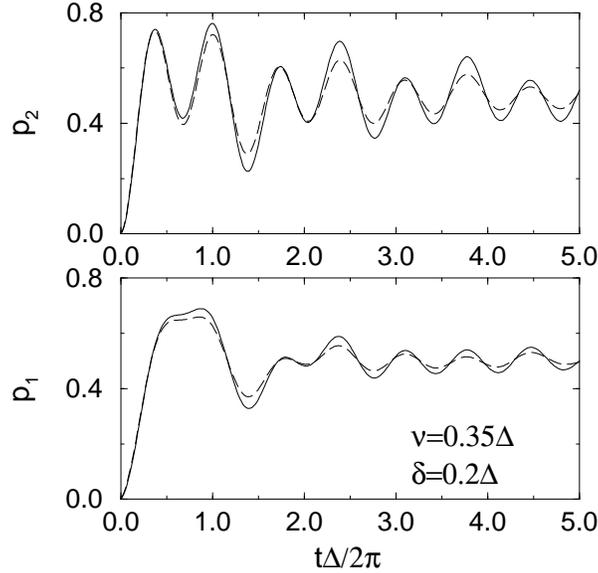}\\
  \caption{Probabilities $p_j$ to find $j$th qubit in the state
$|1\rangle$ in the process of quantum coherent oscillations
starting with the state $|00\rangle$ of two coupled qubits. The
decoherence strength is $S=0.08\Delta$. Solid and dashed lines
correspond, respectively, to the decoherence that is uncorrelated
($S_c=0$) and completely correlated ($S_c=S$) between the two
qubits. } \label{fig2.2}
\end{center}
\end{figure}

The functions $p_{1,2}(t)$ for the qubit parameters, $\delta$,
$\nu$, and $S$, close to those in experiment \cite{Pas03} are
plotted in figure \ref{fig2.2} under additional assumption that
the decoherence is the same at two frequencies, $\omega_+$ and
$\omega_-$. \footnote{The decoherence strength $S$ was taken from
the data for the single-qubit regime in \cite{Pas03}.} The curves
are plotted for the two situations: when decoherence is completely
uncorrelated ($S_{12}=0$) and completely correlated ($S_{12}=S$)
between the two qubits. One can see that the difference between
the two regimes is not very big numerically.  The correlations
between the two reservoirs leade to the effective decoherence rate
that is increased in comparison with the uncorrelated regime by
roughly $30\div50$\%, although the description with a total
correlation is not quantitatively quite appropriate -- see
equations~(\ref{hf28}) and (\ref{hf29}).

The increase of the effective decoherence rate by correlations
illustrated in figure \ref{fig2.2} can be related to the fact that
the initial qubit state, $|00\rangle$, belongs to the subspace
where the correlations increase the decoherence rate, despite the
mixing of this subspace with the DFS\footnote{Here and below we
use the term ``DFS'' for the subspace spanned by the $|01\rangle$
and $|10\rangle$ states, although for interacting qubits it,
strictly speaking, does not fully have the properties of real
DFS.} where the decoherence rate is decreased in the eigenstates
(\ref{hf7}) of the coupled qubit system.  This implies that
increase of decoherence rate by the correlations should be to a
large extent insensitive to the qubit parameters. This conclusion
is supported by the case of identical qubits ($\delta =0$), when
$u_+=v_+=0$ and equation~(\ref{hf28}) is reduced to a very simple
form:
\begin{equation}
p_{1}(t)=\frac{1}{2}-\frac{1}{4}\big[\left(1+\frac{\nu}{
\Omega}\right) e^{-\Gamma_{13}t} \cos \omega_-t  + \left(1-
\frac{\nu}{\Omega}\right) e^{-\Gamma_{32}t} \cos \omega_+t \big]\,
, \label{hf30}
\end{equation} and $p_2(t)=p_{1}(t)$. One can see from
equations~(\ref{hf29}) that both decoherence rates relevant for
equation~(\ref{hf30}), $\Gamma_{13}$ and $\Gamma_{32}$ increase
with increasing correlation strength $S_c$. Equation (\ref{hf30})
shows also that the description of the oscillation decay with a
single decoherence rate can be quite inaccurate: for weak
interaction, $\nu < \Omega$ the amplitudes of the two (high- and
low-frequency) components of the oscillations are nearly the same
while their decoherence rates can be very different.

\subsubsection{Excitation into the DFS}

Now lets discuss decoherence properties of the oscillations in
coupled qubits in the case when they start with the initial qubit
state $|10\rangle$. We note that in the case of experiment similar
to \cite{Pas03}, such an initial condition would require separate
gate control of the two qubits, since the bias change bringing
them into co-resonance is different in this state for the two
qubits. Since the state $|10\rangle$ belongs to the DFS in the
case of completely correlated noise, one can expect that
oscillations with these initial conditions will be more sensitive
to the degree of inter-qubit decoherence correlations than
oscillations with $|00\rangle$ initial condition, and that the
effective decoherence rate will decrease with correlation
strength. All this indeed can be seen from equations~(\ref{hf27})
with the initial conditions (\ref{hf26}) that correspond to the
$|10\rangle$ state. Under the same assumptions as were used in
equation (\ref{hf28}), we get for the now different $p_1(t)$ and
$p_2(t)$:
\begin{eqnarray}
p_{j}(t)=\frac{1}{2}-\frac{(-1)^j}{8}\big\{ \big[A_j
e^{-\Gamma_{42} t} +B_j
(e^{-\Gamma_{13}t}+\frac{(-1)^ju_+}{\Gamma_{13}- \Gamma_{42}}
(e^{-\Gamma_{13}t}-e^{-\Gamma_{42}t}))\big] \cos \omega_-t
\nonumber \\
+ \big[C_j e^{-\Gamma_{32}t} +D_j (e^{-\Gamma_{14}t}-\frac{(-1)^j
v_+}{\Gamma_{14}- \Gamma_{32}}
(e^{-\Gamma_{14}t}-e^{-\Gamma_{32}t})) \big] \cos \omega_+t \big\}
\, , \;\;\; j=1,2 \qquad , \label{hf31} \end{eqnarray} where
\[ A_1=1+\frac{\delta\Delta +\nu^2}{
\epsilon \Omega}+ \frac{\nu}{\Omega}+\frac{\nu}{\epsilon} \, ,
\;\;\; B_1=1+\frac{\delta\Delta +\nu^2}{\epsilon \Omega} -
\frac{\nu}{\Omega} - \frac{\nu}{\epsilon} \, , \]
\[ C_1=1-\frac{\delta\Delta +\nu^2}{ \epsilon \Omega} +
\frac{\nu}{\Omega} -\frac{\nu}{\epsilon} \, , \;\;\;
D_1=1-\frac{\delta\Delta + \nu^2}{\epsilon \Omega} -
\frac{\nu}{\Omega}+ \frac{\nu}{\epsilon} \, , \] and the
amplitudes $A_2,\, B_2\, , C_2\, , D_2$ are given by the same
expressions with $\delta \rightarrow - \delta$.

\begin{figure}
\begin{center}
  \includegraphics[scale=0.5]{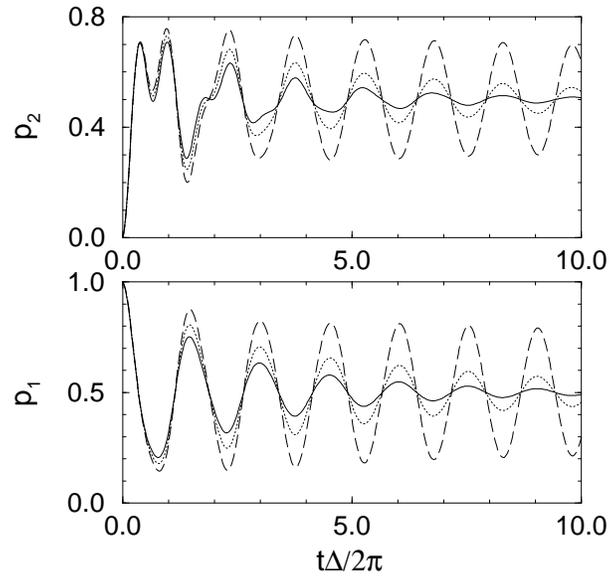}\\
  \caption{Probabilities $p_j$ to find $j^{th}$ qubit in the state
$|1\rangle$ in the process of quantum coherent oscillations
starting with the state $|10\rangle$ of two coupled qubits. Qubit
parameters are the same as in figure \ref{fig2.2}. Solid, dotted,
and dashed lines correspond, respectively, to the decoherence that
is uncorrelated ($S_c=0$), partially ($S_c=0.5S$), and completely
($S_c=S$) correlated between the two qubits. } \label{fig2.3}
\end{center}
\end{figure}

For identical qubits equation (\ref{hf31}) reduces to:
\begin{equation}
p_{j}(t)=\frac{1}{2}-\frac{(-1)^j}{4}\big[\left(1+\frac{\nu}{
\Omega}\right) e^{-\Gamma_{42}t} \cos \omega_-t + \left(1-
\frac{\nu}{\Omega}\right) e^{-\Gamma_{14}t} \cos \omega_+t
\big]\, . \label{hf32}
\end{equation} Which in the conjunction with equations~(\ref{hf29}) show that in
contrast to  equation (\ref{hf30}), the decoherence rate of the
low-frequency component that has larger amplitude is strongly
suppressed by the non-vanishing inter-qubit noise correlations
$S_c$: $\Gamma_{42}= \frac{1}{2} (1+\nu/ \Omega) [S(\omega_-) -
S_c (\omega_-) ]$. This means that the shape of the coherent
oscillations in coupled qubit starting with the state $|10\rangle$
should indeed be more sensitive to the strength of these
correlations than the shape of the oscillations starting with the
$|00\rangle$ state.  The conclusion also remains valid in the case
of not fully symmetric qubits as one can see from figure
\ref{fig2.3} which shows the shape (\ref{hf31}) of the
$|10\rangle$ oscillations for the same set of experimentally
realized parameters as in figure \ref{fig2.2}. Even in this case,
there is a pronounced weakly decaying component of the
oscillations if the decoherence is completely correlated between
the two qubits. For partial correlations, the effective
decoherence rate is reduced.

In summary, we have developed quantitative description of weakly
dissipative dynamics of two coupled qubits based on the standard
Markovian evolution equation for the density matrix. This
description shows that decoherence properties of currently
realized oscillations in coupled qubits are not very sensitive to
inter-qubit correlations of decoherence, while relatively simple
modification of the excitation scheme for the oscillations should
make them sensitive to these correlations.

Before moving on, lets briefly discuss the applicability of our
approach to the realistic Josephson-junction qubits. As we saw
above, one of the main features of equation (\ref{hf11}) is that
the pure dephasing terms disappear at the co-resonance point and
the remaining decoherence is related to the transitions between
the energy eigenstates. Therefore, within the approach based on
equation (\ref{hf11}), the decoherence rates are on the order of
half of the transition rates. On the contrary, the experiments
with charge qubits (see, e.g., \cite{Vio02}) indicate that
decoherence rates are larger than the transition rates even at the
optimum bias point when the pure-dephasing terms should disappear.
Apparently, this is related to the low-frequency charge noise
\cite{Nak02,Vio02} that is coupled to qubit strongly enough for
the lowest-order perturbation theory in coupling (\ref{hf4}) to be
insufficient. This implies that the theory presented so far might
be only qualitatively correct for realistic charge qubits, and
that a more accurate non-perturbative description of the
low-frequency noise is needed in order to achieve quantitative
agreement with experiments.

\chapter{Non-Perturbative Low-Frequency Noise}

The experimentally observed decay time $T_2$ of coherent
oscillations is typically shorter than the energy relaxation time
$T_1$ even at optimal bias points \cite{Vio02,Chi03,Nak02,Wal00}
where the perturbation theory predicts factor-of-two difference
between the two times and no pure dephasing terms. Furthermore,
the observed increase in two-qubit decoherence rates \cite{Pas03}
cannot be explained by the perturbation theory results of the
previous chapter. Qualitatively, the basic reason for discrepancy
between $T_1$ and $T_2$ is the low-frequency noise that can reduce
$T_2$ without changing significantly the relaxation rates.
Mechanisms of low-frequency, or specifically $1/f$ noise exist in
all solid-state qubits: background charge fluctuations for
charge-based qubits \cite{Pal02,Pal03,Ita03,Gal03}, and impurity
spins or trapped fluxes for magnetic qubits \cite{Pro00}.
Manifestations of this noise are observed in the echo-type
experiments \cite{Nak02}. Low-frequency noise for qubits is also
created by the electromagnetic fluctuations in filtered control
lines.

The aim of this chapter is to develop quantitative theory of
low-frequency decoherence by studying single and double qubit
dynamics under the influence of Gaussian, noise with small
characteristic amplitude $v_0$, and long correlation time $\tau$.
In the case of single qubit we will show that the expression
describing the decay times of the coherent qubit oscillations is a
non-perturbative result whose strength is controlled by the
zero-frequency noise spectral density $2v_0^2\tau$. For long
correlation times $\tau \gg \Delta^{-1}$, where $\Delta$ is the
qubit tunnelling amplitude, $2v_0^2\tau$ can be large even for
weak noise $v_0 \ll \Delta$.  Our analytical results are exact as
function of $2v_0^2\tau$ in this limit.

In the  second part of this chapter, the same non-perturbative
technique is applied to double qubit system operating at the
co-resonance point.  It is shown there that the decoherence rates
of the coupled qubits compared to the decoherence rates of the
individual uncoupled qubits qualitatively double. However, the
change in the energy spectrum in going from uncoupled to coupled
system plays an important role that quantitatively varies the
qualitative factor-of-two change of the decoherence rates in much
wider range.

\section{Single Qubit Decoherence by Low-f Noise}

We start with the standard qubit Hamiltonian with the fluctuating
bias $v(t)$,
\begin{equation}
H= -\frac{\hbar}{2}[\Delta\sigma_{x}+(\varepsilon+v(t))
\sigma_{z}], \label{lowf1}
\end{equation}
where the noise $v(t)$ has characteristic correlation time $\tau$.
Therefore, its correlation function and  spectral density can be
taken as
\begin{equation}
\langle v(t) v(t') \rangle = v_0^2 e^{-|t-t'|/\tau} \, , \;\;\;
S_v(\omega) = \frac{2 v_0^2 \tau}{1+(\omega \tau)^2} \, ,
\label{lf2} \end{equation} where $v_0$ is the typical noise
amplitude in units of radial frequency and $\langle \;
\cdots\rangle$ denotes average over different realizations of the
noise. We assume that the temperature $T$ of the noise-producing
environment is large on the scale of the cut-off frequency
$1/\tau$, and  can be treated as classical. \footnote{In the
regime of interest, $1/\tau \ll \Delta$, the temperature can
obviously be still small on the qubit energy scale.}

The decoherence is a decay of the off-diagonal element of the
qubit's density matrix in the interaction basis.  Therefore,
evaluating decoherence is equivalent to evaluating time evolution
of the $\sigma^-$ operator in the same basis. In the path integral
representation this can be expressed as:
\begin{equation}\label{PI}
    \langle\sigma^-(t)\rangle=\int\mathcal{D}[\sigma^-]\left\langle \mathcal{T}^{\dag}
    \exp\left(-i\int_{0}^{t} \hat{H}dt'\right)\sigma^-
     \mathcal{T} \exp\left(-i\int_0^t \hat{H}dt'\right)\right\rangle,
\end{equation}
where $\mathcal{D}[\sigma^-]$ represents integration over all
possible paths $\sigma^-$ can take in going form $\sigma^-(0)$ to
$\sigma^-(t)$, angled brackets represent noise averaging, and $
\mathcal{T}$ and $ \mathcal{T}^{\dag}$ are respectively forward
and reverse time ordering operators.

One contribution to the expression above is from the transitions
between the two energy eigenstates.  The transitions are caused by
the high-frequency part of the noise spectrum and they can be
described by the means of the perturbation theory as shown in the
earlier chapter. The condition of weak noise $v_0\ll \Delta$ makes
the transition rate small compared both to $\Delta$ and $1/\tau$
ensuring that the perturbation theory is sufficient for the
description of transitions. The additional effect of weak noise
dynamics affecting the qubit (\ref{lowf1}) is "pure" or adiabatic
dephasing. As discussed qualitatively earlier, the fact that the
noise correlation time is long, $\tau \gg \Delta^{-1}$, makes the
perturbation theory inadequate for the description of pure
dephasing. For low-frequency noise, a proper (non-perturbative in
$v_0^2\tau$) description is obtained by looking at the
accumulation of the noise-induced phase between the two
instantaneous energy eigenstates in the expression (\ref{PI}).

\begin{figure}
\begin{center}
  \includegraphics[scale=0.6]{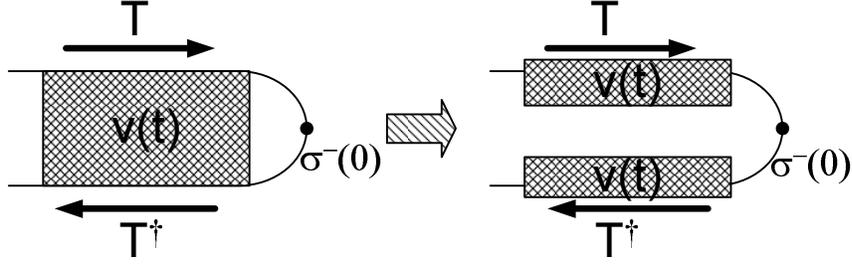}\\
  \caption{The general path integral of $\langle \sigma^-(t) \rangle$
  has a complicated average that connects forward and backward
  Keldyish time contours  (left). Neglecting the transitions, simplifies the integral
  to two independent and equal averages over each of the time directions (right).}\label{fig3.0}
\end{center}
\end{figure}

Since the transitions correspond to averaging over the all
possible states of $\sigma^-$, they can be omitted from the path
integral (\ref{PI}) by neglecting the integration over $\sigma^-$.
This simplification reduces the functional integral over a
complicated Keldysh contour to one that is considerably simpler
(fig. \ref{fig3.0}).
After putting the explicit values for the Hamiltonian
$H=-\sqrt{\Delta^2+(\varepsilon+v(t))^2}\sigma_z/2$, and noting
that $\sigma_z$ and $\sigma^-$ anti-commute, we can define the
factor $F(t)$ that describes the time-dependent, low frequency
suppression of coherence between the two states as

\begin{equation}\label{lf2z}
    F(t)\equiv\frac{\sigma^-(t)}{\sigma^-(0)}=\left\langle\exp
    \left(-i\int_0^tdt'\sqrt{\Delta^2+(\varepsilon+v(t'))^2}\right)\right\rangle.
\end{equation}

If $v_0\ll \Delta$, one can determine the rate of accumulation of
this phase by expanding the energies up to the second order in
noise amplitude $v(t)$, i.e.
$\sqrt{\Delta^2+(\varepsilon+v(t))^2}=\Omega+(\varepsilon v(t'))/
\Omega + (\Delta^2 v^2(t'))/(2\Omega^3), \;\;
\Omega^2=\Delta^2+\varepsilon^2$, and evaluating (\ref{lf2z}) in
the interaction representation. The pure dephasing term is:
\begin{equation}
F(t)= \left\langle \exp \left\{-i \int_0^t \left[\frac{\varepsilon
v(t')}{ \Omega} + \frac{\Delta^2 v^2(t')}{2\Omega^3}
\right]dt'\right\} \right\rangle \, . \label{lf3} \end{equation}

For Gaussian noise, the correlation function (\ref{lf2})
determines the noise statistics completely. In this case, it is
convenient to take the average in equation  (\ref{lf3}) by writing
it as a functional integral over noise. For this purpose, we start
with the ``transition'' probability $p(v_1,v_2,\delta t)$
\cite{Doo42,Bre02} for the noise to have the value $v_2$  time
$\delta t$ after it had the value $v_1$:
\begin{eqnarray}\label{lf3a}
p(v_1,v_2,\delta t)= \frac{1}{\sqrt{2\pi v_0^2(1-e^{-2\delta
t/\tau})}}
 \exp \left\{-\frac{(v_2-v_1e^{-\delta
t/\tau})^2 }{2\pi v_0^2(1-e^{-2\delta t/\tau})} \right\} \, .
\end{eqnarray}
Using this expression we introduce the probability of specific
noise realization as $p_0(v_1)\cdot p(v_1,v_2,\delta t_1)\cdot
p(v_2,v_3,\delta t_2)\cdot ... $, where $p_0(v)=(2\pi
v_0^2)^{-1/2} \exp \{-v^2/2v_0^2 \}$ is the stationary Gaussian
probability distribution of $v$, and $\delta t_i$ are some small
discrete time steps. Putting the explicit values for $p$'s given
above with the same $\delta t$ into the formula for $F(t)$ we
obtain the following expression
\begin{eqnarray*}
F(t) &=&  \frac{\exp \{-v_1^2/2v_0^2 \}}{(2\pi v_0^2)^{(N+1)/2}
(1-e^{-2\delta t/\tau})^{-N/2}} \times \\  \nonumber &
&\prod_{n=1}^N\exp \left\{-\frac{(v_{n+1}-v_{n}e^{-\delta
t/\tau})^2 }{2v_0^2(1-e^{-2\delta t/\tau})} \right\}\exp \left\{-i
\delta t \left[\frac{\varepsilon v_n}{ \Omega} + \frac{\Delta^2
v_n^2}{2\Omega^3} \right] \right\}.
\end{eqnarray*}
After expanding up to second non-vanishing order of $\delta t$ and
taking the limit $\delta t \rightarrow 0$, the above expression
can be written as $F(t)=S(\lambda)/S(1)$, where $S(\lambda)$ is
the following functional integral:
\begin{eqnarray}\label{lf4}
S(\lambda)&=& (const)
\times \frac{1}{\kappa}\int dv(0) dv(t) \mathcal{D}v(t')  \\
\nonumber
 &\times& \exp \left\{ -\frac{v(0)^2+v(t)^2}{\kappa} -
\int_0^z dt (\dot{v}^2+v^2+2i\alpha v) \right\} ,
\end{eqnarray}
with $\kappa$, $\alpha$ and $z$ given as
\begin{equation}\label{lf4a}
    \kappa=\left(1+\frac{2iv_0^2\Delta^2\tau}{\Omega^3}\right)^\frac{1}{2},
    \quad \alpha=\frac{\varepsilon\tau v_0}{\Omega\kappa^{3/2}},
    \quad z=\frac{\kappa t}{\tau}.
\end{equation}
Since the average in equation\ (\ref{lf3}) with the weight
(\ref{lf4}) is now given by the Gaussian integral, it can be
calculated in the usual way (i.e. see \cite{Fey65}):
\begin{eqnarray}
F(t)&=& F_0(t) \exp\left[-\alpha^2\left(\frac{\kappa t}{\tau}-
2[\coth \frac{\kappa t}{2\tau}+\kappa]^{-1} \right)\right],
\label{lf5} \\
F_0(t)&=& e^{t/2\tau} [\cosh(\kappa t/\tau)+
\frac{1+\kappa^2}{2\kappa} \sinh(\kappa t/\tau)]^{-1/2}. \nonumber
\end{eqnarray}

\subsection{Single Qubit Results and Conclusion}

\begin{figure}
\begin{center}
  \includegraphics[scale=0.5]{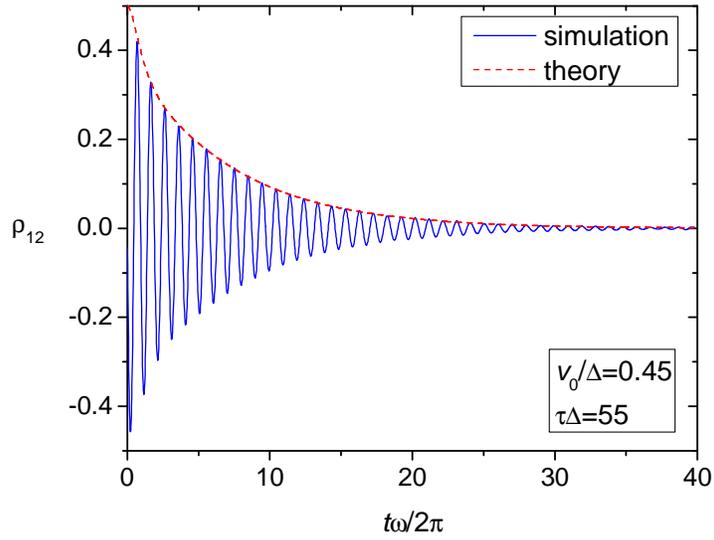}\\
  \caption{The envelope of total decoherence due to adiabatic dephasing and
  the transitions as predicted by equation (\ref{lf5}) and
  the golden rule (dotted line) plotted together with Monte Carlo
  simulation of the quantum oscillations for the same qubit (solid line).  The simulation consists of
  $10^5$ realizations over noise parameters specified on the plot.}\label{fig3.1}
\end{center}
\end{figure}

Equation (\ref{lf5}) is the main analytical result for dephasing
by the Gaussian noise. To analyze its implications, we start with
the zero-bias, (i.e. $\varepsilon =0)$ case, where pure qubit
dephasing vanishes in the standard perturbation theory. Dephasing
(\ref{lf5}) is still non-vanishing and its strength depends on the
noise spectral density at zero frequency $S_v(0)=2 v_0^2 \tau$
expressed through $\kappa= \sqrt{1+is}$, $s\equiv S_v(0)/\Delta$
for convenience. The numerical simulations show that equation
(\ref{lf5}) and the transition contributions fully account for
qubit decoherence as shown in figure \ref{fig3.1}. For small and
large times $t$ equation (\ref{lf5}) simplifies to:
\begin{equation}
F(t)= \left\{\begin{array}{cc} \displaystyle
{\left[\frac{1+t/\tau}{1+t/\tau+ist/2\tau}\right]^{1/2}},
& t \ll \tau\, , \\
2\sqrt{\kappa}e^{-(\gamma+i\delta)t}/(1+\kappa)\, , & t \gg \tau
\, ,
\end{array} \right.
\label{lf6} \end{equation} where
\begin{equation}
\gamma= \frac{1}{2\tau} \left[ \left( \frac{(1+s^2)^{1/2}+1}{2}
\right)^{1/2} -1 \right]\, . \label{lf7}
\end{equation}
Besides suppressing the coherence, the noise also shifts the
frequency of qubit oscillations. The corresponding frequency
renormalization is well defined for $t\gg \tau$:
\begin{equation*}
\delta = \frac{1}{2\tau} \left[\frac{(1+s^2)^{1/2}-1}{2}
\right]^{1/2} \, .
\end{equation*}

\begin{figure}
\begin{center}
\includegraphics[scale=0.5]{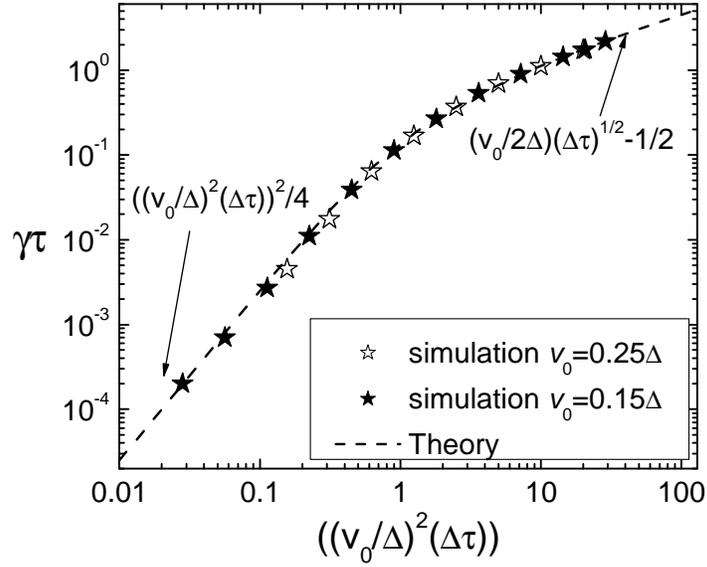}
\caption{The rate $\gamma$ of exponential qubit decoherence at
long times $t\gg\tau$ for $\varepsilon =0$ and  noise with
characteristic amplitude $v_0$ and correlation time $\tau$. Solid
line gives analytical results: \protect Eq.\ (\ref{lf7}) . Symbols
show $\gamma$ extracted from Monte Carlo simulations of qubit
dynamics. Going from left to right along the horizontal axes the
number of Monte Carlo realizations gradually changes from $10^7$
to $10^4$. Inset shows schematic diagram of qubit basis states
fluctuating under the influence of noise $v(t)$. }\label{fig3.2}
\end{center}
\end{figure}

Suppression of coherence (\ref{lf6}) for $t\ll \tau$ can be
qualitatively understood as the result of averaging over the
static distribution of noise $v$. In contrast, at large times
$t\gg \tau$, the noise appears to be $\delta$-correlated, the fact
that naturally leads to the exponential decay (\ref{lf6}). This
interpretation means that the two regimes of decay should be
generic to different models of the low-frequency noise. Indeed,
they exist for the non-Gaussian noise considered
\cite{Rab04,Rab05}, and are also found for Gaussian noise with
$1/f$ spectrum \cite{Mak03}. Crossover between the two regimes
takes place at $t\simeq \tau$, and the absolute value of $F(t)$ in
the crossover region can be estimated as $(1+s^2)^{-1/4}$, i.e.
$s$ determines the amount of coherence left to decay
exponentially. The rate (\ref{lf7}) of exponential decay shows a
transition from the quadratic to square-root behavior as a
function of $S_v(0)$ that can be seen in figure \ref{fig3.2}.  The
figure also shows that the decay rate extracted from numerical
simulations of Gaussian noise agree well with the theoretical
predictions for quite large noise amplitude $v_0$.

Non-zero qubit bias $\varepsilon$ leads to the additional
dephasing $F(t)/F_0(t)$ described by the last exponential factor
in equation (\ref{lf5}). The contribution from $F_0(t)$ is of the
same form as in zero-bias case but now with the replacement
$s\rightarrow s(\Delta/\Omega)^3$.  The additional dephasing
exhibits the crossover at $t\simeq \tau$ from ``inhomogeneous
broadening''(averaging over the static distribution of the noise
$v$) to exponential decay at $t\gg \tau$. In contrast to zero-bias
result, the short-time decay is now Gaussian:
\begin{equation*}
     \ln \left[\frac{F(t)}{F_0(t)} \right] = -
\frac{\varepsilon^2}{\Omega^2} \cdot \left\{\begin{array}{cc}
\displaystyle v_0^2 t^2/2 \, , &  t \ll \tau\, , \\
v_0^2 \tau t/(1+is(\Delta/\Omega)^3) \, , & t \gg \tau \, .
\end{array} \right.
\end{equation*}
We see again that the rate of exponential decay depends
non-trivially on the noise spectral density $S_v(0)$, changing
from direct to inverse proportionality to $S_v(0)$ at small and
large $s$, respectively.

The approach outlined above can be used to calculate the rate of
exponential decay at large times $t$ for Gaussian noise with
arbitrary spectral density $S_v(\omega)$. Such noise can be
represented as a sum of the noises (\ref{lf2}) and appropriate
transformation of the variables in this sum enables one to write
the average over the noise as a functional integral similar to
(\ref{lf4}). For calculation of the relaxation rate at large $t$,
the boundary terms in the integral (\ref{lf4}) can be neglected.
The integral is then dominated by the contribution from the
``bulk'' which can be conveniently written in terms of the Fourier
components
\[ v_n=(2/t)^{1/2} \int_0^t dt' v(t')\sin \omega_nt'\, , \;\;\;
\omega_n=\pi n/t \, ,\] and,  $\langle \cdots  \rangle= \int Dv
\cdots \exp \{ -(1/2) \sum_n |v_n|^2/S_v(\omega_n) \} $. Combining
this equation and equation (\ref{lf3}) we get at large $t$:
\begin{eqnarray}
F(t)=  \exp \left\{- \frac{t}{2} \left[ \frac{\varepsilon^2\Omega
S_v(0) }{\Omega^3+iS_v(0)\Delta^2}+\frac{1}{\pi} \int_0^{\infty} d
\omega \ln \left(1+i\frac{S_v(\omega) \Delta^2}{\Omega^3}\right)
\right] \right\} \, . \label{lf12}
\end{eqnarray} For unbiased qubit, $\varepsilon=0$, this equation
coincides with the one obtained by a more involved diagrammatic
perturbation theory in quadratic coupling \cite{Mak03}.

\section{Double Qubit}

Now we proceed to determine the adiabatic dephasing decoherence
rate for a coupled qubit system operating at the co-resonance
point. As described earlier  the model double-qubit Hamiltonian at
co-resonance point is:
\begin{equation}\label{lf13}
H_0 = \hbar\sum_{j=1,2} \Delta_j \sigma_{x}^{(j)} +\hbar\nu
\sigma_{z}^{(1)}\sigma_{z}^{(2)}+\hbar\sum_{j=1,2} v_j(t)
\sigma_{z}^{(j)} ,
\end{equation}
where again $\sigma$'s denote the Pauli matrices, $\nu$ is the
qubit interaction energy, $\Delta_j$ are the tunnelling amplitudes
of the two qubits, and $v_j$ are two weakly coupled, in general
correlated noise sources.  As it can be seen from figure
\ref{fig2.1}, applying the idea of pure dephasing being dependent
on noise induced variation of the energy splitting between the
levels of interest, is simplified at the co-resonance point by the
vanishing linear dependence between the energy levels and the
noise parameters $v_{1,2}$.

After following the same arguments as for the single qubit case of
neglecting the transition rates and only concentrating on the
variation of energy splitting between the corresponding energy
levels, the general expression for pure dephasing expressed a path
integral in the interaction picture is:
\begin{eqnarray}\label{lf14}
    \frac{\rho_{ij}(t)}{\rho_{ij}(0)}&=&\left\langle \exp\left\{i\sum_{m,n=1,2}\int_0^tdt'v_mc_{mn}^{(ij)}v_n\right\}
    \right\rangle_{v_1,v_2} \\ \nonumber
    (ij)&\in&\{(13),(14),(32),(42)\},
\end{eqnarray}
where $c_{n,m}^{(ij)}$ are elements of  $2\times2$ symmetric
matrix describing the dependence of the energy splitting as a
function of noise sources obtained by Taylor expansion
\begin{equation}\label{lf15}
    c_{n,m}^{(ij)}=\frac{1}{2\hbar}\frac{\partial^2(E_i-E_j)}{\partial v_m\partial
    v_n}.
\end{equation}
The averaging in (\ref{lf14}) is now done over two noise sources
whose dependence on integration variable $t'$ is implied.

The analytic values of the coefficients (\ref{lf15}) can be easily
obtained with the help of the perturbation theory.  As expected
from the vanishing first-order energy variations at the
co-resonance point, the first-order corrections to energy vanish,
i.e.
\begin{equation*}
    \frac{\partial E_i}{\partial v_{1,2}}=\left\langle\psi_i\left|\frac{\partial H}{\partial
    v_{1,2}}\right|\psi_i\right\rangle=0.
\end{equation*}
The second order deviation in energy is non-vanishing, and it can
be shown that
\begin{equation*}
    \frac{\partial^2E_i}{\partial v_n\partial v_m}=2\sum_{j\neq
    i}\frac{\langle\psi_i|\sigma_z^{(n)}|\psi_j\rangle\langle\psi_j|\sigma_z^{(m)}|\psi_i\rangle}{E_i-E_j},
\end{equation*}
where the wave functions are defined in (\ref{hf7}). The
coefficients of four $c^{(ij)}$  matrices are explicitly given in
table \ref{t3.1}.
\begin{table}[h]
  \centering
\begin{tabular}{|c|c|c|c|}
  \hline & & & \\
  $(ij)$ & $c^{(ij)}_{11}$ & $c^{(ij)}_{22}$
  & $c^{(ij)}_{12}$
  \\ & & & \\
  \hline \hline
   & & & \\
  $(13)$ & $\displaystyle  \frac{\epsilon\Omega-\delta\Delta+\nu^2}{\epsilon\Omega(\Omega-\epsilon)}$ &
  $\displaystyle \frac{\epsilon\Omega+\delta\Delta+\nu^2}{\epsilon\Omega(\Omega-\epsilon)}$ &
  $\displaystyle\frac{4\nu}{\Delta^2-\delta^2}$  \\ & & & \\
  \hline & & & \\
  $(42)$ & $\displaystyle \frac{\epsilon\Omega+\delta\Delta-\nu^2}{\epsilon\Omega(\Omega-\epsilon)}$ &
  $\displaystyle \frac{\epsilon\Omega-\delta\Delta-\nu^2}{\epsilon\Omega(\Omega-\epsilon)}$
  & $\displaystyle-\frac{4\nu}{\Delta^2-\delta^2}$ \\ & & & \\
  \hline& & & \\
  $(32)$ & $\displaystyle \frac{\epsilon\Omega+\delta\Delta-\nu^2}{\epsilon\Omega(\Omega+\epsilon)}$ &
  $\displaystyle \frac{\epsilon\Omega-\delta\Delta-\nu^2}{\epsilon\Omega(\Omega+\epsilon)}$
 &  $-\displaystyle\frac{4\nu}{\Delta^2-\delta^2}$ \\ & & & \\
 \hline& & & \\
  $(14)$& $\displaystyle \frac{\epsilon\Omega-\delta\Delta+\nu^2}{\epsilon\Omega(\Omega+\epsilon)}$ &
  $\displaystyle \frac{\epsilon\Omega+\delta\Delta+\nu^2}{\epsilon\Omega(\Omega+\epsilon)}$ &
   $\displaystyle\frac{4\nu}{\Delta^2-\delta^2}$ \\ & & & \\
   \hline
\end{tabular}
\caption{The  coefficients of symmetric $2\times2$ matrix
$c^{(ij)}$ for four different density matrix elements
$\rho_{ij}$.}\label{t3.1}
\end{table}

Lastly, we are left to discus the complications that arise in
evaluation of the expression (\ref{lf14}) due to the correlation
between the noise sources $v_1$ and $v_2$.  In general,
correlation between any two stochastic processes can symmetrically
be expressed through a coefficient $\lambda$ such that
\begin{eqnarray}\label{lf15a}
  v_1(t)
  &=&\frac{1}{\sqrt{2}}\left(\sqrt{1+\lambda}\,\tilde{v}_1(t)+\sqrt{1-\lambda}\,\tilde{v}_2(t)\right),
  \\ \nonumber
  v_2(t) &=& \frac{1}{\sqrt{2}}\left(\sqrt{1+\lambda}\,\tilde{v}_1(t)-\sqrt{1-\lambda}\,\tilde{v}_2(t)\right),
\end{eqnarray} where $\tilde{v}_{1,2}$ are two uncorrelated noise
sources, i.e. $\langle \tilde{v}_1(t) \tilde{v}_2(0)\rangle=0$.
Therefore, after an appropriate orthogonal transformation
(\ref{lf15a}), any averaging over two correlated noise sources can
be expressed as averaging over two uncorrelated sources.

Assuming that the two noise sources $\tilde{v}_{1,2}$ are
identical,  then $\langle\tilde{v}_{1}(t)\tilde{v}_{1}\rangle=
\langle \tilde{v}_{2}(t)\tilde{v}_{2}\rangle=v_0^2e^{-|t|/\tau}$,
and the final expression for adiabatic dephasing of the density
matrix elements needed to calculate the two probability values
$p_{1,2}(t)$ defined in (\ref{hf14z}) is:
\begin{eqnarray}\label{lf16}
    \frac{\rho_{ij}(t)}{\rho_{ij}(0)}&=&\left\langle
    \exp\left(\frac{i}{2}\sum_{m,n=1,2}\tilde{v}_{m}C_{mn}^{(ij)}
    \tilde{v}_{n} \right)\right\rangle_{\tilde{v}_{1,2}}\\ \nonumber
    (ij)&\in&\{(13),(14),(32),(42)\},
\end{eqnarray}
where
\begin{eqnarray*}
  C_{11}^{(ij)} &=& (c^{(ij)}_{11}+c^{(ij)}_{22}+2c^{(ij)}_{12})(1+\lambda), \\
  C_{22}^{(ij)} &=& (c^{(ij)}_{11}+c^{(ij)}_{22}-2c^{(ij)}_{12})(1-\lambda),\\
  C_{12}^{(ij)} &=& C_{21}^{(ij)}=
  (c^{(ij)}_{11}-c^{(ij)}_{22})\sqrt{1-\lambda^2}\; .
\end{eqnarray*}
Each noise source is fully described by conditional probability
(\ref{lf3a}), and the averaging (\ref{lf16}) is straight forward
along the lines of single qubit case.  The un-normalized action
is:
\begin{eqnarray}\label{lf16a}
  &&S(\hat{C}^{(ij)}) \nonumber= (const)\times\\ && \int \prod _{m=1,2}d\tilde{v}_m(0)d\tilde{v}_m(t)
  \mathcal{D}\tilde{v}_m \exp \left\{ -\sum_{m=1,2}\frac{\tilde{v}^2_m(0)+\tilde{v}^2_m(t)}{4 v_0^2}\right\}
  \times \\ \nonumber && \exp\left\{-\frac{\tau}{4v^2_0}\int_0^t dt'
\left[\sum_{m=1,2}\left(\dot{\tilde{v}}_m^2+\frac{\tilde{v}_m^2}{\tau^2}\right)+\frac{2iv^2_0}{\tau}\sum_{m,n=1,2}
\tilde{v}_{m}\hat{C}_{mn}^{(ij)} \tilde{v}_{n}\right] \right\}.
\end{eqnarray}
An orthogonal transformation, $\tilde{v}_{1,2}\rightarrow v_{1,2}$
that diagonalizes the matrix $C^{(ij)}$ decouples (\ref{lf16a})
into two quadratic integrals identical to those of optimally
biased  qubit (\ref{lf5}), where $\kappa^{(ij)}_{1,2}$ are in this
case the functions of $c^{(ij)}_{1,2}$ - i.e. the eigenvalues of
matrix $C^{(ij)}$. The constant term in each of the two integrals
over the paths $\mathcal{D}[\tilde{v}]$ get divided out by the
normalizing terms consisting of interaction free action
(\ref{lf16a}), i.e. $S(\textbf{1}_{2\times2})$. All put together,
the adiabatic contribution to dephasing of double qubit system
operating at co-resonant point is:
\begin{eqnarray}\label{lf17}
\frac{\rho_{ij}(t)}{\rho_{ij}(0)}&=&F_0(t,\kappa^{(ij)}_1)F_0(t,\kappa^{(ij)}_2),
\\ \nonumber F_0(t,\kappa^{(ij)}_k)&=& e^{t/2\tau} \left[\cosh(\kappa^{(ij)}_k t/\tau)+ \frac{1+(\kappa^{(ij)}_k)^2}{2\kappa^{(ij)}_k}
\sinh(\kappa^{(ij)}_k t/\tau)\right]^{-1/2}, \\ \nonumber
\kappa^{(ij)}_{1,2}&=&\sqrt{1+2i\tau v^2_0c^{(ij)}_{1,2}}, \\
\nonumber c^{(ij)}_{1,2}&=&(c^{(ij)}_{11}+c^{(ij)}_{22}+2\lambda
c^{(ij)}_{12})\pm \\
\nonumber &&\left[(c^{(ij)}_{11}+c^{(ij)}_{22}+2\lambda
c^{(ij)}_{12})^2-4(1-\lambda^2)(c^{(ij)}_{11}c^{(ij)}_{22}-(c^{(ij)}_{12})^2)
\right]^{1/2},
\\ \nonumber
    (ij)&\in&\{(13),(14),(32),(42)\}.
\end{eqnarray}
The expressions from the table \ref{t3.1} yield  the following
eigenvalues $c^{(ij)}_{1,2}$:
\begin{eqnarray*}
  c^{(13)}_{1,2} &=& A\pm\sqrt{A^2-4(1-\lambda^2)B}\;,\\
  A &=& 2\left[\frac{1}{\Omega-\epsilon}
  \left(1+\frac{\nu^2}{\epsilon\Omega} \right)
  +\frac{4\lambda\nu}{\Delta^2-\delta^2}\right], \\
  B &=& \frac{1}{(\Omega-\epsilon)^2}\left[\left(1+\frac{\nu^2}{\epsilon\Omega}\right)^2
  -\frac{\Delta^2\delta^2}{\epsilon^2\Omega^2}\right]-\frac{\nu^2}{(\Delta^2-\delta^2)^2},
\end{eqnarray*}
$c^{(42)}_{1,2}\equiv c^{(13)}_{1,2}$ with $\nu\rightarrow -\nu$,
\begin{eqnarray*}
  c^{(14)}_{1,2} &=& a\pm\sqrt{a^2-4(1-\lambda^2)b}\;, \\
  a &=& 2\left[\frac{1}{\Omega+\epsilon}
  \left(1-\frac{\nu^2}{\epsilon\Omega} \right)
  +\frac{4\lambda\nu}{\Delta^2-\delta^2}\right], \\
  b &=& \frac{1}{(\Omega+\epsilon)^2}\left[\left(1-\frac{\nu^2}{\epsilon\Omega}\right)^2
  -\frac{\Delta^2\delta^2}{\epsilon^2\Omega^2}\right]-\frac{\nu^2}{(\Delta^2-\delta^2)^2},
\end{eqnarray*}
and $c^{(32)}_{1,2}\equiv c^{(14)}_{1,2}$ with $\nu\rightarrow
-\nu$.

\subsection{Double Qubit Results and Conclusions}

\begin{figure}
  \begin{center}
  \includegraphics[scale=0.5]{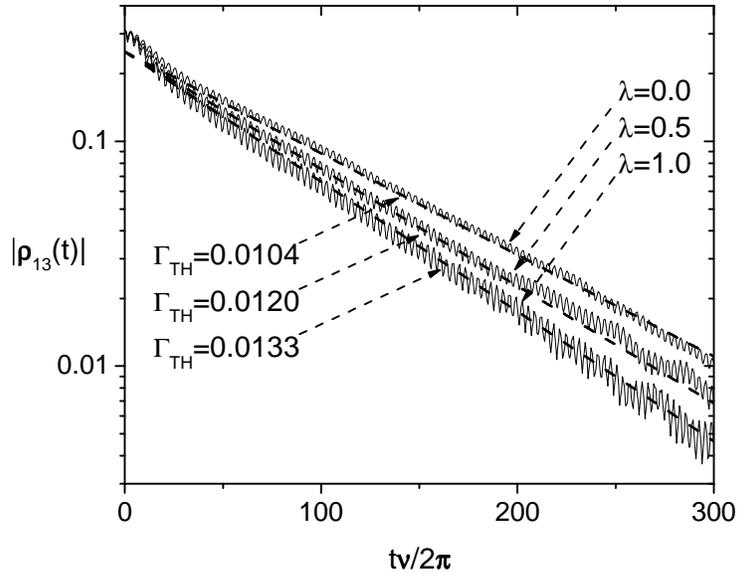}\\
  \caption{Log plot of numerical Monte Carlo simulations (solid lines)
  and theoretical decay predictions $\Gamma_{TH}$(dashed lines) for evolution
  of $|\rho_{13}|$ density matrix element for a coupled qubit
  operating at co-resonance point and with two noises exhibiting
  different levels of correlation ($\lambda=0,0.5,$ and $1)$.  The decay
  rates $\Gamma_{TH}$ are sum of high-f contributions from
  equations (\ref{hf19}) and (\ref{hf20}) and low-f contributions (\ref{lf17})
  in the limit $t\gg\tau$ as given by (\ref{lf7}).
  For the plots $\Delta=5\nu$, $\delta=3\nu$, $\tau=100/\nu$, and
  initial condition $|\Psi(0)\rangle=|00\rangle$,
  while the MC simulations consist of $10^5$ realizations.   }\label{fig3.3}
  \end{center}
\end{figure}

As we see from (\ref{lf17}) the decoherence rates of the double
qubit contain two contributions from the low frequency noise that
are identical to the one obtained for the single qubit
(\ref{lf5}). Therefore, the conclusions of the single qubit carry
over for the double qubit case.  Specifically, there are the two
regions of the decay - one when $t\ll\tau$ that is described by
averaging over the statical distribution of the noise, and the
other, $t\gg\tau$, where the noise appears to be delta-correlated
and that is characterized by exponential decay. The expressions of
the two decay coefficients at opposite time limits are given by
(\ref{lf6}) and (\ref{lf7}). Numerical simulations of the coupled
qubit dynamics show very good agreement with the theoretical
predictions of the decoherence rates as shown in the figure
\ref{fig3.3}.

Although the double qubit system contains twice as many terms that
suppress its coherent oscillations, we have to be careful not to
understate the importance the rearrangement of the energy spectrum
plays in the relative values of these coefficients and naively
assert that the double system decoherence rate doubles as compared
to the system's individual qubit decoherence rates.  This
dependence is rather non-trivial even under all the assumptions
about the system and the noise considered so far.

\begin{figure}
  \begin{center}
  \includegraphics[scale=0.5]{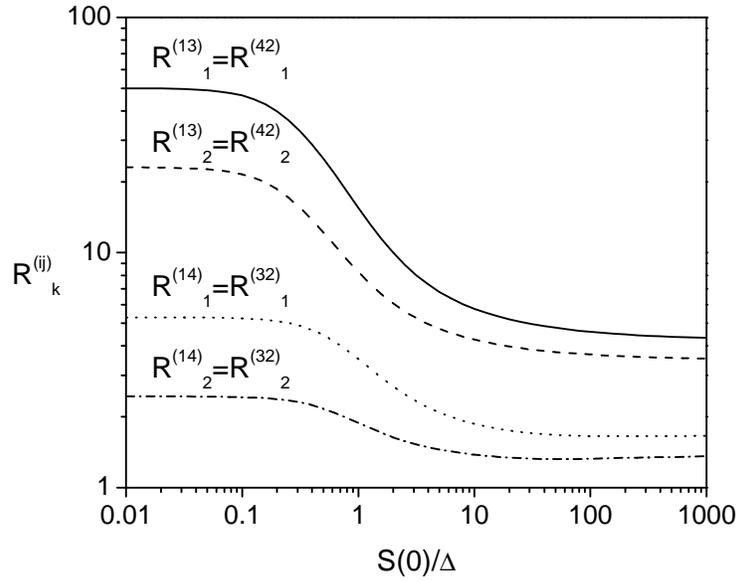}\\
  \caption{The ratio of low-f decoherence rates with the parameters taken
   from experiment \cite{Pas03} and plotted  as function of $S(0)/\Delta$,
   where $\Delta=\Delta_1+\Delta_2.$  The non-trivial relation between the
   change in the ratio of the decoherence rates between  coupled and individual
   qubits in a double qubit Hamiltonian is evident. The expected
   factor-of-two increase of the decoherence rates of the coupled qubit as
   compared to the single qubit is not realized.  Rather, it can
   vary over a much larger range.
  }\label{fig3.4}
  \end{center}
\end{figure}

To illustrate this point we will consider the experiment
\cite{Pas03}.  Under the assumption of uncorrelated, identical
baths,  the low-f dephasing terms will be the same for qubit
density matrix elements (13) and (42) as well as for the terms
(32) and (14). At this point, it is advantageous to define a
coefficient which represent the ratio of the large-time,
($t\gg\tau$), decoherence rates of the coupled qubit density
matrix elements $\rho_{ij}$ to the decoherence rates of the two
individual, uncoupled qubits $k=1,2$:

\begin{eqnarray}\label{lh18}
    R^{(ij)}_k=\frac{\gamma(S(0)c^{(ij)}_1)+\gamma(S(0)c^{(ij)}_2)}{\gamma(S(0)/\Delta_k)},
    \\ \nonumber \gamma(s)= \frac{1}{2\tau} \left[ \left( \frac{(1+s^2)^{1/2}+1}{2}
    \right)^{1/2} -1 \right],
    \\ \nonumber(ij)\in\{(13),(14),(23),(24)\}, \quad k=1,2,
\end{eqnarray}
where $c^{(ij)}_{1,2}$ are given in (\ref{lf17}) and $\Delta_k$
are individual qubit tunneling amplitudes. The the eight ratios
(\ref{lh18}) with the  parameters given in \cite{Pas03} are
plotted in figure \ref{fig3.4} as functions of zero-frequency
noise spectral density, $S(0)$, in the units of
$\Delta=\Delta_1+\Delta_2$. The factor-of-four increase in
decoherence that was observed in the experiment can be explained
by the "numerical coincidence" of zero-frequency noise spectral
density being in the range that enhances the decoherence rates of
the coupled qubits.

The exact value of ratios (\ref{lh18}) for the experiment
\cite{Pas03} can be confirmed by determining $S(0)$.  Since the
tunneling amplitudes of the qubits are different, the value of
$S(0)$  within this limit of the model can be extracted from any
two, out of two $T_1$ and two $T_2$ times.  Furthermore, if only
the times of the same type are known, i.e. both $T_1$ or both
$T_2$, they need to be measured precise enough so that their
difference shows outside the margin of error.  Unfortunately
\cite{Pas03} provides only the two $T_2$ times that are the same
within the margin of error and the exact ratio (\ref{lh18}) cannot
be determined.

In summary, the non-perturbative treatment of the now-f noise from
a fluctuators coupled to the basis-forming degrees of freedom has
provided the analytical expressions which can fully within this
model account for the increase of the decoherence rate beyond the
predictions of Fermi's Golden Rule. In general, this is attributed
to the accumulation of the phase in the qubit due to adiabatic
low-f variations of the bias.  Furthermore, the factor-of-four
increase of decoherence rate of coupled qubits as compared to the
individual qubits \cite{Pas03} can be attributed to changes of the
energy spectrum of the system when the two qubits are coupled as
supposed when they are not.  This supports "numerical coincidence"
explanation to the decoherence rate increase rather than the
existence of new decoherence mechanism.  All the results have been
verified numerically.

\chapter{Realistic Quantum Modelling of Systems Subject to General
Noise}

The ideas  of adiabatic dephasing depending on the variation of
the energy level spacing as the function of coupled weak noise and
the transition rates being influenced by the value of spectral
density of the noise at resonant frequency can be extended to a
general n-qubit case in order to determine the desired decoherence
rates of the system. Never-the-less, obtaining analytical results
forces often a general system to be constrained in many ways such
as operating at particular Hamiltonian constant in time and
subject to specific, analytic noise spectral density.

The realistic applications often call for insights in to the
dynamics of the system with time-varying parameters and subject to
multiple noise sources described by some general spectral
densities. In order to arrive to the result under these quite
general conditions, it is necessary to abandon analytical
expressions that were useful in qualitative understanding of the
noise and resort to numerical simulations.  The statistical nature
of noise makes Monte Carlo (MC) methods ideal in realizing these
simulations.

\section{Simulating the Basic Qubit Dynamics}

In the simulation of the qubit evolution all the time-dependent
qubit ($\varepsilon,\Delta$) and noise ($v,u$) parameters of the
system Hamiltonian
\begin{equation}\label{ne2}
    H(t)=-\frac{1}{2}\left[(\varepsilon(t)+v(t))\sigma_z +
    (\Delta(t)+u(t))\sigma_x\right],
\end{equation}
are considered constant during a given time step $\delta t$, and
the evolution of the qubit's density matrix $\rho$ can be
calculated exactly during the time  interval:
\begin{equation}\label{ne3}
    \rho(t_n+\delta t)=\exp(-i\delta t H_n/\hbar)\rho(t_n)\exp(i\delta t
    H^{\dagger}_n/\hbar),
\end{equation}
where $H_n\equiv H(t_n)$.  At the boundary, the final values of
density matrix are used as initial values of density matrix for
the next interval with up-to-date parameters,
\begin{equation}
    \rho(t_{n+1},H_{n+1})=\rho(t_n+\delta t,H_n).
\end{equation}
Repeating this process eventually evolves an initially specified
density matrix of a qubit to some final value along the path
specified by deterministic Hamiltonian and a random realization of
each of the noises.  Restarting this process over many different
realizations of the noises and averaging the density matrix of the
system at each time point yields complete time-evolved
MC-evaluated density matrix of the system.

The uncertainty of the averaged $\rho(t)$ can be attributed to
three sources: (1) computer truncation, (2) the assumption of
constant parameters at the discrete time steps and, (3)
statistical uncertainty due to the Monte Carlo averaging.

Since the density matrix parameters are all on the order of unity
the truncation error is at least sixteen orders of magnitude
smaller. Its accumulation does not pose any serious problems since
it is not compounded over the different realizations  and at the
extreme one realization has $10^{10}$ calculations.  Thus even in
the worst case, the result will contain error that is six orders
of magnitude smaller and therefore negligible.

The error due to the discretization  of the qubit parameters
during the interval $\delta t$ is more serious because of the
possibility of having unstable time step $\delta t$. This would
lead  to numerical result diverging away from the exact, unknown
value. The most direct way to improve on this error is to reduce
$\delta t$ or use a mid-point or some more  elaborate
extrapolating method when converting the continuous parameters to
discrete ones. In the case of unstable time step $\delta t$, the
simulation of the system without noise would produce density
matrix exhibiting non-unitary behavior. For this reason, a very
good check for an adequate choice of $\delta t$ consists of
evolving the system without the noise forward in time from
$\rho(0)$ to $\rho(t_{final})$, then reversing the time parameter,
(i.e. $\delta t\rightarrow -\delta t$), and evolving
$\rho(t_{final})$ backward in time to $\tilde{\rho}(0)$. The
unitarity of quantum mechanics implies that the two matrices
$\rho(0)$ and $\tilde{\rho}(0)$ should be identical if $\delta t$
is chosen appropriately.

With $\delta t$ chosen such that $\rho(0)=\tilde{\rho}(0)$ the
only error in the result comes form the uncertainty due to Monte
Carlo averaging. This error  is given as
\begin{equation}\label{ne1}
    \Delta(\rho_{ij}(t_n))=\sqrt{\frac{\langle\langle \rho^2_{ij}(t_n)
    \rangle\rangle-\langle\langle\rho_{ij}(t_n)\rangle\rangle^2}{N}},
\end{equation}
where $N$ is the number of realizations, and the double-angled
brackets denote arithmetic average over the realizations
\cite{Pre92}. Taking into account that the elements of density
matrix are on the order of unity, the error due MC averaging can
be made reasonably small by increasing the number of realizations.

The non-trivial nature of generating the noise form a given
spectral density of its correlator does not allow the improvement
to the variance (\ref{ne1}) by utilization of stratified or
importance sampling techniques.  The $1/\sqrt{N}$ reduction of the
variance can be improved by using the antithetic technique or by
employing quasi-random sampling.  The antithetic technique
\cite{Rub81} is implemented by using a single realization of the
noise twice - once as generated by random precess and second time
reused but with opposite sign. The quasi-random sampling assures
better covering of the sample space by generating the random
numbers with the help of number theory that assures the "filling"
of the empty space between already sampled "random" points.  This
method can reduce the $1/\sqrt{N}$ dependence to $1/N$
\cite{Pre92}.

\section{Generating Noise}

Unlike in the case of previous qubits, the numeric Hamiltonian
(\ref{ne2}) has noises affecting  both the tunneling and  the bias
of the qubit. This is often the case in the realistic situations.
For instance in the experiment \cite{Fri00} the tunneling noise
can be attributed to the impurities in the layer separating the
superconducting electrodes of the Josephson junctions or to the
variations in the modulating flux $\Phi_{xd.c.}$ as defined in
figure \ref{fig1.6}. In the same experiment the bias noise
originates form the variations in the applied flux $\Phi_x$ and
from the dc-SQUID magnetometer.  With exception to the impurity
caused noise, the noises caused by the sources above are in
general specified by some noise spectral densities $S_i(\omega)$
obtained form the known impedances of the circuits $Y_i(\omega)$
and the fluctuation-dissipation theorem \cite{Cal51}. Thus, the
random noise in the qubit simulation must be generated in a way
that it matches the given spectral density. Since the significant
portion  of noise comes from the detector, which is usually
coupled to the basis forming degree(s) of freedom of the qubit,
the discussion that follows assumes that the only noise affecting
qubit is coupled through $z$ component so that $u\equiv0$ in
(\ref{ne4}).

As discussed in \cite{Bil90}, the random variable $v(t_m)$
generated from some given $S(\omega)$ as
\begin{equation}\label{ne4}
    v(m\Delta t)=\sum_{n=0}^{M}\sqrt{4S(n\delta\omega)\delta\omega}
    \cos(n\delta\omega m\Delta t+\phi_n),
    \quad m=0..M,
\end{equation}
is a Gaussian, and in the limit $N\rightarrow\infty$ it has a
correlator whose spectral density is $S(\omega)$. In the equation
(\ref{ne4}), $\delta\omega=1/(M\Delta t)$ is discrete frequency
interval while $\phi_n$ are uniformly distributed random phases
taken randomly from the interval $[0,2\pi)$. In the numerical
implementation of the same equation, the Fourier transform is
replaced by fast Fourier transform (FFT).  In doing so, it is
important to choose $M$  and $\Delta t$ so that the cutoff time
$(M\Delta t)$ is larger than the desired simulation time, and the
cutoff frequency $\omega_{c}$ exceeds the qubits maximal
transition frequency.  It is easily shown that
$\omega_{c}=1/\Delta t$ \cite{Pre92}. Furthermore, in order to
avoid numerical aliasing caused by approximating Fourier transform
by FFT it is necessary to set the cutoff to be
factor-of-two\footnote{This is not a strict limit. It was arrived
to, through various computer simulations as well with the help
from the references \cite{Pre92,Bri74}.} larger than the maximal
level spacing of the qubit $\Omega_{max}$ \cite{Pre92,Bri74}.

\begin{figure}
\begin{center}
  \includegraphics[scale=0.6]{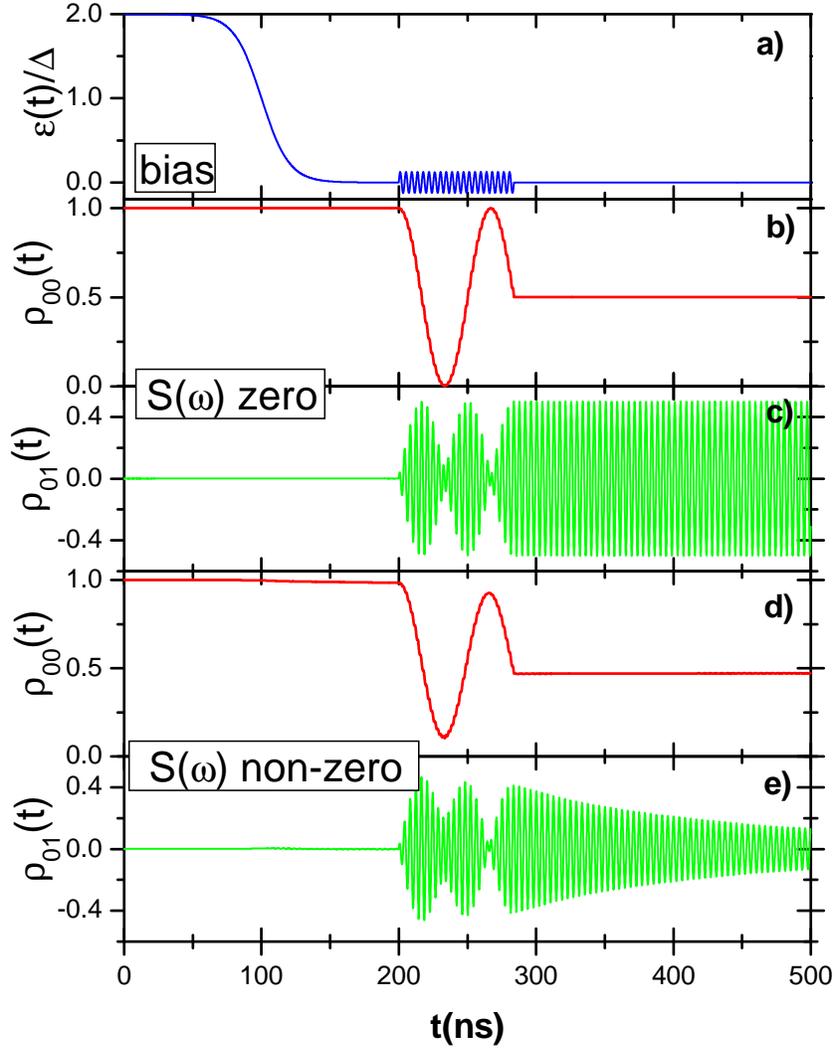}\\
  \caption{Hadamard transformation modelled by varying the bias as depicted in
  the graph a).  The plots b) and c) show the variation of the respective density
  elements without noise while the plots d) and e) show the same evolution subject to
  noise.  The simulation assumes $ \Delta$ to be constant, while
  $S(\omega)$ was supplied numerically.}\label{fig4.2}
  \end{center}
\end{figure}

The discrete Fourier transform given in (\ref{ne4}) can be
expressed as a real part of more general complex transform
\begin{equation}\label{ne5}
    z(m\Delta t)=\sum_{n=0}^{M}\left[\sqrt{4S(n\delta\omega)\delta\omega}
    \exp(i\phi_n)\right]\exp(-inm/M),
    \quad m=0..M,
\end{equation}
where the random phase is now included in the harmonic
coefficient.  This way written transform is more favorable for FFT
computation because the random phase can be factored out.  Also,
real and imaginary parts of noise $z(m\Delta t)$ are uncorrelated
since $\cos(nx)$ and $sin(mx)$ are two orthogonal functions.
Preforming one complex FFT yields two realizations of the noise,
thus speeding up the simulation significantly.  The antithetic
variance reduction technique discussed earlier is used at this
point in which case there are total of four noise realizations
after single fourier transform (\ref{ne5}).

Lastly the noise realization given in the time steps of $ \Delta
t$ needs to be expressed in the steps of the simulation $\delta
t$.  This can be achieved along the same lines of discretization
of the continuous parameters of the Hamiltonian (\ref{ne4})
discussed at the beginning of this chapter.

The approach outlined so far is very powerful, and it can produce
results impossible to produce by analytical methods. Hadamard
transform of a qubit state is a such example, and it's simulation
shown in the figure \ref{fig4.2}.

\section{Inclusion of Temperature}

Up to this point the temperature has not entered into the
discussion, so the noise affecting the qubit is high-temperature
noise with $k_BT/\hbar\gg\Delta, \varepsilon$. This implies that
noise-induced up and down transition rates are the same.
Realistically this is not the case since qubit experiments are
performed on temperature scales of the order of $10mK\ll\Delta,
\varepsilon$, where the up rate is severely suppressed.

\begin{figure}
\begin{center}
  \includegraphics[scale=0.5]{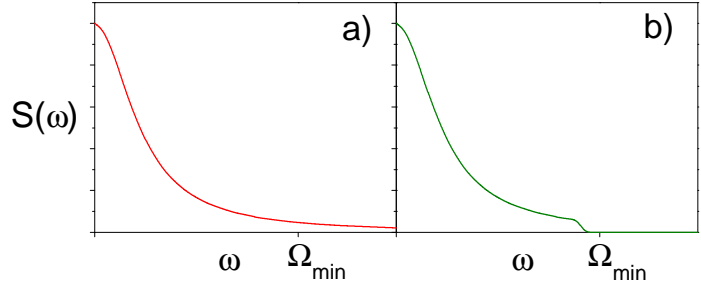}\\
  \caption{In order to eliminate transitions from the simulation the original
  spectral density a) needs to be truncated at the higher frequencies
  $\omega_{tr}\lesssim\Omega_{min}$ b).}\label{fig4.3}
\end{center}
\end{figure}

\begin{figure}
\begin{center}
  \includegraphics[scale=0.65]{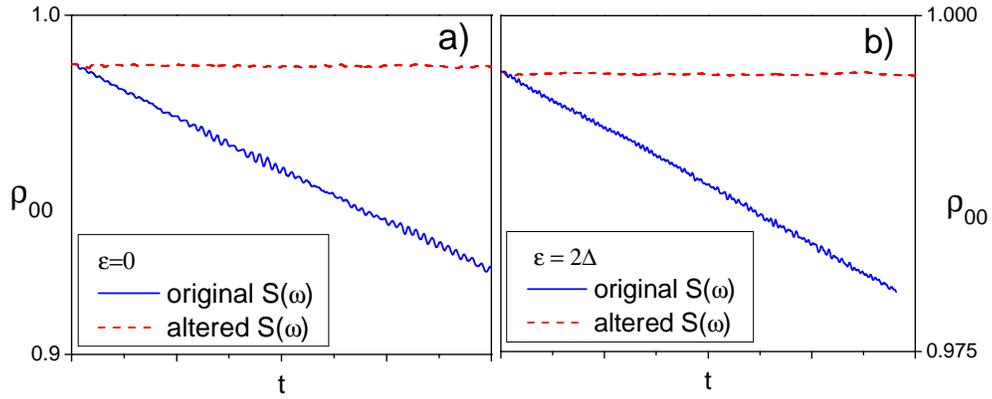}\\
  \caption{Two simulations with bias a) $\varepsilon=0$  and b) $\varepsilon=2\Delta$ each run
  with original (solid line) and truncated (dashed line) spectral densities.  The  lack of the
  relaxation, a phenomenon solely caused by transitions, is apparent
  for truncated spectral density, while the relaxation is present in the simulation with original
  spectral density. }\label{fig4.4}.
\end{center}
\end{figure}

As shown in the second chapter,  the transitions are easily
treated by perturbation theory and they depend on the
high-frequency part of the spectral density.  Thus truncating the
spectral density at the frequencies larger than some frequency
$\omega_{tr}$ that is on the order and also below minimal qubit
level spacing $\Omega_{min}$ (figure \ref{fig4.3}) eliminates the
transitions from the simulation as shown in figure \ref{fig4.4}.
It is important to realize here that although the values of
spectral density above the truncation frequency are zero, the
value $\Delta t=1/\omega_c$ in (\ref{ne5}) stays fixed. If this
was not the case, the removal of the transition frequencies form
the generated noise will not be complete because it will contain
aliased contribution. Aliasing is a serious problem in the signal
processing \cite{Bri74}. The best way to assure that the aliasing
is negligible while at the same time keeping in mind that
increasing cut-off is prolonging the computation time, is to pick
up lowest cut-off frequency of the noise spectrum for which the
truncated spectral density produces no energy relaxation, i.e.
$T_1\rightarrow \infty$.

\begin{figure}
\begin{center}
  \includegraphics[scale=0.5]{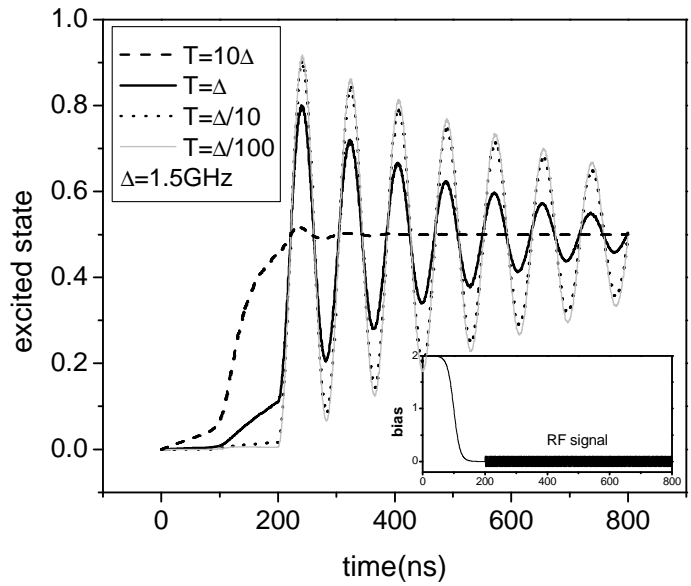}\\
  \caption{Plot of Rabi oscillations at the optimal point $(\varepsilon=0)$
  simulated with different temperature values and the same rf-varied bias (insert).}\label{fig4.5}
\end{center}
\end{figure}

Since the analytic expressions for the decay rates that obey the
detailed balance relation are known (e.g. see \cite{Ave00a}), the
temperature dependent transitions are accounted for after adding-
in "by hand" the appropriate coefficients in the equation
(\ref{ne3}).
Working explicitly in the interaction basis of the qubit, the
diagonal density matrix elements  acquire "by hand" inserted decay
that restores the transitions removed by the truncation of the
spectral density. The same can be said about the high-f decay
component of the off-diagonal matrix element. Unlike before, the
restored transition rates can be made to obey the detailed balance
relation with an arbitrary temperature. The explicit modifications
to the equation (\ref{ne3}) are:
\begin{eqnarray}
  \rho^d_{00}(t_n+\delta t) &=& \frac{\gamma_d}{\gamma_u+\gamma_d}+\rho^d_{00}(t_n)\exp[-(\gamma_u+\gamma_d)\delta t] \\
  \rho^d_{01}(t_n+\delta t) &=& \rho^d_{01}(t_n)\exp\left[-\left(i\Omega+\frac{\gamma_u+\gamma_d}{2}\right)\delta
  t\right].
\end{eqnarray}
The rates are given as
\begin{equation}
    \gamma_d=\frac{\Delta}{2\Omega}S(\Omega), \quad
    \gamma_u=e^{-k_BT(\Omega)/\Omega\hbar}\gamma_d,
\end{equation}
where $T(\Omega)$ is effective temperature defined through FDT as
\begin{equation}\label{ne6}
    S(\omega)=Re(Y(\omega))\frac{\hbar \omega}{2\pi}\coth\frac{\hbar
    \omega}{2k_BT(\omega)}.
\end{equation}
The inclusion of temperature requires additional specification of
effective temperature $T(\omega)$ or environmental resistance
$Re(Y(\omega))$ over the transition frequency range.
The identical simulations run at various temperatures  exhibit
increasing decoherence rates with the increase of the effective
temperature.  The figure \ref{fig4.5} shows temperature dependence
of  Rabi oscillations of excited qubit state caused by the
rf-pumping (insert). The direct relation between temperature and
the degradation of coherent oscillations is obvious.

\section{Generalization to Larger Systems of Qubits}

The method outlined above for simulating the  dynamics of qubit
can be extended to more complex qubit systems under the influence
of classical noise.  Inclusion of the temperature dependent
transitions like as done above is impossible since the
perturbative analytic expression for time decay of the system's
density matrix elements do not exist.  In addition, the dimension
of the Hilbert space scales exponentially $(2^N)$ to the number of
the qubits $N$, and the density matrix approach is increasingly
inefficient due to need for multiple, large-matrix
multiplications. These shortcomings could be overcome by use of
stochastic wave function approach
\cite{Dal92,Dum92,Gar92,Dum92a,Car93} that is unfortunately beyond
this work.

\part{Quantum Measurement}

\chapter{Continuous Weak Measurement with Mesoscopic Detectors}

\section{Quantum Measurement and Mesoscopic Detectors}

Textbook examples of quantum measurements often recall the
Stern-Gerlach  experiment where the measured phenomenon is quantum
mechanical in nature, but the measurement itself is purely
classical and could only determine the amplitudes of the system's
wave function if repeated on a large number of identically
prepared systems. This is a direct consequence of the design of
the measuring equipment where the measured system, i.e. the beam
of the electrons, is absorbed by the detector - a photographic
plate or florescent screen. For this reason, the quantum state is
always destroyed by the measurement.

These destructive measurements are not general representative of
the quantum measurement process  as formulated by von-Neuman and
the others \cite{Whe83,Mey83,Mil90,Bra92,Men00}. The postulate of
quantum measurement asserts that the measurement of an observable
quantity $q$ which gives a value $q_j$ leaves the density matrix
of the measured system $\rho$ in the state
\begin{equation}\label{qm1}
    \frac{1}{Tr\{|q_j\rangle\langle q_j|\rho\}}
    |q_j\rangle\langle q_j|\rho|q_j\rangle\langle q_j|,
\end{equation}
where $|q_j\rangle$ is the  wave function of the corresponding
measured value $q_j$ and the tracing in the denominator is done
over both systems.  Thus, measuring a general quantum state of a
system and obtaining some result $q_j$ reduces the density matrix
of the measured system in a way that the obtained result is now
incorporated into the total, updated density matrix, where
$|q_j\rangle$ are defined by the detector-system coupling.


In practice, the detector needs to be capable of measuring the
state of the system while being at the same time  minimally
detrimental to the measured system's coherence. Since the
coherence is purely quantum phenomenon, its preservation imposes
necessity of measuring the quantum system by entangling it with
another system that also exhibits some form of dynamics on the
same energy scale. Furthermore, the need of having the result in
the form of a classical, measurable signal makes mesoscopic
phenomenon such as Coulomb blockade or mesoscopic transport in
single or multichannel conductors prime candidates for a such
task.

By entangling a such mesoscopic detector with the measured system,
the observation of a detector outcome does not need to destroy the
measured system.  If the detector is perfect, the reduction of the
measured system's matrix is given by (\ref{qm1}). In general, the
detector can act back onto the quantum system and the general
state after the measurement can be represented as

\begin{equation}\label{qm3}
    \frac{1}{Tr\{|q_j\rangle\langle q_j|\rho_{sys}\}}
   \hat{\Omega}(q_j)\rho_{sys}\hat{\Omega}^{\dag}(q_j),
\end{equation}
where $\Omega(q_j)$ describes the total back action of the
detector on the measured system. If  $\Omega(q_j) \equiv
|q_j\rangle\langle q_j|$ density matrix of the measured system
will remain pure and the detector will be preforming in ideal or
quantum-limited regime. The exact form of the operators
$\Omega(q_j)$ depends on the type of the detector.

If the measurement outcomes are orthogonal, i.e.
$|q_i\rangle\langle q_j|=\delta_{ij}$ then $|q_j\rangle$ are
eigenvectors of the measurement basis and the measurement
preformed by the detector will be projective or von-Neumann
measurement. Assuming the quantum limited regime for now, the
state of the quantum system after the measurement of $q_j$ will be
projected into state given in (\ref{qm1}).  As a consequence of
reduction of the density matrix and the orthogonality of
$|q_i\rangle\langle q_j|$, any further measurements of the same
quantity will give the identical outcome.

Orthogonality condition implies the direct correlation between the
quantum system and the detector.  Achieving this requires strong
coupling between the two, which in general induces strong detector
back-action beyond the limit of the ideal measurement. Reducing
the back-action of the detector on the quantum system can be
achieved by making the detector-system coupling weak. This leads
to imperfect correlation between the detector output and the
measured observable which is reflected in non-orthogonality of the
operators $|q_i\rangle\langle q_j|$. Relaxing the orthogonality
constraint also implies that two successive measurements of the
static quantum system do not need to give the same outcome, since
two or more different outcomes can be implied by the same signal,
i.e. the states $|q_j\rangle$ are now some superpositions of the
measurement basis states. In this case the output does not tell
what state was measured, but it reflects the likelihood of the
system being in the states that all could with some weights belong
to the observed output. The successive weak measurements of the
quantum system collapse its wave function to one of the states
determined by the system-detector coupling. The slow reduction of
the wave function enables continuous measurement of the quantum
system and avoids the possibility of abrupt energy relaxation that
is characterized by projective measurements which can result in
total destruction of the quantum state due to the secondary
reasons \cite{Sid04}.

A way to study weak measurement is in the framework of linear
response theory where the interaction of weak detector force $f$
and the quantum system coordinate $x$ is assumed to be linear.
\footnote{In certain situations the lowest order detector-system
coupling is not linear but the analysis is carried along the
similar perturbative steps outlined below, e.g. see \cite{Mao04}.}
The Hamiltonian is specified as $H=H_q+H_d+xf$, which in addition
to the interaction coupling contains individual Hamiltonians of
the quantum system $H_q$ and the mesoscopic detector $H_d$. Large
size of the detector justifies the assumptions that the detector
is static and that the changes in the detector are observable in
the form of some classical output signal $o(t)$.  To the first
order correction, the detector output can be written as consisting
of the detector's noise $v(t)$ and measurement signal superimposed
over it,
\begin{equation}\label{qm4}
    o(t)=v(t)+\frac{i}{\hbar}\int d\tau[f(\tau),v(t)]x(\tau).
\end{equation}
Since different times correspond to different and independent
measurements, the noise of the detector is white and  is
completely specified by its correlator $\langle
v(t)v(t')\rangle_d=2\pi S_v\delta(t-t')$, where $S_v$ is the noise
spectral density that is assumed constant at the relevant
low-frequency range of the output signal. The delta function is a
consequence of the instantaneous detector response. The detector
acts on the measured system through the detector force $f(t)$
which induces stochastic motion to the quantum system. In this
light, the correlator of the force describes the back-action of
the detector, and it as well is characterized by some spectral
density $S_f$ constant in the relevant frequency range, it
responds instantaneously and is independent of its values at the
different times, thus $\langle f(t)f(t')\rangle_d=2\pi
S_f\delta(t-t')$. The average detector force component is assumed
to be zero, $\langle f \rangle_d\equiv0$, since any non-vanishing
value can be absorbed by renormalizing the detector Hamiltonian
$H_d$. The commutator of the cross-correlator between the noise
and the detector force $i\langle
[f(t'),v(t)]\rangle_d=\hbar\lambda\delta(t-t'-0)$ essentially
describes the information extraction from the measured system by
the detector.  The infinitesimal shift in the detector response is
required in order to impose the causality in (\ref{qm4}).

In the case of the continuous measurement, the extraction of
information from the measured system is not instantaneous but it
is spread out over some time $\tau_m$. Complementary to the
extraction of the information from the quantum system the
detector's back-action is dephasing the measured system's
wave-function.  In the case of static quantum system
($H_q\equiv0$), the density matrix $\rho$ of the measured system
evolves in the eigenbasis of the observable $q$ as:

\begin{equation}\label{qm4a}
    \dot{\rho}_{jj'}=-\frac{i}{\hbar}(q_j-q_{j'})f(t)\rho_{jj'},
\end{equation}
whose solution is given by
$\rho_{jj'}(t)=\rho_{jj'}(0)\exp(-i(q_j-q_{j'})/\hbar\int^t
f(t')dt')$. After the averaging over the different realizations of
$f(t)$ and using the well known result, $\langle \exp(x)
\rangle=\exp\langle x^2/2 \rangle$, the  decay of the density
matrix elements due to the detector back-action is:

\begin{equation}\label{qm4b}
    \rho_{jj'}(t)=\rho_{jj'}(0)\exp(-\Gamma_dt),\quad
    \Gamma_d=\frac{\pi S_f (q_j-q_{j'})^2}{\hbar^2}.
\end{equation}

At the same time, the dc component of the detector output signal
changes by the value $\delta o=\lambda(q_j-q_{j'})$ where
$\lambda$ is the detector response coefficient defined earlier. In
order to read-out the output signal, the background noise needs to
subside to at least the half value of the $\delta o$.  Since  the
average background detector noise $\Delta o$ averaged over some
time interval $\Delta t$ decays inversely to it as $\Delta o=(2\pi
S_v/\Delta t)^{1/2}$, the measurement time $\tau_m$ is:
\begin{equation}\label{qm4c}
    \tau_m=\frac{8\pi S_v S_f}{[\lambda(q_j-q_{j'}))]^2}.
\end{equation}
From (\ref{qm4b}) and (\ref{qm4c}) the measurement time and
back-action dephasing rate $\Gamma_d$ are related as

\begin{equation}\label{qm5}
    \tau_m\Gamma_d=8(\pi/\hbar\lambda)^2S_qS_f.
\end{equation}

As detailed in \cite{Ave03}, the detector force and noise spectral
densities can be defined as vector products since they are assumed
constant at low frequencies:

\begin{eqnarray*}
    S_{l}(\omega)=\frac{1}{2}\left[S_{ll}(\omega)+S_{ll}(-\omega)\right]\\
    S_{lk}\equiv\langle l|k\rangle,\quad l,k\in\{v,f\}.
\end{eqnarray*}
Then the response coefficient can be expressed as $\lambda=4\pi Im
(S_{fv})/\hbar$ and the Schwartz inequality $S_{vv}S_{ff} \ge
|S_{fv}|^2$, yields that
\begin{equation}\label{qm5a}
    |\lambda|\le\frac{4\pi}{\hbar}\left[S_fS_v-Re(S_{fv})^2\right]^{1/2}.
\end{equation}
Combining (\ref{qm5}) and (\ref{qm5a}) establishes a fundamental
relation between the measurement time and back-action decoherence
rate,
\begin{equation}\label{qm5b}
    \tau_m\Gamma_d\ge\frac{1}{2}.
\end{equation}
Equation (\ref{qm5b}) implies that the acquisition of information
is never faster than the dephasing due to back-action. At best,
the quantum system can be measured as fast as it is being
dephased.  This occurs for the detectors for which
$Re[S_{fv}(\omega)]=0$, where explicitly $S_{fv}=\int d\tau
exp(i\omega t)\langle fv(\tau)\rangle_d$.  The vanishing real part
of spectral density function between the noise and the back-action
implies that the detector noise is not correlated with the
detector force thus the detector back-action is solely caused by
exchange of information and not by the detector excitations of the
measured system. This-way optimized detector is known as
quantum-limited or optimal detector.

\section{Quantum Point Contact as a Detector}

The discussion above can be illustrated with a specific example of
quantum point contact (QPC) detector situated in the proximity of
the Cooper pair box qubit. QPC is a one dimensional channel whose
resistive properties in the tunneling regime are very sensitive to
electric fields \cite{Fie93,Gur97,Kat99,Kor99,Spr00}. Therefore,
the current through biased QPC is modulated by charge present in
the box.

If the variations in the transmission of QPC  are linear, then the
detector and the interaction Hamiltonians are respectively:

\begin{eqnarray}\label{qm6}
   H_d &=&
  \sum_{j=l,r}\sum_{i}E_{i}\hat{a}^{\dag}_{ij}\hat{a}_{ij}
  \\ \nonumber
  H_i &=& \frac{U}{2}\sigma_z \qquad
  U=\sum_{j,j'=l,r}U_{jj'}\sum_{i,i'}
  \hat{a}^{\dag}_{ij}\hat{a}_{i'j'},
\end{eqnarray}
where $\hat{a}^{\dag}_{l,r}$ and $\hat{a}_{l,r}$ are electron
creation and annihilation operators of the left and the right QPC
electrodes, and $E_{i}$ are energies of electrons populating $i$th
level. The coupling $U$  is caused by the changes in the electron
scattering potential $\pm U(x)$ of QPC, where $\pm$ depends on
weather the cooper pair occupies the box or not. Since the qubit
is in a stationary state, its Hamiltonian is equivalent to zero.
Left and right wave function of the electrons in the electrodes,
$\Psi(x)_{l,r}$, define the scattering matrix elements as
$U_{ij}=\int\Psi^*_i(x)U(x)\Psi_j(x)dx$.

Under the assumption of the QPC bias $V$ larger than the qubit
parameters and the environment temperature
$(eV>>\hbar\varepsilon,\hbar\Delta,k_BT)$, the average current
passing through QPC is classical and in the zeroth order the
correlators are
\begin{eqnarray}\label{qm6y}
  \langle U(t)U(t+\tau)\rangle &=& \frac{e\hbar V}{4\pi}\frac{(\delta T)^2+u^2}{TR}\delta(t) \\
  \langle U(t)I(t+\tau)\rangle &=& \frac{e^2V}{2\pi}(i\delta T+u)\delta(\tau-0) \\
  \langle I(t)I(t+\tau)\rangle &=&
  \frac{e^3VTR}{\pi\hbar}\delta(\tau)+\frac{(\delta I)^2}{4}\langle
  \sigma_z\sigma_z(t)\rangle, \label{qm6z}
\end{eqnarray}
where $T$ and $R$ are transmission and reflection probabilities,
$\delta T$ is the change in transmission due to location of the
cooper pair, while $u$ is a dimensionless parameter that does not
affect the current through the contact. Rather, it reflects
asymmetry of the coupling between the box and QPC \cite{Kor01}.

The back-action dephasing can be obtained by solving an equation
of motion for the reduced qubit density matrix.  Since the qubit
is static, only its off-diagonal elements are of interest.  In
this case, the transition rate through the point contact depends
on the qubit state and after the averaging over $U$ as done in
going form  (\ref{qm4b}) to (\ref{qm4c}), the off-diagonal density
matrix element of the qubit evolves according to
\begin{equation}\label{qm7}
    \dot{\rho}_{01}=-\Gamma_d\rho_{01}, \quad
    \Gamma_d=\frac{eV}{\hbar}\frac{(\delta T)^2+u^2}{8\pi TR}.
\end{equation}
The asymmetry factor $u$ does not affect the current response,
therefore it has an effect of increasing decoherence rate without
any increase of the data acquisition rate $\Gamma_m=(\delta
I)^2/4S_0$, where $S_0\equiv 2e^3VTR/\pi\hbar$ is shot-noise
spectral density and $\delta I=e^2\delta TV/\hbar
\pi$\cite{Kor01,Kor03}. The ratio of two decoherence rates,
\begin{equation}\label{qm7a}
    \frac{\Gamma_m}{\Gamma_d}=\frac{(\delta
    T)^2}{(\delta T)^2+u^2}\le 1,
\end{equation}
again is such that information extraction is always limited from
the above by the qubit decoherence.

Since the asymmetry factor does not affect the current response,
it has no effect on information extraction from the qubit. On the
other hand, the qubit decoherence (\ref{qm7}) is increased by
non-zero $u$ and the  ratio (\ref{qm7a})  is the most optimal for
$u=0$. This illustrates the interplay of back-action decoherence
and extraction of information - only ideally optimized QPC $(u=0)$
preforms quantum measurement that extracts information from the
qubit as fast as it is dephasing the quantum state of the qubit by
its back-action. For non-optimal case $(u \ne 0)$,  a part of the
information about the qubit is contained in the phase differences
between the forward and backward scattering electrons in the QPC.
Since this information cannot be extracted by measuring the
ballistic properties of the QPC but by rather some "elaborate", or
better said un-achievable interference between the forward and
backward scattering channels, the information is lost and the QPC
detector will dephase the qubit before it is able to measure its
state completely. It is important to point out that the limitation
(\ref{qm7a}) is independent of the strength of the qubit-detector
coupling, rather it is an universal limitation of quantum
mechanics.

\section{Conditional Measurement}

In the regime of the quantum-limited detection, the overall
evolution of the detector and the measured system is
quantum-coherent and the only source of the information loss is
averaging over the detector. For a detector, different outcomes of
the evolution are classically distinguishable, and it is
meaningful to ask how the measured system evolves for a given
detector output. In the quantum-limited regime, specifying
definite detector output eliminates all losses of information, and
as a result there is no back-action dephasing present in the
dynamics of the measured system conditioned on specific detector
output.

Conditional description in the quantitative form is obtained (see,
e.g., \cite{Kor03,Dal92,Bre02}) by separating in the total wave
function the terms that correspond to a specific classical outcome
of measurement and renormalizing this part of the wave function so
that it corresponds to the total probability of 1.

In the conditional measurement approach the measured system is
specified by its wave function
$|\Psi\rangle=\sum_i\alpha_i|q_i\rangle$.  Conditional that the
single detector particle was transmitted/reflected the amplitudes
$\alpha_t$ are updated as

\begin{equation}\label{qm7b}
    \alpha_i\rightarrow\frac{\alpha_i
    t_i}{\sqrt{\sum_j|\alpha_jt_j|^2}},\quad \alpha_i\rightarrow\frac{\alpha_i
    r_i}{\sqrt{\sum_j|\alpha_jr_j|^2}},
\end{equation}
where $t_i,r_i$ are the values of the detector
transmission/reflection as functions of the state $|q_i\rangle$ of
the measured system.

It is important to stress that the changes in the coefficients
$\alpha_i$ for a system with vanishing Hamiltonian (as we assumed
from the very beginning) is unusual from the point of view of
Schr\"{o}dinger equation, and provides quantitative expression of
reduction of the wave function in the measurement process.

If the dynamics of the detector is not quantum limited, then the
information is lost and the dephasing is non-vanishing even in
conditional evolution. To generalize equations (\ref{qm7b}) to the
case of finite dephasing, we need to look at the change of the
density matrix of the measured system due to a single detector
event.  If the detector particle is specified as some free wave
superposition $|\Psi\rangle=\int dkb(k)|k\rangle$, then after
tracing out the detector degrees of freedom from the total,
system-plus-detector density matrix,  the density matrix
$\rho_{ij}$ of the measured system changes as
\begin{eqnarray}\label{qm7c}
  \rho_{ij} & \rightarrow & \int dk|b(k)|^2(t_i(k)t^*_j(k))/\sum_{j'}\rho_{j'j'}T_{j'} \\
  \rho_{ij} & \rightarrow & \int
  dk|b(k)|^2(r_i(k)r^*_j(k))/\sum_{j'}\rho_{j'j'}R_{j'} \label{qm7d}
\end{eqnarray}
where
\begin{equation*}
    T_j=\int dk|b(k)|^2(t_j(k)t^*_j(k)), \quad R_j=1-T_j
\end{equation*}
are total transmission and reflection probabilities for a given
state $|q_j\rangle$ of the measured system. From these relations
it is straight forward to see that quantum-limited detection
requires the density matrix elements to remain pure in the
conditional evolution.

\section{Measurement of Quantum Coherent Oscillations}

In order to study the measurement of the quantum coherent
oscillations, the output signal correlator $\langle
o(t+\tau)o(t)\rangle$ needs to be evaluated explicitly for a
measured system in non-static case.  In the specific example of
QPC detector, the correlator is the current-current correlator
given in (\ref{qm6z}).  The first part of the expression is the
contribution form the shot noise while the second part is the
detector response to the qubit oscillations.  To evaluate this
$\langle\sigma_z\sigma_z(\tau)\rangle$ correlator, it is necessary
to solve the equation of motion of the reduced  density matrix of
the qubit, which after substitution $\rho=1/2+\sigma$ yields:
\begin{equation}\label{qm7u}
    \dot{\sigma}_{00}=\Delta Im[\sigma_{01}],\quad \dot{\sigma}_{01}=
    (i\varepsilon-\Gamma_d)\sigma_{01}-i\Delta\sigma_{11},
\end{equation}
where the back-action decoherence rate is given in (\ref{qm7}) and
the normalization condition $\sigma_{11}=-\sigma_{00}$ was used to
simplify the second equation.  This equation is in general
solvable, and in the case of zero bias ($\varepsilon=0$), the
output signal spectral density $S(\omega)=2\int\langle I(0)
I(\tau)\rangle e^{i\omega\tau}d\tau$ is:
\begin{equation}\label{qm8}
    S(\omega)=S_0+\frac{\Gamma\Delta^2(\delta I)^2}
    {(\omega^2-\Delta^2)^2+\Gamma^2\omega^2},
\end{equation}
where $S_0\equiv 2e^3VTR/\pi\hbar$ is  the shot noise spectral
density.

The zero-bias point is the most optimal for the  qubit
oscillations.  For this value, the spectral density is peaked at
the qubit oscillation frequency $\Delta$.  The maximal value of
the peak above the shot noise is $S_{max}=(\delta I)^2/ \Gamma$,
and $S(\omega)$ is limited from above such that signal-to-noise
ratio is
\begin{equation}\label{qm9}
    \frac{S_{max}}{S_0}=\frac{4(\delta T)^2}{(\delta T)^2+u^2}\le 4.
\end{equation}
This again illustrates the limitation between the information
extraction and the back-action decoherence.  In the static case
the limit was exhibited through the relation between the rate of
decoherence and the rate of information extraction, while in the
dynamic case it is exhibited through the signal-to-nose ratio that
is limited to four from above.  As before, the limitation is
fundamental (i.e. independent on the qubit-detector coupling).  It
represents the fact that in the time domain the oscillations are
drowned in the detector's shot noise.

The analytic expression for the output spectral density is
possible after solving (\ref{qm7u}) with Cardain's formula, but
the cumbersome expressions are not very insightful. Numerical
evaluations of the spectral density for finite bias qubit can be
found in \cite{Kor03}.

\section{Quantum Non-Demolition Measurement}

Even if ideally optimized, the quantum measurements discussed so
far do not exceed the standard quantum limit given by (\ref{qm7a})
or (\ref{qm9}). This is a consequence of the general choice for
the system-detector coupling.  For this reason, the measurement
signal contains information about both conjugate variables,
$[p,q]=\pm i\hbar$ of the quantum system. The Heisenberg
uncertainty about them $(\Delta p\Delta q\ge \hbar/4)$, is
contained in the output signal and it prevents the signal from
reflecting one of them exactly.

The need to measure the mesoscopic observables beyond the quantum
limit has produced a special group of measuring schemes known as
quantum non-demolition (QND) measurements
\cite{Cav80,Bra96,Gra98,Pei99}. They all are based on the idea of
measuring only one, i.e. $q$, of the conjugate variables of the
system. This way constructed detector still acts back and perturbs
the unobserved conjugate variable $p$, but since the detector is
only coupled to $q$ the output signal will not exhibit any
consequences of the back-action on $p$.  In other words the
detector is coupled to the constant of motion of the measured
system.  Mathematically this implies that the measured observable
and the  operator $\hat{\Omega}(q_i)$ defined in (\ref{qm3})
commute \cite{Bra92},
\begin{equation}\label{qm16}
    [q,\hat{\Omega}(q_i)]=0.
\end{equation}

Specifically, in the qubit measurement discussed so far the
signal-to-noise limit (\ref{qm9}) of monitoring quantum coherent
qubit oscillations demonstrates the inability of the measurement
scheme to obtain the information about the qubit beyond the
quantum limit. The origin of this limit can be traced to the
nature of the qubit-detector coupling that is only sensitive to
$\sigma_z$ projection of the qubit in some general, non-diagonal
basis.  For this reason, the detector is measuring directly the
oscillating coordinate and consequently it localizes it.  This is
reflected through appearance of additional "quantum-limiting
dephasing" which constrains the signal-to-noise spectral density
of the output signal to four from above. Two known ways of
overcoming this obstacle and realizing the QND measurement of
qubit dynamics revolve around an idea of making the qubit-detector
coupling time-dependent \cite{Ave02,Jor05}.

The first of the designs solves this problem on the most
straight-forward way - by coupling the detector to the unbiased
qubit such that the detector  follows the qubit as it oscillates
in $zy$-plane. If the  coupling is oscillating with some frequency
$\omega_c$ the system is described by Hamiltonian:
\begin{equation}\label{qm18}
    H=H_d-\frac{1}{2}\Delta\sigma_x-
    \frac{f}{2}\left[\cos(\omega_ct)\sigma_z+\sin(\omega_ct)\sigma_y\right].
\end{equation}
In the regime of small detuning $\gamma$  between the measurement
frame oscillation $\omega_c$  and the oscillation of the qubit
$\Delta$, linear response language yields the spectral density of
the output signal to be \cite{Ave02}:
\begin{equation}\label{qm19}
    S(\omega)=S_0+\frac{\lambda^2}{2\pi}\frac{\Gamma_e+\gamma}
    {\omega^2+(\Gamma_e+\gamma)^2},
    \quad \gamma \equiv \frac{(\omega_c-\Delta)^2}{\Gamma},
\end{equation}
where $\lambda$ is the linear response coefficient, $\Gamma_e$ is
the environment induced suppression of the coherence, and $\Gamma$
is the detector back-action decoherence rate.  In the case of zero
detuning, $\gamma= 0$ and zero environment decoherence,
$\Gamma_e=0$, the signal to noise ratio diverges for dc values of
the output signal and the limit of standard quantum measurement
(\ref{qm7a}) is surpassed.

A somewhat different scheme introduced in \cite{Jor05} achieves
the QND measurement by employing the kicked-qubit technique where
the charge qubit is coupled to QPC with the periodically switched
coupling. In this case the detector has access to the qubit only
at approximately discrete times.  Ideally, the coupling part of
the total Hamiltonian is a sequence of periodic delta functions
separated by $\tau$ and of strength $f/2$,
\begin{equation}\label{qm20}
    H=-\frac{1}{2}\left[\varepsilon\sigma_z+\Delta\sigma_x\right]+
    \sum_{n=-\infty}^{\infty}\delta(t-n\tau)\frac{f(t)}{2}\sigma_z+H_d.
\end{equation}
The density matrix of initially pure quantum state
$|\Psi\rangle=\alpha|R\rangle+\beta|L\rangle$ after $n$ kicks with
$\tau=2\pi/\Omega$ is
\begin{equation}\label{qm21}
    \rho(n)=\left(%
\begin{array}{cc}
  |\alpha|^2 & Z\alpha\beta^* \\
  Z^*\alpha^*\beta & |\beta|^2 \\
\end{array}%
\right),
\end{equation}
where $Z=exp[in\langle f\rangle-n\langle (\delta f)^2\rangle/2]$
is decay factor obtained by central limit theorem \cite{Fel45}. It
is obvious that the measurement scheme is preserving the
information about the $\sigma_z$ projection of the qubit, while
the information about the remaining two projections
$\sigma_x,\sigma_y$ is lost. The spectral density of the detector
output is discrete, but again it diverges at dc frequency value.

If the "kicked"  QND measurement technique is applied to unbiased
qubit, ($H_q\equiv\Delta\sigma_x/2$), then it is possible to
synchronize the measurement pulses so that the detector interacts
with the qubit only when the off-diagonal elements are identical
to zero. Ideally in this case, the measurement will not affect the
qubit evolution at all.  Realistically, the coupling is not
instantaneous, and the detector interacts with the qubit when the
off-diagonal elements are nearly zero.  In this process, the
detector induces some small relaxation still beyond quantum limit
decay rate that eventually collapses the qubit. More about this
will be said in the following chapter.

\chapter{Rapid Single Flux Detector}

Many flux qubits employ the measurement scheme based on the
ability of a flux qubit to modulate the rate of tunneling out of
the stationary supercurrent carrying states  of a Josephson
junction or a SQUID. This process can be viewed as tunneling of a
magnetic flux quantum and has several attractive features as the
basis for measurement. Most important one is the sufficiently
large sensitivity which comes from the strong dependence of the
tunneling amplitude on the parameters of the tunneling potential
controlled by the qubit.  There is  an important disadvantage to
this approach. In simple few-junction systems, the supercurrent
decay brings the system into finite-voltage state characterized by
large energy dissipation.  This strongly perturbs the system and
the detector itself, and makes it impossible to repeat the
measurement sufficiently quickly.  This strong perturbation
destroys the state of the measured quantum system by changing it
on an uncontrolled way. Both of these factors prevent the
realization of the non-trivial quantum measurement strategies
introduced in the previous section.

The goal of this chapter is to suggest and analyze the flux
detector based on the tunnelling of individual magnetic flux
quanta, which avoids the transition into the dissipative state
after fluxon-qubit entanglement by utilizing their ballistic
motion\footnote{Another way of avoiding the transition into the
dissipative state in the course of measurement is to detect the
variations in the junction impedance caused by the measured
system.  Such "impedance -measurement" schemes \cite{Ill03,Wal05}
enable one to perform measurements continuously, but they
typically require narrow-band coupling between the detector and
the measured system which in turn limits their time resolution.}.
The detector has quantum-limited back-action on the qubit and time
resolution sufficient for realization of non-trivial quantum
measurements of superconducting qubits.

\section{Josephson Transmission Line as a Flux Detector}

\begin{figure}
\begin{center}
  \includegraphics[scale=0.6]{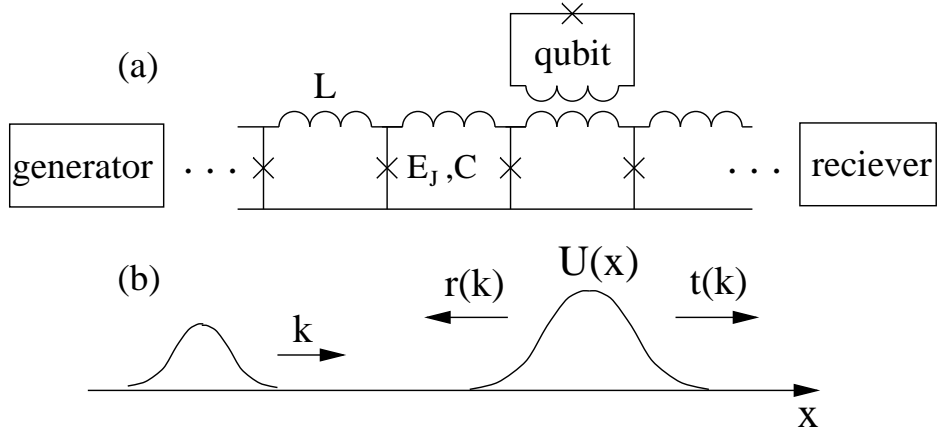}\\
  \caption{(a) Equivalent circuit of the flux detector based on the Josephson
  junction transmission line (JTL) and (b) diagram of scattering of the fluxon
  by the potential U(x) that is controlled by the measured qubit.  The fluxons are
  periodically injected into JTL by the generator and their scattering characteristics
  (transmission and reflection coefficients t(k), r(k)) are registered by the
  receiver.}\label{fig6.1}
  \end{center}
\end{figure}

The detector is based on the ballistic motion of the fluxons in
the Josephson transmission line (JTL) formed by unshunted
junctions with critical currents $I_c$, and capacitances $C$
coupled by inductances $L$ as shown in  figure \ref{fig6.1}.  The
detector can be viewed as the flux analog of QPC detector
introduced in the previous chapter.  Both detectors are based on
the ability of qubit to control the ballistic motion of the
independent particles (electrons in QPC and fluxons in JTL
detector) through one-dimensional channel - the role played by JTL
in the case of JTL detector as discussed in the introduction.
Since the injection of fluxons can be controlled, the detector can
be optimized to perform new measurement schemes (e.g. time-delay
measurement) otherwise impossible with QPC.

In the flux detector, the flux $\phi^{(e)}(x)$ generated by the
measured system creates potential $U(x)$ for the fluxons moving
along JTL (figure \ref{fig6.1}).  The potential $U(x)$ is
localized in some area on the JTL which acts as a scattering
region for the incident fluxons.  The fluxons are one-by-one,
periodically with period $1/f$, injected by the generator  and
their scattering characteristics (transmission probability through
$U(x)$ or the time delay associated with the same potential) are
registered by receiver. Since the scattering properties of the
fluxons depend on $U(x)$ that is controlled by the measured
system, the scattered fluxons contain the information about the
state of the system. The injection frequency is set sufficiently
low so that only one fluxon at a time moves inside JTL.

In what follows, we are interested in the regime of low segment
inductances $L\ll\Phi_0/I_c$ where $\Phi_0=\pi\hbar/e$ is the
magnetic flux quantum.  This-way optimized JTL acts as a uniform
ballistic fluxon channel.  The phase difference across each
segment is small, and the junction can be viewed as distributed
structure described by standard sine-Gordon Lagrangian,
\begin{eqnarray}\label{jtl1}
    L=\sqrt{2\epsilon_J \epsilon_L}\int dx\left\{
    \frac{1}{2}(\partial_{\tau}\phi)^2
    -\frac{1}{2}(\partial_{x}\phi)^2+\cos(\phi)-\phi^{(e)}(x)\partial_x\phi\right\},
\end{eqnarray}
where $x$ is expressed in the units of Josephson penetration depth
$\lambda_J=\sqrt{2\epsilon_L/\epsilon_J}$.  The Josephson energy
is replaced by Josephson energy density
$\epsilon_J=\Phi_0J_c/2\pi$ (i.e. $I_c \rightarrow J_c$), while
the  segment inductance $L$ becomes junction inductance density
$l_0$ which gives inductive energy $\epsilon_L =
(\Phi_0/2\pi)^2/2l_0$ units of "energy" $\times$ "distance".  The
time is conveniently expressed in the units of inverse plasma
frequency $\omega_p^2=\epsilon_j/c_0$ where $c_0$ is the
capacitance density of the line. The commutation relation for the
charges and phases of individual junction give the following
equal-time commutation relations for the field $\phi(\tau,x)$:
\begin{equation}\label{jtl2}
    [\phi(x),\partial_{\tau} \phi(x')]=\beta^2\delta(x-x'),
\end{equation}
where $\beta^2\equiv(4e^2/\hbar)/\sqrt{l_0/c_0}$ measures the wave
impedance $\sqrt{l_0/c_0}$ of the JTL in the absence of Josephson
tunneling relative to quantum resistance.

Known results for quantum sine-Gordon model (e.g. see
\cite{Raj82}) show that for $\beta^2\ge 8\pi$, i.e.
$\sqrt{l_0/c_0}\ge \hbar/e^2\simeq 25k\Omega$, quantum
fluctuations of the field $\phi$ completely destroy the
quasiclassical excitations of the junction.  This situation is
similar to that in small Josephson junction \cite{Sch83}, and the
dynamics of the supercurrent flow in JTL should then be described
in terms of tunneling of individual Cooper pairs \cite{Ave85}.
While this regime might be reachable in very narrow JTLs of
sub-micron width \cite{Kov99} we assume in this work the more
typical situation when the impedance $\sqrt{l_0/c_0}$ is on the
order of $100\Omega$, and $\beta^2\ll1$.  In this case, the JTL
supports a number of quasiclassical  excitations including most
importantly for this work topological solitons that each carry
precisely one quantum of magnetic flux.  The dynamics of such
"fluxons" is equivalent to that of stable, in general relativistic
particles \cite{Raj82} with terminal velocity
$c=\lambda_j\omega_p$ and mass
\begin{equation}\label{jtl3}
    m=\sqrt{2\epsilon_J\epsilon_L}/\lambda_J^2\omega_p^2.
\end{equation}
Another type of quasiclassical excitations in JTL are the
small-amplitude plasmon waves with frequency
\begin{equation}\label{jtl4}
    \omega(k)=(\omega_p^2+c^2k^2)^{1/2},
\end{equation}
where $k$ is the plasmon wave vector.

In this work we are interested in "non-relativistic" regime of
fluxon dynamics, when the fluxon velocity $u$ is small in
comparison to terminal velocity $c$, i.e. $u\ll c$. The equations
(\ref{jtl3}) and (\ref{jtl4}) show that in this regime, the fluxon
terminal energy $\epsilon=k^2/2m$ can be made smaller than the
lowest plasmon energy $\hbar\omega_p$,
\begin{equation}\label{jtl5}
    \epsilon/\hbar\omega_p=(2u/c\beta)^2,
\end{equation}
so that for $u<c\beta$ the fluxon cannot emit a plasmon even when
it is scattered by non-uniformities of the JTL potential
\cite{Shn97}.  Intrinsic dissipation associated with emission of
plasmons is then suppressed, and the fluxon motion in JTL is
elastic, provided that other, "extrinsic", sources of dissipation
are sufficiently weak. Another case when plasmon dissipation is
reduced including the cases of fast fluxons, is a situation of
sufficiently smooth JTL potential. In this case the plasmon
emission precesses are suppressed even when they are energetically
possible. Although the JTL operation as the flux detector should
be possible for any strength of fluxon dissipation, significant
dissipation would prevent the detector from reaching the
quantum-limited regime.

The shape of the scattering potential $U(x)$ for fluxons created
by the measured system is determined by the convolution of the
distribution of the flux $\Phi^{(e)}$ with the distribution of
current in each fluxon \cite{Fog76,Mcl78}, and can be written as:
\begin{equation}\label{jtl6}
    U(x)=\frac{\Phi_0}{2\pi l_0}\int dx'\frac{\partial \Phi^{(e)}(x')}{\partial x}
    \phi_0(x'-x),
\end{equation}
where $\phi(x)$ is the shape of the fluxon that in general can be
distorted by the potential $U(x)$ itself.  If the potential is
smaller than $\omega_p$, or does not vary appreciably on the scale
on the size of the fluxon given by $\lambda_J$, the changes in the
fluxon shape are negligible and one can use in equation
(\ref{jtl6}) the regular fluxon shape in the uniform case.  In
non-relativistic limit the shape of the fluxon is given by
sine-Gordon kink soliton $\phi_0(x)=4 \arctan[\exp(x/\lambda_J)]$.
One of the implications of equation (\ref{jtl6}) for the JTL
detector is that the width of the scattering potential $U(x)$
cannot be made smaller than $\lambda_J$.

\begin{figure}
\begin{center}
  \includegraphics[scale=0.6]{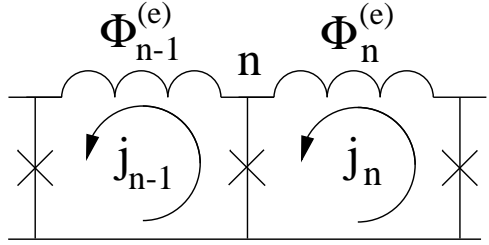}\\
  \caption{Equivalence between the magnetic and galvanic coupling to
the detector. External fluxes $\Phi^{(e)}_n$ through the JTL cells
induce the circulating currents $j_n$ in them, which in turn
create the currents $j^{(e)}_n=j_n-j_{n-1}$ through the JTL
junctions. The situation would be the same if the currents
$j^{(e)}_n$ are injected directly into the JTL junctions by
external system.}\label{fig6.1a}
\end{center}
\end{figure}

If the measured system is not coupled to JTL inductively as in
figure \ref{fig6.1}, but rather galvanically, and injects the
current with density $j^{(e)}(x)$ in the nodes of JTL as show in
in figure \ref{fig6.1a}, the potential created in the junction is
still given by the equation (\ref{jtl6}) if one substitutes
\begin{equation}\label{jtl7}
    (1/l_0)\partial \Phi^{(e)}/\partial x=j^{(e)}(x).
\end{equation}

\begin{figure}
\begin{center}
  \includegraphics[scale=1.33]{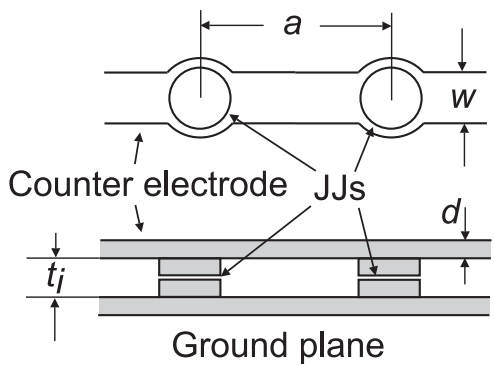}\\
  \caption{Geometry of one cell of the multi-layer JTL: top view and
vertical cross-section. Internal layers of the structure are used
to create Josephson junctions (JJs), while the two external
layers, the ground plane and the counter-electrode, separated
vertically by the distance $t_i$, define inductance $L$. The
length $a$ of the cell is set by the distance between the
junctions.}\label{fig6.1b}
\end{center}
\end{figure}

Another important remark is that although the assumed condition
$\beta^2\ll 1$ makes quantum fluctuations of the fluxon shape
small, the dynamics of the fluxon as whole can still be completely
quantum, as recently observed experimentally \cite{Wal03}.
Achieving the limit of small $\beta^2$ demands low capacitance of
the junction. This can be achieved by in reality constructing the
JTL consisting of small junctions and inductances, rather than out
of uniform electrodes (see figure \ref{fig6.1b}). As discussed in
\cite{Ave05}, the above description is still valid if the cell
size is made smaller than the fluxon length $\lambda_J$.

Operation of the JTL detector is based on free, non-relativistic
dynamics of fluxons of mass (\ref{jtl3}) on the potential
described by (\ref{jtl6}) and (\ref{jtl7}).  It also requires that
the fluxons are injected by the generator into one end of the JTL
and detected by the detector on the other end (figure
\ref{fig6.1}).  Both of these circuits can be designed following
the general principles of SFQ electronics \cite{Lik91} and will
not be explicitly discussed here\footnote{Adaption of SFQ
electronics to qubit applications \cite{Sem03,Sav05} is an
important and not yet fully understood part of the development of
scalable superconducting qubits}.  It is simply assumed that the
ends of JTL are matched appropriately to these circuits  so that
the fluxons can enter or leave the JTL without reflection and are
injected into the JTL in appropriate quantum state.

Initial quantum state $\psi(x,t=0)$ of an injected fluxon is
characterized by the average fluxon velocity $u_0$ and the shape
of the wave packet $\psi_0(x)$ defining its position:
\begin{equation}\label{jtl8}
    \psi(x,t=0)=\psi_0(x)e^{ik_0x},\quad k_0=mu_0.
\end{equation}
As it will be seen form the discussion in the next section, many
properties of the JTL detector are independent of the specific
shape of the wave packet $\psi_o(x)$, as long as it is well
localized both in coordinate and momentum space of the fluxon.
These properties depend on the wave packet width $w$ in coordinate
space and the corresponding width $\delta k\sim1/w$ in momentum
space.  These parameters should satisfy two obvious conditions:
(1) $w\ll l $, where $l$ is the total length of JTL, and (2)
$\delta k \ll k_0$.  We assume a stronger version of the latter
condition that follows from the requirement that the broadening of
the wave packet by $\delta x \sim \delta kt/m$ due to momentum
uncertainty and travel time $t \sim l/u_0$ is negligible in
comparison to the original width: $\delta x \ll w$.  The
discussion below shows that this requirement is necessary for the
quantum-limited operation of the detector in the time-delay mode.
The two conditions put together limit the width $w$ such that:
\begin{equation}\label{jtl9}
    (l/k_0)^{1/2}\ll w \\ l.
\end{equation}

In the cases when it is necessary to specify the shape of the wave
packet $\psi_0(x)$ we will take it to be Gaussian:
\begin{equation}\label{jtl10}
    \psi_0(x)=(\pi
    w^2)^{-1/4}\exp\left[-\frac{(x-\bar{x})^2}{2w^2}\right],
\end{equation}
where $\bar{x}$ is the initial fluxon position in the JTL. Besides
being well localized both in coordinate and momentum space, the
wave pocket (\ref{jtl10}) can be obtained as the result of fluxon
generation process naturally implementable with SFQ circuits. The
fluxons satisfying (\ref{jtl9}) and (\ref{jtl10}) can be obtained
in two steps: (1) relaxation of the fluxon to the ground state of
weakly damped and nearly quadratic Josephson potential with
required width $w$ of the wave function of the ground state, and
(2) rapid acceleration to velocity $u$ after the potential is
switched off.  The fluxons injected in JTL in the state
(\ref{jtl8}) can be used for measurements as described in the
following section.

\section{Measurement Dynamics of the JTL Detector}

The measurement by the JTL detector consists qualitatively in
scattering of the fluxons in the JTL of the potential controlled
by the measured system.  This process has the simplest dynamics if
the measured system is stationary.  In this case, it is convenient
to consider it in the basis of the eigenstates $|q_j\rangle$ of
the system operator (e.g. magnetic flux in the qubit loop in the
example shown in figure \ref{fig6.1}) which couples to the JTL. In
each state $|q_j\rangle$ the system creates different potential
$U_j(x)$ for the fluxons propagating through JTL. Different
realizations of $U_j(x)$ produce different scattering coefficients
for injected fluxons: the amplitude $t_j(k)$ of transmission and
the amplitude $r_j(k)$ of reflection back to toward the generator.
Both amplitudes depend on fluxon momentum $k$.  Since the
scattering amplitudes depend on the state $|q_j\rangle$ of the
measured system, scattered fluxons carry information about
$|q_j\rangle$.

For the JTL detector, the detector-system entanglement arises as a
result of fluxon scattering, and the rates of information
acquisitions and back-action dephasing can be expressed in terms
of the scattering parameters.  In this respect JTL is very similar
to QPC detector and its characteristics can be obtained following
the similar derivation \cite{Ave05} for QPC.

Evolution of the density matrix $\rho$ of the measured system in
scattering of one fluxon can be obtained by considering first the
time dependence of the total wave function of the fluxon injected
into JTL and the wave function $\sum_j\alpha_j|q_j\rangle$ of the
measured system:
\begin{equation}\label{jtl11}
    \psi(x,t=0)\cdot \sum_j \alpha_j|q_j\rangle\rightarrow
    \sum_j \alpha_j\psi_j(x,t)|q_j\rangle.
\end{equation}
Here the initial fluxon wave function $\psi(x,t=0)$ is given by
equation (\ref{jtl8}) and its time evolution $\psi(x,t)$ depends
on the realization  $U_j(x)$ of the scattering potential created
by the measured system.  Evolution of $\rho$ is then obtained by
taking the trace over the fluxon part of the wave functions
(\ref{jtl11}):
\begin{equation}\label{jtl12}
    \rho_{ij}=\alpha_i\alpha_j^* \rightarrow \alpha_i\alpha_j^*
    \int dx \psi_i(x,t)\psi_j^*(x,t).
\end{equation}

Qualitatively, the time evolution in equation (\ref{jtl11})
describes the propagation of the initial wave packet (\ref{jtl8})
towards the scattering potential and then separation of this wave
packet in coordinate space into the transmitted and reflected
parts that are well-localized on the opposite sides of the
scattering region.  If we assume that the scattering potential
$U(x)$ has a simple shape (e.g. without narrow quasi-bound states)
and is non-vanishing only in some small region of size $a\ll l$,
the time $t_{sc}$ from the fluxon injection to completion of the
scattering is not much different from the time $l/u_0$ of free
fluxon propagation through JTL.  Then at time $t>t_{sc}$, the
separated wave packets move in the region free from $j$-dependent
scattering potential and the unitarity of the quantum mechanical
evolution of $\psi_j(x,t)$ implies that the overlap of the fluxon
wave functions in (\ref{jtl12}) becomes independent of $t$.

This overlap can directly be found in momentum representation:
\begin{equation}\label{jtl13}
     \int dx \psi_i(x,t)\psi_j^*(x,t)=\int dk |b(k)|^2
     \left[t_it_j^*+r_ir_j^*\right].
\end{equation}
Here $b(k)$ is the probability amplitude for the fluxon to have
momentum $k$ in the initial state (\ref{jtl8}), e.g. in the case
of the Gaussian wave packet (\ref{jtl10})
\begin{equation}\label{jtl14}
    b(k)=(w^2/\pi)^{-1/4}\exp\left[\frac{(k-k_0)^2w^2}{2}-i(k-k_0)\bar{x}\right].
\end{equation}

Equations (\ref{jtl12}) and (\ref{jtl13}) show that the diagonal
elements of the density matrix $\rho$ do not change in the process
of scattering of one fluxon, while the off-diagonal elements are
suppressed by the factor
\begin{equation}\label{jtl15}
    \eta=\left|\int
    dk|b(k)|^2\left[t_i(k)t_j(k)^*(k)+r_i(k)r_j^*(k)\right]\right|\le
    1 \;,
\end{equation}
where the inequality is the consequence of Schwartz inequality as
mentioned in the previous chapter.  After summing  the suppression
factors for all fluxons injected at a frequency $f$, the rate of
back-action dephasing is obtained as
\begin{equation}\label{jtl16}
    \Gamma_{ij}=-f\ln\left|\int
    dk|b(k)|^2\left[t_i(k)t_j(k)^*+r_i(k)r_j^*(k)\right]\right|.
\end{equation}

Equation (\ref{jtl16}) is similar, but not identical, to the
back-action dephasing rate by the QPC detector.  The main
difference is at the stage of momentum summation, since different
momentum electrons scatter independently, while this is not the
case for fluxons.  They are constrained by the condition that
whole fluxon is either transmitted or reflected by the potential,
since the scattering events of different fluxons are well
separated in time.  As we will see below, this difference makes it
possible to operate the JTL detector in time-delay mode that is
not possible with the dc-biased QPC detector.

As examples of applications of equation (\ref{jtl16}) we consider
several specific cases motivated by the measurement regimes
discussed below.  In one, the phases of the scattering amplitudes
are assumed to cancel out form equation (\ref{jtl16}), while the
variation of the absolute values of the amplitudes is small. Then,
the dephasing rate (\ref{jtl16}) can be expressed in the terms of
variations $\delta T_j(k)$ of the fluxon transmission probability
in different states $|j\rangle$ around some average transmission
$T(k)$: $|t_j(k)|^2=T(k)+\delta T_j(k)$, $\delta T_j(k)\ll T(k)$.
To the lowest non-vanishing order in $\delta T_j(k)$ we get:
\begin{equation}\label{jtl17}
    \Gamma_{ij}=f\int dk |b(k)|^2\frac{(\delta T_i(k)-\delta T_j(k))^2}{8T(k)(1-T(k))}
\end{equation}
This equation describes "linear response" regime of operation of
the JTL detector, when its properties follow from the general
theory of linear detectors \cite{Ave03}.  In particular the
dephasing (\ref{jtl17}) can be understood as being caused by the
back-action noise arising from the randomness of the fluxon
transmission/reflecion process.

In the "tunnel" limit of weak transmission $|t_j(k)|\ll 1$, and
still under the assumption of cancelling scattering phases, the
equation (\ref{jtl16}) reduces to:
\begin{equation}\label{jtl18}
    \Gamma_{ij}=(f/2)\int dk
    |b(k)|^2\left(|t_i(k)|-|t_j(k)|\right)^2.
\end{equation}
The last two equations have corresponding analogues in the case of
QPC detectors \cite{Ale97,Lev97,Gur97}.

As the last example, that does not have an analogue in the QPC
physics, we consider the situation when the reflection is
negligible, $|r_k(t)|\equiv 0$, and the system-JTL interaction
modifies only the phases $\chi_j(k)$ of the transmission
amplitudes: $t_j(k)=e^{i\chi_j(k)}$, so that
\begin{equation}\label{jtl19}
    \Gamma_{ij}=-f\ln\left|\int
    dk|b(k)|^2\exp[\chi_i(k)-\chi_j(k)]\right|.
\end{equation}
In coordinate representation this means that the scattering only
affects the position and the shape of the transmitted wave packet.
If the width $\delta k$ of initial fluxon state in the momentum
representation is sufficiently narrow, and the phases $\chi_j(k)$
do not vary strongly over the momentum range, they can be
approximated as
\begin{equation}\label{jtl20}
    \chi_j(k)=\chi_j(k_0)-(k-k_0)x_j,\quad x_j\equiv \chi_j'(k_0).
\end{equation}
This approximation neglects the distortion of the wave packet's
shape during the scattering process, while it does take into
account the potential induced shift $x_j$ along the coordinate
axis.  This shift can directly be related to the "time delay"
$\tau_j=x_j/u_0$ due to scattering. The value of the delay can
take on positive and negative values depending on the scattering
potential. The dephasing rate equation (\ref{jtl19}) is
conveniently expressed in coordinate representation in the terms
of initial fluxon wave packet $\psi_0(x)$:
\begin{equation}\label{jtl21}
    \Gamma_{ij}=-f\ln \left|\int dx
    \psi_0(x-x_i)\psi^*_0(x-x_j)\right|.
\end{equation}
This expression explicitly shows that the back-action dephasing by
the JTL detector arises form the entanglement between the measured
system and the scattered fluxons which are shifted in time by an
interval dependent on the state of the measured system.  The
degree of suppression of coherence between the different states of
the measured system in scattering of one fluxon is determined then
by the magnitude of the relative shift of the fluxon in these
states on the scale of wave packet width.  For instance, if the
wave packet of initial fluxon is Gaussian (\ref{jtl10}) the
equation (\ref{jtl21}) gives:
\begin{equation}\label{jtl22}
    \Gamma_{ij}=f\frac{(x_i-x_j)^2}{4w^2}.
\end{equation}

Back-action dephasing represents only a part of the measurement
process.  The other part is the information acquisition by the
detector about the state of the measured system. In the case of
the JTL detector, this information is contained in the scattering
characteristic of fluxons, and the rate of its acquisitions
depends on the specific characteristics recorded by the fluxon
receiver.  There are at least two possibilities in this respect.
One is to detect the probability of fluxon transmission through
the scattering region (or, equivalently, the corresponding
probability of the fluxon reflection back to the generator).
Another possible detection scheme is for the receiver to measure
the time delay associated with the fluxon propagation through the
JTL.  Even if the measured system changes the potential $U_j(x)$
in such way that the fluxon transmission probability is not
affected, the potential can still change the fluxon's propagation
time, which will contain the information about the state of the
measured system.  In general, one can have a situation when the
information is contained both in the changes of propagation time
and transmission probability, and in order not to lose any
information one would need to detect both scattering
characteristics. In this work, we consider only the two "pure"
cases of transmission and time-delay detection modes.

\subsection{Transmission Detection Mode}

If the fluxon receiver of the JTL detector records only the fact
of fluxon arrival at the recever, then only the modulation of the
fluxon transmission probability by the measured system conveys
information about the system.  The information contained in all
other features of the scattering amplitudes (e.g. their phases
\cite{Ave03} or propagation time) is lost in the receiver.  In
this case the rate of information acquisition can be calculated
simply by starting with the probabilities of
transmission/reflection $T_j$ and $R_j$, when the measured system
is in the state $|q_j\rangle$:
\begin{equation}\label{jtl23}
    T_j=\int dk|b(k)|^2|t_j(k)|^2,\quad R_j=1-T_j.
\end{equation}

The independence among the different fluxon events implies that
the probability to have $n$ transmitted out of $N$ incident
fluxons is given by binomial distribution
\begin{equation}\label{jtl24}
    p_j(n)=C^n_NT^n_jR^{N-n}_j.
\end{equation}
The task of distinguishing different states  $|q_j\rangle$ of the
measured system is transformed into distinguishing the probability
distribution  $p_j(n)$ for different $j$'s.  If the fluxons are
injected into the JTL periodically with frequency $f$,  the number
of the scattering events increases with time, $N=ft$, and the
probability distribution (\ref{jtl24}) becomes more and more
strongly peaked  around the average number $NT_j$ of transmitted
fluxons.  Thus the states with different $T_j$ are distinguishable
with increasing certainty.   The rate of the increase of this
certainty can be characterized quantitatively by some overlap of
different probability distributions $p_j(n)$.  While there are
different ways to characterize the overlap of different
probability distributions  \cite{Nie00}, the characteristic which
is appropriate in quantum measurement context \cite{Woo81,Kor03}
is closely related to "fidelity" in quantum information
\cite{Nie00}: $\sum_n[p_i(n)p_j(n)]^{1/2}$.  The rate of
information acquisition is then naturally defined as
\begin{equation}\label{jtl25}
    W_{ij}=-1/t\ln\sum_n[p_i(n)p_j(n)]^{1/2},
\end{equation}
which after substitution of (\ref{jtl24}) becomes:
\begin{equation}\label{jtl26}
    W_{ij}=-f\ln\left[\sqrt{T_iT_j}+\sqrt{R_iR_j}\right],
\end{equation}
where the transmission and reflection probabilities are given in
(\ref{jtl23}).

The equation (\ref{jtl26}) characterizes the information
acquisition rate of the JTL detector in transmission detection
mode.  For an arbitrary detector the information acquisition is
smaller or equal to the back-action dephasing, and the regime when
the two rates are equal is quantum-limited.  Comparing equations
(\ref{jtl16}) and (\ref{jtl26}) one can see that for our detector,
indeed
\begin{equation}\label{jtl27}
    W_{ij}\le\Gamma_{ij},
\end{equation}
and that the equality golds if several conditions are satisfied.
First two conditions require that there is no information in the
phases of the transmission amplitudes:
\begin{eqnarray}
  \phi_i(k)=\phi_j(k), && \phi_j(k)\equiv\arg(t_j(k)/r_j(k)) \label{jtl28}\\
  \chi_j(k)&-&\chi_i(k)=const. \label{jtl29}
\end{eqnarray}
Equation (\ref{jtl29}) means that the difference of the phases
$\chi_j(k)$ of the transmission amplitudes should be effectively
independent of the momentum $k$ in the relevant range set by the
fluxon distribution $|b(k)|^2$ over $k$.  The two conditions
(\ref{jtl28}) and (\ref{jtl29}) have different physical origin.
The first implies that the scattered states contain no information
on $|q_j\rangle$ that can be obtained in principle by arranging an
interference between the transmitted and reflected parts of the
wave function \cite{Ave03}.  In the practical terms, the safest
way to satisfy this requirement is to make the scattering
potential symmetric,i.e. $U_j(-x)=U_j(x)$ for all states
$|q_j\rangle$.  The unitarity of the scattering matrix for the
fluxon scattering in the JTL implies in this case that
$\phi_j=\pi/2$ for all $j$'s.

Condition (\ref{jtl29}) means that no information on $|q_j\rangle$
is contained in the shape or position of the transmitted wave
packet that would be lost in the fluxon receiver operating on the
transmission-detection mode.  The similar condition for the
reflected fluxon is implied by both requirements (\ref{jtl28}) and
(\ref{jtl29}).  In general, condition (\ref{jtl29}) requires that
the spread of the initial wave packet $\delta k$ in the momentum
space gives the rise to the energy uncertainty of the fluxon
$\delta \epsilon\simeq u_0\delta k$, that is much smaller than the
energy scale $\Omega$ of the transparency variation of the
scattering potential $U_j(x)$.

One more condition of the quantum-limited operation of the JTL
detector in the transmission-detection mode is that the fluxon
transmission probabilities are effectively momentum and therefore
energy independent in the respective  relevant ranges:
\begin{equation}\label{jtl30}
    |t_j(k)|^2=T_j.
\end{equation}
This condition requires that $\delta \epsilon\ll\Omega$.  It is
more restrictive than the corresponding condition for the QPC
detector which can be quantum-limited even in the case of
energy-dependent transmission probability
\cite{Ave03,Pil02,Cle03}.  To obtain equation (\ref{jtl30}) one
starts from the back-action dephasing rate (\ref{jtl16}) which can
be written as
\begin{equation}\label{jtl31}
    \Gamma_{ij}=-f\ln\int
    dk|b(k)|^2\left[\left|t_i(k)t_j^*(k)\right|+\left|r_i(k)r_j^*(k)\right|\right]
\end{equation}
under the conditions (\ref{jtl28}) and (\ref{jtl29}).  Schwartz
inequality for the functions $|t_j(k)|$:
\begin{equation*}
    \sqrt{T_iT_j}\ge\int dk|b(k)|^2|t_i(k)t_j(k)|,
\end{equation*}
and the similar inequality for $|r_j(k)|$ show that $\Gamma_{ij}$
from (\ref{jtl31}) and the information acquisition rate $W_{ij}$
satisfy the inequality (\ref{jtl27}).  Furthermore, the equality
is reached only when
\begin{equation}\label{jtl32}
    |t_j(k)|=\xi_jt(k),\quad |r_j(k)|=\xi'_jr_j(k),
\end{equation}
and the ratios $|t_i(k)|/|t_j(k)|$ and $|r_i(k)|/|r_j(k)|$ are
independent of $k$.  Similarly to (\ref{jtl29}) equations
(\ref{jtl32}) demand that no information about the state of the
measured system is contained in the shape of transmitted or
reflected wave packets.  In general, when both transmission and
reflection probabilities are not small the two relations in
(\ref{jtl32}) are incompatible.  They have only a trivial solution
in which all amplitudes are independent of $k$ in the relevant
range of fluxon momentum distribution, thus proving the equation
(\ref{jtl30}) for $T_j \sim R_j\sim 1/2$.  The transmission and
reflection probabilities have the roughly the same magnitude  when
the fluxon energy $\epsilon$ is close to the maximum $U$ of the
scattering potential.  In this case, small spread of $\epsilon$:
$\delta\epsilon\ll\Omega$ implies that the characteristic range of
the scattering potential should be small: $a\ll w$.

The condition (\ref{jtl30}) of the quantum-limited operation is
not necessary when either $T_j\ll 1$ or $R_j\ll 1$.  In this case
one of the relations in (\ref{jtl32}) applies to the linear
response regime, when the variations $\delta T_j$ of JTL
transparencies between the different states $|q_j\rangle$ are
small and the back-action dephasing is given by (\ref{jtl17}).
Expanding (\ref{jtl26}) in $\delta T_j$: $T_j=T+\delta T_j$, where
all transparencies are defined as in (\ref{jtl23}), we get:
\begin{equation}\label{jtl33}
    W_{ij}=f\frac{(\delta T_i-\delta T_j)^2}{8T(1-T)}.
\end{equation}
This equation differs from the equation (\ref{jtl17}) only by the
order in which the integration over the momentum is performed.
This means that the information acquisition and dephasing rates
satisfy in general the inequality (\ref{jtl27}) and are equal only
if the transparency of JTL is constant in the relevant momentum
range.

In the linear response regime, each individual fluxon carries only
small amount of measurement information,  and it is convenient to
employ the quasi-continuous description in which the fluxon
receiver acts as the voltmeter registering not the individual
fluxons, but the rate of arrival of many fluxons represented by
voltage $V(t)$ across the segment junctions of JTL.  The average
voltage for state $|q_j\rangle$ is
\begin{equation}\label{jtl34}
    \langle V(t)\rangle=fT_j\Phi_0,
\end{equation}
where $\langle\cdots\rangle$ implies average over the scattering
outcomes and time $t$ of the fluxon injection cycle.  Because of
the randomness of the fluxon scattering, the actual voltage
fluctuates around the average value (\ref{jtl34}) even at low
temperatures.  These fluctuations can be attributed to the shot
noise of fluxon and its spectral density

\begin{equation}\label{jtl35}
    S_V(\omega)=\int d\tau e^{-i\omega\tau}
    \left(\langle V(t+\tau)V(t)\rangle -\langle
    V(t)\rangle^2\right),
\end{equation}
which is constant at frequencies $\omega$ below the fluxon
injection frequency $f$.    Straightforward calculation similar to
that for regular shot noise shows that the constant value of the
spectral density is
\begin{equation}\label{jtl36}
    S_0=fT(1-T)\Phi_0^2,
\end{equation}
where in the linear response regime we can neglect the small
differences $\delta T_j$ of transparencies between the different
states $|q_j\rangle$ in the expression for the noise.  The
equation shows that in accordance with the general theory of
linear quantum measurements \cite{Ave03}, the information rate
(\ref{jtl33}) as the rate with which  one can distinguish dc
voltage values (\ref{jtl34}) in the presence of white noise with
spectral density (\ref{jtl36}).

To conclude this subsection, we arrive the the same conclusions by
following the conditional evolution approach. This is easily
accomplished by looking at the evolution of the density matrix
elements in the conditional evolution as given in (\ref{qm7c}) and
(\ref{qm7d}). After applying the requirement that the qubit stays
"pure" during the ideal measurement (i.e.
$\rho_{00}\rho_{11}=|\rho_{01}|^2$) the cross-integrals of  the
transmission/reflection  rates  need to satisfy:
\begin{eqnarray*}
  \int dk|b(k)|^2t_i(k)t^*_j(k) \cdot \int dk|b(k)|^2t_j(k)t^*_1(k) &=& 1, \\
  \int dk|b(k)|^2r_i(k)r^*_j(k) \cdot \int dk|b(k)|^2r_j(k)r^*_1(k) &=&
  1.
\end{eqnarray*}
There two conditions are only satisfied if the rates $t_i,r_i$ are
momentum independent in the range of the injected fluxons and if
the phases of the rates  are the same. This is equivalent to the
conditions (\ref{jtl28}) and (\ref{jtl32}), which leads to the
main conclusion of this section that in transmission-detection
mode the most relevant regime $T_j\simeq 1/2$ of maximal detector
response to the input signal, the conditions of quantum-limited
operation are given by conditions (\ref{jtl28}) and (\ref{jtl30}).

\subsection{Time-Delay Detection Mode}

Since the range $a$ of the scattering potential (\ref{jtl6})
cannot be smaller than the fluxon size $\lambda_J$ it might be
difficult in practice to satisfy the $a\ll w$ condition needed for
the quantum-limited operation of the JTL detector in the
transmission-detection mode.  For quasiclassical potential
barriers that are smooth on the scale of the fluxon wave packet,
$a\ge w$, the "transition" region of energies near the top of the
barrier, where the reflection and transmission amplitudes have
comparable magnitude, is narrow.   If the interval $\delta
\epsilon$ of the energies avoids this narrow region, then either
the transmission or reflection coefficient can be neglected.
Ballistic motion of the fluxons in this region still contains
information about the potential $U_j(x)$ that can be used for
measurement.  This information is contained in the time shift
$\tau_j$ caused by the propagation though the region with
non-vanishing potential rather than in transmission/reflection
probabilities. Quantum mechanically, the time-shift information is
contained in the phases of the scattering amplitudes.  To be
specific, we discuss here the regime where the JTL detector is
operated in time-delay mode using the transmitted fluxons,
$|t_j(k)|= 1$.  In the energy range where $|r_j(k)|=1$, the same
detection process is possible using the reflected fluxons.  The
only advantage of the transmission case is the possibility to use
the fill range of values of the scattering potential:
$U_j(x)\lessgtr 0$.

For sufficiently smooth $U_j(x)$ the phase $\chi_j(k)$ of the
transmission amplitude can be calculated in quasi-classical
approximation:
\begin{equation}\label{jtl37}
    \chi_j(k)=\int dx \left[2m(\epsilon(k)-U_j(x))\right]^{1/2}.
\end{equation}
If the potential is weak, $U_j(x)\ll \epsilon$, the
potential-induced contribution to the phase (\ref{jtl37}) can be
expressed as
\begin{equation}\label{jtl38}
    \chi_j(k)=-\frac{1}{v(k)}\int dx U_j(x), \quad
    v(k)\equiv\sqrt{\frac{2\epsilon(k)}{m}}.
\end{equation}

Under the adopted assumption of quasiclassical potential and
vanishing reflection, the condition (\ref{jtl9}) of the negligible
broadening of the fluxon wave packet during free propagation
through the JTL is still walid in the presence of the potential.
In this case one can use the approximation (\ref{jtl20}) for the
phases $\chi_j(k)$ which implies that the wave packet shifts as
whole and without distortion.  The potential induced part of the
shift $x_j=-\chi_j'(k_0)$ can be obtained from the equation
(\ref{jtl37}) and has the classical form:
\begin{equation}\label{jtl39}
     x_j=\int dx \left[1-\frac{u_0}{u_j(x)}\right], \quad
     u_j(x)=\sqrt{2[\epsilon(k)-U_j(x)]/m}.
\end{equation}
For weak potential, $U_j(x)\ll\epsilon$:
\begin{equation}\label{jtl40}
    x_j=-\frac{1}{2\epsilon}\int dx U_j(x).
\end{equation}

\begin{figure}
\begin{center}
  \includegraphics[scale=.66]{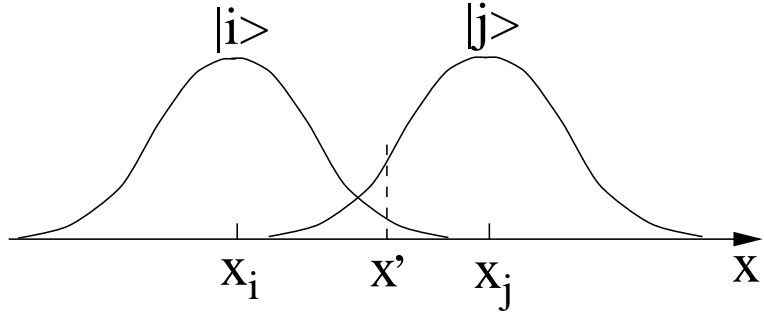}\\
  \caption{Illustration of the information acquisition process by the JTL detector
  in the time-delay mode.  The fluxon wave packet is shifted by the distance $x_j$
  dependent on the state $|q_j\rangle$ of the measured system.  In conditional description,
  observation of the fluxon with position $x'$ changes the system wave function according
  to equation (\ref{jtl45}).}\label{fig6.2}
\end{center}
\end{figure}

Back-action dephasing rate by the JTL detector in this regime is
given by equations (\ref{jtl21}) and (\ref{jtl22}).  The
information about the states $|q_j\rangle$ contained in the shift
$\tau_j$ of the fluxon in time or equivalently in coordinate
$x_j=u_0\tau_j$, can be read-off by distinguishing different
shifts $x_j$ against the background of finite width $w$ of the
fluxon wave packet $\psi_0(x)$.  Since $|\psi_0(x)|^2$ gives the
probability of finding the fluxon at coordinate $x$, this task is
equivalent to the task of distinguishing two shifted probability
distributions  (see figure \ref{fig6.2}) that was discussed above
for the transmission-detection mode.  Similarly to equation
(\ref{jtl25}) we can write the information acquisition rate of the
JTL detector in the time-delay mode as follows:
\begin{equation}\label{jtl41}
    W_{ij}=-f\int dx |\psi_0(x-x_i)\psi_0(x-x_j)|.
\end{equation}
Comparing this to the dephasing rate (\ref{jtl21}), we see that in
general the two rates satisfy the inequality (\ref{jtl27}) as they
should.  The rates are equal if the phases of the initial packet
$\psi_0(x)$ of the injected fluxon is independent of x, i.e. if
$\psi_0(x)$ is real.  In particular case of the Gaussian wave
packet, the detector is quantum-limited ($W_{ij}=\Gamma_{ij}$),
and the two rates are given by (\ref{jtl22}). These considerations
also imply that the JTL detector in the time-delay mode would
loose the property of being quantum-limited if the fluxon wave
packet spreads noticeably during the propagation through JTL. This
process creates non-trivial $x$-dependent phase of the wave packet
and makes the information acquisition rate smaller than the
back-action rate.

The  derivation of the conditional evolution equations introduced
in the last chapter can be repeated with only minor modifications
in time-delay mode of detector operation.  In this case,the
different classical outcomes of measurements are the observed
instances of time when the fluxon reaches the receiver, that for
convenience can be directly translated into different fluxon
positions $x$ at some fixed time. If the fluxon is observed at the
position $x'$ in a given injection cycle (figure \ref{fig6.2}),
the evolution of the amplitude $\alpha_j$  of the measured system
due to interaction with the fluxon is:
\begin{equation}\label{jtl45}
    \alpha_j\rightarrow \frac{\psi_0(x'-x_j)\alpha_j}{\sqrt{\sum_i|\alpha_i\psi_0(x'-x_i)|^2}}
\end{equation}
Qualitatively, and similarly to the conditional evolution in the
transmission-detection mode, the sequence of transformations
(\ref{jtl45}) describes weak measurement: gradual localizationof
the system wave function to one of the states $|j\rangle$ with
increasing number of fluxon scattering events which lead to the
gradual accumulation of the information about the system.  In
contrast to the transmission-detection mode, conditional evolution
mode (\ref{jtl45}) always remains pure under the symplifying
assumptions adopted in this work for the JTL detector: coherent
propagation of fluxons with wave packet of fixed form $\psi_0(x)$
injected into the JTL with the same energy.

\section{Non-Quantum-Limited Detection}

Quantum-limited operation of the JTL detector discussed in the
preceding sections requires quantum coherent dynamics of fluxons
in the JTL. While this dynamics can be observed experimentally
\cite{Wal03}, the task of realizing it is very difficult.  From
this prospective it is important that several attractive features
of the JTL detector, e.g. large operating frequency and reduced
parasitic dephasing during the time intervals between the fluxon
scattering, remain even in the non-quantum-limited regime.  Our
discussion above was limited on the fundamental back-action
dephasing which in its quantum limit is unavoidable part of
measurement process.  Although the Josephson junctions of JTL and
those of "external" parts of the JTL detector (generator and
receiver) can give rise to dissipation and dephasing not related
directly to measurement, in the JTL geometry (figure \ref{fig6.1})
parasitic dissipation is suppressed due to the screening by the
supercurrent flow in the JTL junctions \cite{Sem03,Sav05}.

The dominant deviation from the quantum limited detection should
be associated then with the fluctuations in the fluxon motion.
These fluctuation make the dephasing factor $\eta$ (\ref{jtl15})
due to scattering larger than the amount of the information
conveyed by the fluxon scattering.  Some interesting measurement
strategies are still possible with  the JTL non-ideality of this
type.  The most natural example is QND measurement briefly
discussed in the previous chapter (e.g. see \cite{Ave02,Jor05})
which in principle make the detector back-action irrelevant. The
possibility of timing the fluxon scattering in the JTL detector
makes it particularly suitable for the "kicked" version
\cite{Jor05} of the QND qubit measurement.

Consider the qubit Hamiltonian

\begin{equation}\label{jtl46}
    H=-\frac{\hbar\Delta}{2}\sigma_x
\end{equation}
which performs quantum coherent oscillations with frequency
$\Delta$ . The qubit is coupled to the JTL through its $\sigma_z$
operator, i.e. the states $|q_j\rangle$ used in the discussion in
the preceding sections, are the two eigenstates of $\sigma_z$. We
assume that the Hamiltonian (\ref{jtl46}) already includes
renormalization of parameters due to the qubit-detector coupling.
If the qubit oscillations are weakly dephased at the rate $\gamma
\ll \Delta$ (e.g., by residual parasitic dissipation in the JTL
detector), the time evolution of the qubit density matrix $\rho$
during the time intervals between the successive fluxon scattering
processes can be written in the following form in the
$\sigma_z$-representation:
\begin{equation}\label{jtl47}
    \rho(t)=\frac{1}{2}\left[1+e^{-\gamma t} \left(%
\begin{array}{cc}
  x & -iy \\
  iy & -x \\
\end{array}%
\right)\right],
\end{equation}
\begin{equation}\label{jtl48}
    \dot{r}=-i\Delta r,\quad r=x+iy,
\end{equation}
where $r(t=0)=\pm 1$ depending on weather the qubit starts at
$t=0$ from the $\sigma_z=1$ or $\sigma_z=-1$ state.

\begin{figure}
\begin{center}
  \includegraphics[scale=0.55]{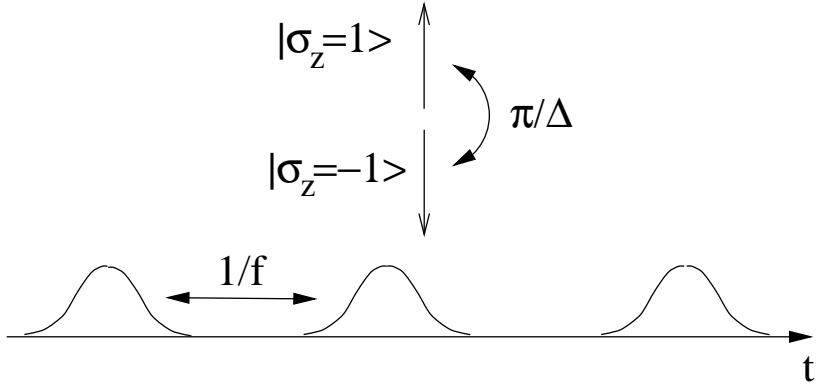}\\
  \caption{Schematics of the QND fluxon measurement of a qubit which suppresses the
  effect of back-action dephasing on its coherent quantum oscillations.  The fluxon
  injection frequency $f$ is matched to the qubit oscillation frequency $\Delta$:
  $f\simeq \Delta$, so that the individual acts of measurement are done when the
  qubit density matrix is nearly diagonal in the $\sigma_z$ basis, and the
  measurement back-action does not introduce dephasing in the oscillation dynamics.}
  \label{fig6.3}
\end{center}
\end{figure}

The fluxon scattering at times $t_n=n/f$ leads to suppression of
the off-diagonal elements of $\rho$:
\begin{equation}\label{jtl49}
    y(t_n^+)=\eta y(t_n^-).
\end{equation}
However, if the fluxons are injected in the JTL with the time
interval $1/f$ close to the half-period $\pi/\Delta$ of the qubit
oscillations (figure \ref{fig6.3}), the qubit density matrix
(\ref{jtl48}) is nearly diagonal at the moments of fluxon
scattering, $y(t_n)\ll 1$, and suppression (\ref{jtl49}) does not
affect strongly the qubit oscillations.

Such a QND qubit measurement is possible with the JTL detector
operating in either detection mode; to be specific we assume the
transmission mode. The dependence of the fluxon transmission
probability on the qubit state can be written then as $T +
\sigma_z\delta T$ . We consider the linear-response regime,
$\delta T\ll T$, when the scattering of one fluxon provides only
small amount of information about the qubit oscillations. For
quantum-limited detection, the linear-response condition $\delta
T\ll T$ implies that $\eta \rightarrow 1$. In the
non-quantum-limited case, the back-action can be stronger, and we
take $\eta$ to be an arbitrary factor within the $[0, 1]$
interval.

If the frequency $f$ is matched precisely to the qubit
oscillations, $f = \Delta/\pi$, the matrix (\ref{jtl48}) is
diagonal at the times of measurement, $y(t_n) = 0$, and detector
does not affect at all the qubit dynamics. If the mismatch is non-
vanishing but small, $\delta=\Delta/f-\pi\ll 1$, diagonal elements
of $\rho$ (\ref{jtl48}) evolve quasi-continuously even if
suppression factor $\eta$ is not close to $1$. Equations
(\ref{jtl48}) and (\ref{jtl49}) give the following equation for
this quasi-continuous evolution:
\begin{equation}\label{jtl50}
    \dot{x}=-\left(\frac{1+\eta}{1-\eta}\right)\frac{f\delta^2}{2}x.
\end{equation}

In the assumed linear-response regime, the qubit oscillations
manifest themselves as a peak in the spectral density $S_V
(\omega)$ (\ref{jtl35}) of the voltage $V$ across the JTL
junctions. For $f\simeq \Delta/\pi$ the oscillation peak in $S_V
(\omega)$ is at zero frequency. Equations (\ref{jtl47}) and
(\ref{jtl50}) describing the decay of correlations in the qubit
dynamics in the $\sigma_z$ basis imply that the oscillation peak
has Lorentzian shape:

\begin{equation}\label{jtl51}
    S_V(\omega)=S_0+\frac{2\Gamma f^2(\delta
    T)^2\Phi_0^2}{\omega^2+\Gamma^2},
\end{equation}
and the oscillation line width $\Gamma$ is
\begin{equation}\label{jtl52}
    \Gamma=\gamma+\frac{1+\eta}{1-\eta}\frac{(\Delta-\pi
    f)^2}{2f}.
\end{equation}

For the quantum-limited detection, $\eta \rightarrow 1$, the
equation (\ref{jtl52}) reproduces previous results for the QND
measurement \cite{Jor05} if one introduces the back-action
dephasing rate $\Gamma_d=f(1-\eta)$. In this case, equation
(\ref{jtl52}) is valid for sufficiently small mismatch between the
measurement and oscillation frequencies, $|\delta|\ll1-\eta$. For
larger $\delta$, the oscillation peak in the detector output $S_V
(\omega)$ moves to finite frequency $\Delta-\pi f$ and the QND
nature of the measurement is lost \cite{Ave02}. Equation
(\ref{jtl52}) shows also that in the limit of "projective"
measurements $\eta = 0$, broadening of the oscillation peak is
weaker, and the peak remains at zero frequency for all reasonable
values of the detuning parameter $|\delta|\ll1$. Therefore, the
stronger back-action of the JTL detector is advantageous for the
QND measurements of coherent qubit oscillations. The last remark
is that although our discussion here assumed that the fluxon
arrival times are spaced exactly by $1/f$,  equations
(\ref{jtl51}) and (\ref{jtl52}) should remain valid even in the
presence of small fluctuations of the measurement times. These
fluctuation can be described by taking into account that the
detuning $|\delta|$ cannot be made smaller that the relative line
width of the fluxon generator.

This concludes the discussion about the ballistic JTL detector for
the flux qubits - the last part of this thesis.

\bibliographystyle{unsrt}
\bibliography{thesis-bib}

\end{document}